%% file: JSTSP_OPT_SAM.tex
\documentclass[journal,transmag]{IEEEtran}
\input{preamble}

\ifCLASSINFOpdf
\else
\fi
\begin{document}
\title{Learning-Based Compressive Subsampling} 

\author{
    \IEEEauthorblockN{Luca Baldassarre, 
    Yen-Huan Li, 
    Jonathan Scarlett, 
    Baran G\"ozc\"u,
    Ilija Bogunovic and
    Volkan Cevher
    }
    \IEEEauthorblockA{Laboratory for Information and Inference Systems \\
    \'Ecole Polytechnique F\'ed\'erale de Lausanne, Switzerland}
%     
%    \thanks{This work was supported in part by the European Commission under grants MIRG-268398 and ERC Future Proof and by the Swiss Science Foun- dation under grants SNF 200021-146750 and SNF CRSII2-147633.  J.S.~acknowledges support from the ‘EPFL Fellows’ fellowship programme co-funded by Marie Skłodowska-Curie, Horizon2020 Grant agreement no. 665667.
%    }
} 

\markboth{IEEE JOURNAL OF SELECTED TOPICS IN SIGNAL PROCESSING}
{Shell \MakeLowercase{\textit{et al.}}: Bare Demo of IEEEtran.cls for Journals}

\maketitle

\begin{abstract}
    The problem of recovering a structured signal $\x \in \C^p$ from a set of dimensionality-reduced linear measurements $\b = \A\x$ arises in a variety of applications, such as medical imaging, spectroscopy, Fourier optics, and computerized tomography.  Due to computational and storage complexity or physical constraints imposed by the problem, the measurement matrix $\A \in \C^{n \times p}$ is often of the form $\A = \P_{\Omega}\bPsi$ for some orthonormal basis matrix $\bPsi\in \C^{p \times p}$ and subsampling operator $\P_{\Omega}: \C^{p} \rightarrow \C^{n}$ that selects the rows indexed by $\Omega$.  This raises the fundamental question of how best to choose the index set $\Omega$ in order to optimize the recovery performance.  Previous approaches to addressing this question rely on non-uniform \emph{random} subsampling using application-specific knowledge of the structure of $\x$.  In this paper, we instead take a principled learning-based approach in which a \emph{fixed} index set is chosen based on a set of training signals $\x_1,\dotsc,\x_m$.  We formulate combinatorial optimization problems seeking to maximize the energy captured in these signals in an average-case or worst-case sense, and we show that these can be efficiently solved either  exactly or approximately via the identification of modularity and submodularity structures.  We provide both deterministic and statistical theoretical guarantees showing how the resulting measurement matrices perform on signals differing from the training signals, and we provide numerical examples showing our approach to be  effective on a variety of data sets.
\end{abstract}


\IEEEdisplaynontitleabstractindextext
\IEEEpeerreviewmaketitle

\input{intro}
\input{optimal_sampling}

\input{gen_bounds}

\input{experiments}

\section{Conclusion and Discussion} \label{sec:CONCLUSION}
We have provided a principled learning-based approach to subsampling an orthonormal basis for structured signal recovery, thus providing a powerful alternative to traditional approaches based on parametrized random sampling distributions.  We proposed combinatorial optimization problems based on the average-case and worst-case energy captured in the training signals, obtained solutions via the identification of modularity and submodularity structures, and provided both deterministic and statistical generalization bounds.  Our numerical findings reveal that our approach performs well on a variety of data sets, while having the desirable features of avoiding any reliance on randomization, and permitting the use of an efficient and scalable linear decoder. 

While our numerical results support the use of the simple linear decoder in a variety of settings, we note that there are cases where it is unsuitable \cite{candes2006modern,candes2006robust}.  In particular, to perform well, we need to capture most of the energy in $\x$ using only $n$ measurements, and hence we require a degree of \emph{compressibility in the measurement basis $\bPsi$}.

A simple example where the linear decoder is not suitable is when $\x$ is sparse in the canonical basis, whereas the measurements are taken in the Fourier basis.  In this case, a large number of measurements is required to capture most of the energy, whereas for non-linear decoders, several results on the exact recovery of $s$-sparse signals are known using only $n = O(s\log^{\nu} p)$ measurements for some $\nu > 0$ \cite{foucart2013mathematical}.  See \cite{candes2006robust} for a related example with radial Fourier sampling, where for a particular set of measurements it is observed that non-linear decoding is highly preferable.  Based on these examples, extending our learning-based approach to non-linear decoders is a very interesting direction for further research.

A variety of other extensions would also be valuable, including the consideration of measurement noise, imperfectly-known training signals, more general recovery criteria beyond the $\ell_2$-error, and further measurement constraints motivated by applications (e.g., see \cite{boyer2014algorithm}).

\section*{Acknowledgment}

The authors gratefully acknowledge Marwa El Halabi for helpful discussions regarding submodularity and discrete optimization, the summer interns Vipul Gupta and Gergely Odor for providing insightful comments and helping with the experiments, and Dr. Bogdan Roman for providing the indices used in Figure 2. 

This work was supported in part by the the European Commission under grant ERC Future Proof, by the Swiss Science Foundation under grants SNF 200021-146750 and SNF CRSII2-147633, and by the `EPFL Fellows' programme under Horizon2020 grant 665667.

% \appendix
% \input{appendix}

\ifCLASSOPTIONcaptionsoff
  \newpage
\fi

\bibliographystyle{IEEEtran}
\bibliography{references_optimal_sampling}

%\end{IEEEbiography}

\end{document}

%% file: preamble.tex
\usepackage{graphicx}
\usepackage[cmex10]{amsmath}
\usepackage{amssymb,amsthm}
\usepackage{array}
\usepackage{url}
\usepackage{cite}
\usepackage{multirow}
\usepackage{color}
\usepackage{algorithm}
\usepackage[noend]{algpseudocode}

\DeclareMathOperator*{\argmin}{arg\,min}
\DeclareMathOperator*{\argmax}{arg\,max}

\usepackage{tikz,pgfplots}
\def\x{{\mathbf x}}

\def\b{{\mathbf b}}
\def\z{{\mathbf z}}
\def\P{{\mathbf P}}
\def\X{{\mathbf X}}

\def\V{{\mathbf V}}

\def\R{{\mathbb R}}
\def\C{{\mathbb C}}

\def\A{{\mathbf A}}

\def\P{{\mathbf P}}

\def\w{{\mathbf w}}

\def\V{{\mathbf V}}

\def\I{{ \mathbf I }}

\def\bpsi{{\boldsymbol{\psi}}}
\def\bPsi{{\boldsymbol{\Psi}}}

\def\bPhi{{\boldsymbol{\Phi}}}
\def\bone{{\boldsymbol{1}}}

\def\fmin{{f_{\mathrm{min}}}}
\def\favg{{f_{\mathrm{avg}}}}
\def\fgen{{f_{\mathrm{gen}}}}

\def\calA{{\mathcal A}}
\def\diag{{ \mathrm{diag} }}
\def\cmin{{ c_{\mathrm{min}} }}
\def\cmax{{ c_{\mathrm{max}} }}
\def\Omegabest{{ \Omega_{\mathrm{best}} }}
\def\GPC{{ \mathrm{GPC} }}
\def\xnew{{ \mathbf{x} }}
\def\PP{{ \mathbb P }}
\def\EE{{ \mathbb E }}

\def\EEhat{{ \hat{\mathbb E} }}

\def\calA{{ \mathcal A }}

\newtheorem{mydef}{Definition}
\newtheorem{thm}{Theorem}

\newtheorem{rem}{Remark}

%% file: intro.tex
%!TEX root = JSTSP_OPT_SAM.tex

\section{Introduction}

In the past decade, there has been a tremendous amount of research on the problem of recovering a structured signal $\x \in \C^{p}$ (e.g., a signal with few non-zero coefficients) from a set of dimensionality-reduced  measurements of the form
\begin{equation}
    \b = \A\x + \w,
\end{equation}
where $\A \in \C^{n \times p}$ is a known measurement matrix, and $\w \in \C^n$ represents unknown additive noise.  

This problem is often referred to as compressive sensing (CS), and is ubiquitous in signal processing, having an extensive range of applications including medical resonance imaging (MRI), spectroscopy, radar, Fourier optics, and computerized tomography \cite{candes2008intro}.  The key challenges in CS revolve around the identification of structure in the signal (e.g., sparsity), computationally efficient algorithms, and measurement matrix designs.  In this paper, we focus primarily on the latter. 

In the classical CS setting where $\x$ is sparse, the best theoretical recovery guarantees are obtained by random matrices with independent and identically distributed (iid) entries, along with sparsity-promoting convex optimization methods \cite{foucart2013mathematical}. However, such random matrices impose a significant burden in the data processing pipeline in terms of both storage and computation as the problem dimension $p$ increases.  

To overcome these limitations, significant attention has been paid to subsampled structured matrices of the form \cite{foucart2013mathematical}
\begin{equation}
    \A = \P_{\Omega} \bPsi, \label{eq:subsampling}
\end{equation}
where $\bPsi \in \C^{p \times p}$ is an orthonormal basis,  and $\P_{\Omega}: \C^{p} \rightarrow \C^{n}$, where $[\P_{\Omega} \x]_l = \x_{\Omega_l}$, $l= 1,\ldots,n$, is a subsampling operator that selects the rows indexed by the set $\Omega$, with $|\Omega| = n$.  Popular choices for $\bPsi$ include Fourier and Hadamard matrices, both of which permit efficient storage and computation.  

Measurements of the form \eqref{eq:subsampling} are also of direct interest in applications, such as MRI and Fourier optics, where one cannot choose the basis in which measurements are taken, but one does have the freedom in choosing a subsampling pattern.  Constraining the measurement matrix in this manner raises the following fundamental question: \emph{How do we best choose the index set $\Omega$?}  The answer depends on the structure known to be inherent in $\x$, as well as the recovery algorithm used.

For general sparse vectors, uniform random subsampling provides a variety of recovery guarantees under suitable assumptions on the incoherence of $\bPsi$ with the sparsity basis of the signal to be recovered \cite{foucart2013mathematical}.  However, it has been observed that this approach can be significantly outperformed in applications via \emph{non-uniform} random subsampling techniques, in which one takes more samples in the parts of the signal that tend to contain more energy, thus exploiting structured sparsity in the signal as opposed to sparsity alone \cite{lustig2007sparse}.  Explicit distributions on $\Omega$ are proposed in \cite{lustig2007sparse,roman2014asymptotic}, containing parameters that require tuning for the application at hand.

From a practical perspective, the randomization of $\Omega$ is typically not desirable, and perhaps not even possible. %; for example, one would expect that the indices need to be fixed at the time of designing an MRI machine.  
Moreover, there is strong evidence that in real-world applications there is a ``best'' choice of $\Omega$ that one should aim for. For instance, while the compression standard \cite{pennebaker1993jpeg} does not require it, the subsampling for JPEG in the discrete cosine transform (DCT) domain is typically chosen independently of the image being compressed, in order to simplify the encoder.  The indices for MRI systems, which are based on Fourier measurements, are often randomly chosen based on matching image statistics and performing manual tuning, and are then fixed for later scans \cite{lustig2007sparse,adcock2013breaking,roman2014asymptotic,boyer2014algorithm}.  Despite its importance, principled approaches to the ``best'' index selection problem appear to be lacking. % since the problem is combinatorial \cite{lustig2007sparse} and we do not hope to solve it in general. 

To bridge this gap, our paper takes a \emph{learning-based} approach to subsampling via training signals and combinatorial optimization problems.  The high-level idea is simple: We  select the indices that preserve as much energy as possible in a set of training signals, either in a worst-case or average-case sense, and we show that this is also equivalent to minimizing the $\ell_2$-error achieved by a simple linear decoder. By identifying combinatorial optimization structures such as submodularity,  we show that we can find exact or near-exact solutions to these optimization problems in polynomial time.  % that are endowed with strong theoretical generalization capabilities. 
%In this paper, we provide a \emph{learning-based} approach to optimizing $\Omega$ based on a set of training signals. 
% These notions lead to combinatorial optimization problems, and in each case we  identify discrete structures that permit an approximate or exact solution to be found efficiently, such as submodularity.  
We then provide both deterministic and statistical theoretical guarantees characterizing how well the selected set of indices perform when applied to new signals differing from the training set.  Finally, we demonstrate the effectiveness of our approach on a variety of data sets, showing matching or improved performance compared to \cite{lustig2007sparse,adcock2013breaking,roman2014asymptotic} in several imaging applications. 

Both our theory and our recovery results are based on the use of highly efficient {linear} encoder \emph{and decoder} pairs, and we provide some examples that challenge the conventional wisdom in CS that non-linear decoding methods, such as basis pursuit, are necessary for reliable recovery \cite{candes2006modern}. Indeed, we find that although such methods are needed when considering \emph{arbitrary} sparse or compressible signals, we may avoid them in certain settings where the signals and the sampling mechanisms exhibit more specific structures.  Nevertheless, non-linear extensions of our work would also be of considerable interest; see Section \ref{sec:CONCLUSION} for further discussion.

Our main theoretical contribution is our statistical generalization bound, which states that when the training and test signals are independently drawn from \emph{any} common distribution, the average $\ell_2$-norm error of the above-mentioned linear decoder for the test signal can be made arbitrarily close to the best possible, including randomized strategies, when the number of training signals exceeds a value depending on the complexity of the set constraints, e.g., $O\big(n \log \frac{p}{n}\big)$ training signals suffice when there are no constraints.

% \vspace{-3mm}
\subsection{Problem Statement} \label{sec:SETUP}
For clarity of exposition, we focus on the noiseless case throughout the paper, and consider the measurement model
\begin{equation}
    \b = \P_{\Omega} \bPsi \x \label{eq:model}
\end{equation}
for some orthonormal basis matrix $\bPsi \in \C^{p \times p}$, and some subsampling matrix $\P_{\Omega}$ whose rows are canonical basis vectors. 

Given a set of $m$ training signals $\x_1,\dotsc,\x_m$, we seek to select an index set $\Omega$ permitting the recovery of further signals that are in some sense similar to the training signals; see Section \ref{sec:GEN_BOUNDS} for some formal notions of similarity. We assume without loss of generality that $\|\x_j\|_2 = 1$ for all $j$.  While our focus is on the selection of $\Omega$, it is clear that the recovery algorithm also plays a role towards achieving our goal.  

\textbf{Linear Decoding:} The main procedure that we consider simply expands $\b$ to a $p$-dimensional vector by placing zeros in the entries corresponding to $\Omega^c$, and then applies the adjoint $\bPsi^* = \bPsi^{-1}$:
\begin{equation}
    \hat{\x} = \bPsi^* \P_{\Omega}^T \b. \label{eq:linear_dec}
\end{equation}
Note that this is a \emph{linear} decoder, and can be implemented highly efficiently even in large-scale systems for suitably structured matrices $\bPsi$ (e.g., Fourier or Hadamard).  In fact, it is easily shown to be equivalent to the \emph{least-squares} decoder, i.e., the pseudo-inverse of $\P_{\Omega}\bPsi$ is $\bPsi^* \P_{\Omega}^T$ for unitary $\bPsi$.

\textbf{Basis Pursuit Decoding:} If $\x$ is known to be approximately sparse in some known basis, i.e., $\x = \bPhi^*\z$ for some approximately sparse vector $\z$ and basis matrix $\bPhi$, then stable and robust recovery is possible using standard CS algorithms.  A particularly popular choice is basis pursuit (BP), which estimates
\begin{equation}\label{eq: BP}
    \hat{\z} = \argmin_{\tilde{\z} \,:\, \b = \P_{\Omega} \bPsi \bPhi^* \tilde{\z}} \|\tilde{\z}\|_1
\end{equation}
and then sets $\hat{\x} = \bPhi^* \hat{\z}$. We can also replace the basis pursuit recovery with other convex programs that leverage additional ``structured'' sparsity of the coefficients, e.g., see \cite{elhalabi2014tu}.

We will use both decoders \eqref{eq:linear_dec} and \eqref{eq: BP}  in our numerical results in Section \ref{sec:NUMERICAL}, and in fact see that they behave similarly in all of the examples therein, despite BP having a significantly higher computational complexity.  

\subsection{Related Work}

\textbf{Variable-density Subsampling:} To the best of our knowledge, all previous works on variable-density subsampling have considered randomized choices of $\Omega$ and sought the corresponding distributions, rather than considering fixed choices.  A common approach is to design such distributions using empirically-observed phenomena in images \cite{lustig2007sparse,roman2014asymptotic} or adopting specific signal models such as generalized Gaussian \cite{wang2010variable}. %or generalized Pareto \cite{volkanNIPS2009,compressibledistributions}.  
In each of these works, the proposed designs involve parameters that need to be learned or tuned, and we are not aware of any efficient principled approaches for doing so.  An alternative approach is taken in \cite{puy2011variable} based on minimizing coherence (see also \cite{elad2007optimized} for a related work with unstructured matrices), but the optimization is only based on the measurement and sparsity bases, as opposed to training data.

\textbf{Learning-based Measurement Designs:} Several previous works have proposed methods for designing \emph{unstructured} measurement matrices based on training data, with a particularly common approach being to seek the best restricted isometry property (RIP) constant \cite{foucart2013mathematical} with respect to the training data.  In \cite{hedge2015numax}, this problem is cast as an affine rank minimization problem, which is then relaxed to a semidefinite program.  A different approach is taken in \cite{sadeghian2013embeddings}, where the affine rank minimization problem is reformulated and approximated using game-theoretic tools.  An alternating minimization approach is adopted in \cite{bah2014metric} in the context of metric learning.  A drawback of all of these works is that since the measurement matrices are unstructured, they suffer from similar storage and computation limitations to those of random matrices with iid entries.

% \vspace{-4mm}
\subsection{Contributions}
The main contributions of this paper are as follows:
\begin{itemize}
    \item We present a class of optimization problems for selecting the subsampling indices based on training data, seeking to choose those that capture as much energy as possible with respect to the average case or worst case, or equivalently, that minimize the $\ell_2$ error achieved by a simple linear decoder (see Section \ref{sec:MOTIVATION}).  For each of these, we identify inherent modularity and submodularity structures, thus permitting exact or approximate solutions to be obtained efficiently using existing algorithms.
    \item For the average-case criterion, we show that a simple sorting procedure achieves the exact solution.  Moreover, we show that several additional constraints on the subsampling pattern can be incorporated while still permitting an efficient solution, including matroid constraints \cite{schrijver2003combinatorial} and wavelet-tree based constraints \cite{baranuik2010model}.
    \item We provide theoretical results stating how the selected indices perform when applied to a test signal not present in the training set.  For both the average-case and worst-case criteria, this is done from a deterministic perspective, assuming the new signal to be sufficiently close to the training data in a sense dictated by $\Omega$.  For the average-case criterion, we also provide bounds from a statistical perspective, assuming the data to be independently drawn from an unknown probability distribution, and drawing connections with empirical risk minimization.
    \item We demonstrate the effectiveness of our approach on a variety of data sets, in particular showing matching or improved performance compared to \cite{roman2014asymptotic,lustig2007sparse,wang2010variable} in certain imaging scenarios.
\end{itemize}

\vspace{-3mm}
\subsection{Organization of the Paper}

In Section \ref{sec:STRATEGIES}, we formally introduce our optimization problems, our techniques for solving them, and the resulting optimality guarantees.  Our theoretical results are presented in Section \ref{sec:GEN_BOUNDS}, namely, the deterministic and statistical generalization bounds.  In Section \ref{sec:NUMERICAL}, we demonstrate the effectiveness of our approach on a variety of data sets, and provide comparisons to previous approaches.  Conclusions are drawn in Section \ref{sec:CONCLUSION}.

%% file: optimal_sampling.tex
\section{Learning-Based Subsampling Strategies} \label{sec:STRATEGIES}

In this section, we motivate our learning-based approach to subsampling, formulate the corresponding optimization problems in terms of average-case and worst-case criteria, and propose computationally efficient algorithms for solving them exactly or approximately.

\subsection{Motivation} \label{sec:MOTIVATION}

The idea behind our subsampling strategy is simple: Given the training signals $\x_1,\dotsc,\x_m$, we seek a subsampling scheme that preserves as much of their energy as possible.  Perhaps the most intuitively simple optimization problem demonstrating this idea is the following:
\begin{equation}
    \hat{\Omega} = \argmax_{\Omega\,:\,|\Omega|=n} \min_{j=1,\dotsc,m} \| \P_{\Omega} \bPsi \x_j\|_2^2. \label{eq:opt_energy}
\end{equation}
Beyond the natural interpretation of capturing energy, this can also be viewed as optimizing the worst-case performance of the linear decoder proposed in \eqref{eq:linear_dec} with respect to the error in the $\ell_2$-norm.  Indeed, substituting \eqref{eq:model} into \eqref{eq:linear_dec}, we obtain
\begin{align}
    \|\x - \hat{\x}\|_2^2
        &= \|\x - \bPsi^* \P_{\Omega}^T \P_{\Omega} \bPsi \x\|_2^2 \\
        &= \|\bPsi\x - \P_{\Omega}^T \P_{\Omega} \bPsi \x\|_2^2 \label{eq:dec_err_2} \\
        &= \|\P_{\Omega^c}^T \P_{\Omega^c} \bPsi \x\|_2^2 \label{eq:dec_err_3} \\
        &= \|\P_{\Omega^c} \bPsi \x\|_2^2 \label{eq:dec_err_4} 
\end{align}
where \eqref{eq:dec_err_2} follows since $\bPsi$ is an orthonormal basis matrix and thus $\bPsi\bPsi^* = \I$, \eqref{eq:dec_err_3} follows since $\P_{\Omega}^T \P_{\Omega} + \P_{\Omega^c}^T \P_{\Omega^c} = \I$, and \eqref{eq:dec_err_4} follows since a multiplication by $\P_{\Omega^c}^T$ simply produces additional rows that are equal to zero.  By decomposing the energy according to the entries on $\Omega$ and $\Omega^c$, we have
\begin{equation}
    \| \P_{\Omega} \bPsi \x\|_2^2 + \| \P_{\Omega^c} \bPsi \x\|_2^2 = 1. \label{eq:pythag}
\end{equation}
whenever $\|\x\|_2 = 1$, and substitution into \eqref{eq:dec_err_4} yields
\begin{equation}
    \|\x - \hat{\x}\|_2^2 = 1 - \| \P_{\Omega} \bPsi \x\|_2^2. \label{eq:error}
\end{equation}
Thus, maximizing the objective in \eqref{eq:opt_energy} amounts to minimizing the $\ell_2$-norm error for the decoder in \eqref{eq:linear_dec}.

Recalling that $\bPsi$ is an orthonormal basis matrix, we have $\| \P_{\Omega} \bPsi \x_j\|_2 \le \|\x_j\|_2$.  Thus, defining $\X := [\x_1,\dotsc,\x_m]$ and $\V := \bPsi\X$, we can equivalently write \eqref{eq:opt_energy} as
\begin{equation}
    \hat{\Omega} = \argmin_{\Omega\,:\,|\Omega|=n} \| \bone - \diag( \V^T \P_{\Omega}^T \P_{\Omega} \V ) \|_{\infty}, \label{eq:opt_energy2}
\end{equation} 
where $\bone$ is the vector of $m$ ones, and $\diag(\cdot)$ forms a vector by taking the diagonal entries of a matrix.  In this form, the optimization problem can also be interpreted as finding the subsampling pattern providing the best restricted isometry property (RIP) constant \cite{foucart2013mathematical} with respect to the training data, analogously to the optimization problems of \cite{bah2014metric,sadeghian2013embeddings} for unstructured (rather than subsampled) matrices.

% \vspace{-3mm}
\subsection{Optimization Criteria and Constraints}\label{sec: example constraints}

Generalizing \eqref{eq:opt_energy}, we study the following class of problems: %a class of optimization problems of the form
\begin{equation}
    \hat{\Omega} = \argmax_{\Omega \in \calA} F(\Omega), \label{eq:opt_gen}
\end{equation}
where $\calA \subseteq \{\Omega\,:\,|\Omega| = n\}$ is a cardinality constrained subset of $\{1,\dotsc,p\}$, and the set function $F(\cdot)$ is given by 
\begin{equation}
    F(\Omega) := f\big( \| \P_{\Omega} \bPsi \x_1\|_2^2, \dotsc, \| \P_{\Omega} \bPsi \x_m\|_2^2 \big)
\end{equation}
for some function $f$.  For example, \eqref{eq:opt_energy} is recovered by setting $f(\alpha_1,\dotsc,\alpha_m) = \min_{j=1,\dotsc,m} \alpha_j$ and $\calA = \{\Omega\,:\,|\Omega| = n\}$.  While this choice of $\calA$ is perhaps the most obvious, one may be interested in more restrictive choices imposing structured constraints on the subsampling pattern; some examples are given below.

The optimization problem in \eqref{eq:opt_gen} is combinatorial, and in general finding the exact solution is NP hard (e.g., this is true for the special case in \eqref{eq:opt_energy}; see \cite{krause2008submod}).  The key idea in all of the special cases below is to identify advantageous combinatorial structures in the problem in order to efficiently obtain near-optimal solutions. In particular, we will see that \emph{submodularity} structures in the objective function, and \emph{matroid} structures in the constraint set, play a key role.  We proceed by presenting these definitions formally. 

\begin{mydef}
    A set function $h(\Omega)$ mapping subsets $\Omega \subseteq \{1,\dotsc,p\}$ to real numbers is said to be \emph{submodular} if, for all $\Omega_1,\Omega_2 \subset \{1,\dotsc,p\}$ with $\Omega_1 \subseteq \Omega_2$, and all $i \in \{1,\dotsc,p\} \backslash \Omega_2$, we have
    \begin{equation}
        h(\Omega_1 \cup \{i\}) - h(\Omega_1) \ge h(\Omega_2 \cup \{i\}) - h(\Omega_2).
    \end{equation}
    The function is said to be \emph{modular} if the same holds true with equality in place of the inequality.
\end{mydef}
This definition formalizes the notion of diminishing returns: Adding an element to a smaller set increases the objective function more compared to when it is added to a larger set.  Our focus in this paper will be on submodular functions that are also \emph{monotone}, i.e., $h(\Omega_2) \ge h(\Omega_1)$ whenever $\Omega_1 \subseteq \Omega_2$. 

Submodular or modular set functions often allow us to efficiently obtain near-optimal solutions with matroid constraints \cite{schrijver2003combinatorial} which we now define.

\begin{mydef} \label{def:matroid}
    Given $V = \{1,\dotsc,n\}$ and a non-empty set of subsets (of $V$) $\calA$, the pair $(V,\calA)$ is said to be a matroid if (i) for any $A \in \calA$, all subsets $A' \subseteq A$ are also in $\calA$; (ii) given $A \in \calA$ and $B \in \calA$ with $|B| > |A|$, there exists an element $v \in B \backslash A$ such that $A \cup \{v\} \in \calA$.
\end{mydef}

We will see in Section \ref{sec:STATISTICAL} that the cardinality of the constraint set $\calA$ plays a key role in determining the number of training signals needed to obtain statistical guarantees on near-optimality. 

In the context of structured signal recovery, a particularly notable example of a matroid constraint is \emph{multi-level subsampling} \cite{adcock2013breaking}, where the indices $\{1,\dotsc,p\}$ are split into $K$ disjoint groups with sizes $\{p_k\}_{k=1}^K$, and the number of measurements within the $k$-th group is constrained to be $n_k$, with $\sum_{k=1}^K n_k = n$.  This corresponds to a matroid known as the \emph{partition matroid} \cite{schrijver2003combinatorial}. In this case, the total number of possible sparsity patterns is $\prod_{k=1}^K {p_k \choose n_k}$.  As opposed to the \emph{random} multi-level subsampling scheme in \cite{adcock2013breaking}, our framework also optimizes the samples within each level given the sparsity constraints. % based on the training signals.

We can also go beyond matroid constraints; we mention one additional example here. In the context of image compression with image-independent subsampling, $\bPsi$ may correspond to the wavelet basis, and a suitable choice for $\calA$ forces the coefficients to form a rooted connected subtree of the wavelet tree of cardinality $n$ \cite{baldassarre2013group,baranuik2010model}.  In this case, the total number of subsampling patterns is the Catalan number, $\frac{1}{n+1} {2n \choose n}$ \cite{baranuik2010model}. This does not correspond to a matroid constraint, but it can nevertheless be handled using dynamic programming \cite{baldassarre2013group}. 

% For instance, modular objectives with this constraint set corresponds to a weighted maximum cover problem, which is in general NP-hard. However, we can obtain pseudo-polynomial time solutions via % with mild dependence on the dimensions \cite{baldassarre2013group}. 

% \vspace{-3mm}
\subsection{Average-case Criterion ($f = \favg$)} \label{sec:OPT_AVG}

We first consider the function $\favg(\alpha_1,\dotsc,\alpha_m) := \frac{1}{m} \sum_{j=1}^m \alpha_i$, yielding the optimization problem
\begin{equation}
    \hat{\Omega} = \argmax_{\Omega \in \calA} \frac{1}{m}\sum_{j=1}^m \sum_{i \in \Omega} |\langle \bpsi_i, \x_j\rangle|^2, \label{eq:avg_recast}
\end{equation}
where $\bpsi_i$ is the transpose of the $i$-th row of $\bPsi$.  This corresponds to maximizing the \emph{average} energy in the training signals, which may be preferable to the worst-case criterion in \eqref{eq:opt_energy} due to an improved robustness to outliers, e.g., corrupted training signals.

Since the sum of (sub)modular functions is again (sub)modular \cite{krause2012submodular}, we see that \eqref{eq:avg_recast} is a modular maximization problem, thus permitting an exact solution to be found efficiently in several cases of interest.

\textbf{Case 1 (No Additional Constraints):} In the case that $\calA = \{ \Omega\,:\,|\Omega|=n \}$, the exact solution is found by sorting:  \emph{Select the $n$ indices whose values of $\frac{1}{m}\sum_{j=1}^m |\langle \bpsi_i, \x_j\rangle|^2$ are the largest}.  The running time
is dominated by the pre-computation of the values $\frac{1}{m}\sum_{j=1}^m |\langle \bpsi_i, \x_j\rangle|^2$, and behaves as $O(mp^2)$ for general matrices $\bPsi$, or $O(mp\log p)$ for suitably structured matrices such as Fourier and Hadamard.

\textbf{Case 2 (Matroid Constraints):}  In the case that $\calA$ corresponds to a matroid, we can also find the exact solution by a simple greedy algorithm \cite{schrijver2003combinatorial}: \emph{Start with the empty set, and repeatedly add the item that increases the objective value by the largest amount without violating the constraint, terminating once $n$ indices have been selected.}  The values $|\langle \bpsi_i, \x_j\rangle|^2$ can be computed in $O(mp^2)$ (general case) or $O(mp\log p)$ time (structured case), and the greedy algorithm itself can be implemented in $O(nmp)$ time, with the factor of $m$ arising due to the summation in \eqref{eq:avg_recast}.

In some cases, the complexity can be further improved; in particular, for the above-mentioned multi-level sampling constraint, the exact solution is found by simply performing sorting within each level.

\textbf{Case 3 (Other Constraints):} As hinted above, the types of constraints that can be efficiently handled in modular optimization problems are not limited to matroids.  Choosing the set $\calA$ that forces the coefficients to form a rooted connected wavelet tree \cite{baldassarre2013group,baranuik2010model}, there exists a dynamic program for finding the optimal solution in $O(nmp)$ time \cite{baldassarre2013group}.  We can also obtain exact solutions for \emph{totally unimodular} constraints via linear programming (LP) relaxations; see \cite{elhalabi2014tu} for an overview in the context of sparse recovery.

\subsection{Generalized Average-case Criterion ($f = \fgen$)} \label{sec:F_GEN}

We generalize the choice $f=\favg$ by considering $\fgen(\alpha_1,\dotsc,\alpha_m) := \frac{1}{m}\sum_{j=1}^m g(\alpha_j)$, yielding %the optimization problem
\begin{equation}
    \hat{\Omega} = \argmax_{\Omega\in\calA} \frac{1}{m}\sum_{j=1}^m  g\bigg( \sum_{i \in \Omega} |\langle \bpsi_i, \x_j\rangle|^2 \bigg), \label{eq:conc_recast}
\end{equation}
for some function $g \,:\, [0,1] \to \R$. We consider the case that $g$ is an increasing concave function with $g(0) = 0$.   

Note that our framework also permits weighted averages of the form $\frac{1}{\sum_j \beta_j}\sum_{j}\beta_j g(\alpha_j)$, which may be relevant when some training signals are more reliable than others.  More generally, we can allow for weights $\beta_{ij}$ on the terms $|\langle \bpsi_i,\x_j\rangle|$, which may be desirable when is is more important to capture the energy in certain parts of the signal than in others.  In this paper, we focus on uniform weights for clarity of exposition.

By simple rearrangements, we can see that the choice $g(\alpha) = 1-(1-\alpha)^q$ ($q \ge 1$) in \eqref{eq:conc_recast} is equivalent to considering \eqref{eq:opt_energy2} with $\|\cdot\|_{\infty}$ replaced by $\|\cdot\|_q^q$.  In particular, the case $q=1$ recovers \eqref{eq:avg_recast}, and $q=2$ is another reasonable choice that is in a sense ``in between'' the choices $q=1$ and $q=\infty$.  Section \ref{sec:NUMERICAL} shows that the choice $q=2$ performs well numerically.

We established above that the argument to $g$ in \eqref{eq:conc_recast} is a modular function of $\Omega$.  Recalling that $g$ is concave and increasing by assumption, it follows that \eqref{eq:conc_recast} is a submodular optimization problem \cite[Sec.~1.1]{krause2012submodular}.  While finding the exact solution is hard in general, we can efficiently find an approximate solution with rigorous guarantees in several cases.

\textbf{Case 1 (No Additional Constraints):} In the case that $\calA = \{\Omega \,:\, |\Omega| = n\}$, a solution whose objective value is within a multiplicative factor of $1 - \frac{1}{e}$ of the optimum can be found via the greedy algorithm \cite{wolsey2014integer}.

\textbf{Case 2 (Matroid Constraints):} With matroid constraints, the greedy algorithm is only guaranteed optimality to within a factor of $\frac{1}{2}$ \cite{schrijver2003combinatorial}.  However, there also exist polynomial-time algorithms for obtaining solutions that are within a factor of $1 - \frac{1}{e}$ of the optimum; for example, see \cite{calinescu2011maximizing,filmus2012tight} and the references therein.

% \vspace{-3mm}
\subsection{Worst-case Criterion ($f=\fmin$) }

Finally, we consider the choices $\calA = \{ \Omega\,:\,|\Omega| = n \}$ and $\fmin(\alpha_1,\dotsc,\alpha_m) := \min_{j=1,\dotsc,m} \alpha_j$, yielding %the optimization problem 
\begin{equation}
    \hat{\Omega} = \argmax_{\Omega\,:\,|\Omega|=n} \min_{j=1,\dotsc,m} \sum_{i \in \Omega} |\langle \bpsi_i, \x_j\rangle|^2, \label{eq:min_recast}
\end{equation}
which coincides with \eqref{eq:opt_energy}.  This can be thought of as seeking robustness with respect to the ``worst'' image in the training set, which may be desirable in some cases.  To our knowledge, the algorithm that we consider below has not been extended to more general choices of $\calA$, and we thus leave such cases for future work.

% For this criterion, general $\calA$ requires a deeper review of approximation algorithms for cover problems, which we omit due to lack of space.  
%, though developing an algorithm for such choices is an interesting direction for further work.

The objective function in \eqref{eq:min_recast} is the minimum of $m$ modular functions.  This form of optimization problem was studied in \cite{krause2008submod}, where an algorithm called Saturate was proposed, which has guarantees for the general template % optimization problems of the form
\begin{equation}\label{krause_problem}
    \max_{\Omega\,:\,|\Omega| \le n} \min_{j=1,\dotsc,m}~f_j(\Omega),
\end{equation} 
where $f_1,\cdots,f_m$ are monotone submodular functions with $f_j(\emptyset) = 0$.  The algorithm takes a parameter $\alpha$ representing how much the set size may exceed the ``target'' value $n$, and a parameter $\epsilon$ used as a stopping criterion.  The details are shown in Algorithm \ref{alg:saturate}, and the greedy partial cover (GPC) subroutine is shown in Algorithm \ref{alg:GPC}.

\begin{algorithm} 
    \caption{Saturate($f_1,\cdots,f_m, n,\alpha,\epsilon$) \cite{krause2008submod} \label{alg:saturate}}
    
    \begin{algorithmic}[1]
    \State $\cmin \leftarrow 0; \cmax \leftarrow \min_{j} f_j ( \{1,\dotsc,p\} ); \Omegabest \leftarrow \emptyset$ 
    \While{$(\cmax - \cmin ) > \epsilon$}
        \State $c \leftarrow (\cmin + \cmax)/2$
        \State $\bar{f}_c(\Omega) \leftarrow \frac{1}{m} \sum_{j=1}^m \min\{f_j(\Omega), c\}$
        \State $\hat{\Omega} \leftarrow \GPC(\bar{f}_c, c)$
        \If {$|\hat{\Omega}| > \alpha n$}
            \State  $\cmax \leftarrow c$
        \Else
            \State  $\cmin \leftarrow c; \Omegabest \leftarrow \hat{\Omega}$
        \EndIf
    \EndWhile
    \end{algorithmic}
\end{algorithm} 

\begin{algorithm} 
    \caption{GPC$(\bar{f}_c, c)$ (Greedy Partial Cover) \cite{krause2008submod} \label{alg:GPC}}
    
    \begin{algorithmic}[1]
    \State $\Omega \leftarrow \emptyset$
    \While{$\bar{f}_c(\Omega) < c$} 
        \State $\Delta_j \leftarrow \bar{f}_c(\Omega \cup \{j\}) - \bar{f}_c(\Omega)$
        \State $\Omega \leftarrow \Omega \cup {\argmax_j \Delta_j }$
    \EndWhile
    \end{algorithmic}
 \end{algorithm} 
 
It is shown in \cite[Thm.~5]{krause2008submod} that for integer-valued functions $\{f_j\}_{j=1}^m$ the Saturate algorithm finds a solution $\hat{\Omega}$ such that
\begin{equation}
     \min_{j} f_j(\hat{\Omega}) \ge \max_{|\Omega|\le k} \min_{j} f_j(\Omega), \quad \text{ and } \quad |\hat{\Omega}| \le \alpha k
\end{equation}
with $\alpha = 1 + \log\big(\max_{i=1,\dotsc,p} \sum_{j=1}^m f_j(\{i\})\big)$.  While we do not have integer-valued functions in our setting, we may use the observation from \cite[Sec.~7.1]{krause2008submod} that in the general case, analogous guarantees can be obtained at the expense of having an additional term in $\alpha$ depending linearly on the number of bits of accuracy.  We observe that when applied to \eqref{eq:min_recast}, we have $\log\big(\max_{i=1,\dotsc,p} \sum_{j=1}^m f_j(\{i\})\big) = \log\big(\max_{i=1,\dotsc,p} \sum_{j=1}^m |\langle \bpsi_i, \x_j \rangle | \big)$, which is in turn upper bounded by $\log m$ since both $\bpsi_i$ and $\x_j$ have unit norm.  Thus, even with this crude bounding technique, this term only constitutes a logarithmic factor.

Moreover, it was observed empirically in \cite{krause2008submod} that the Saturate algorithm can provide state-of-the-art performance even when the functions are not integer-valued and $\alpha$ is set to one.  Thus, although the theory in \cite{krause2008submod} does not directly capture this, we expect the algorithm to provide a good approximate solution to \eqref{eq:min_recast} even without the logarithmic increase in the number of measurements.

The running time is $O(p^2 m \log m)$ in the worst case \cite{krause2008submod}. In practice, we observe the algorithm to run much faster, as was also observed in \cite{krause2008submod}. Moreover, we found the total number of samples returned by the algorithm to be very close to its maximum value $\alpha n$ (e.g., within 1\%).

%% file: gen_bounds.tex
\section{Generalization Bounds} \label{sec:GEN_BOUNDS}

The optimization problems in the previous section seek to capture as much of the signal energy as possible on the training signals $\x_j$, which also corresponds to minimizing the $\ell_2$-norm error of the decoder in \eqref{eq:linear_dec} (\emph{cf.}, Section \ref{sec:MOTIVATION}).  However, it is not immediately clear to what extent the same will be true on a new signal $\xnew$.  In this section, we take two distinct approaches to characterizing this theoretically. 

% \vspace{-3mm}
\subsection{Deterministic Approach}

This subsection provides simple bounds showing that the energy loss remains small on any new signal  $\xnew$ that is ``close'' to the training data in a sense dictated by $\Omega^c := \{1,\dotsc,p\} \backslash \Omega$.  We assume that $\xnew$ is normalized in the same way as the training signals, i.e., $\|\xnew\|_2 = 1$; this is without loss of generality, since general signals may be normalized to satisfy this condition, and this normalization constant does not need to be known by the recovery algorithm.

We first consider the generalized average-case criterion, which includes the average-case criterion as a special case.

\begin{thm} \label{thm:gb_avg}
    \emph{(Deterministic generalization bound for $f = \fgen$)} Fix $\delta>0$ and $\epsilon > 0$, and suppose that for a set of training signals $\x_1,\dotsc,\x_m$ with $\|\x_j\|_2 = 1$, we have a sampling set $\Omega$ such that 
    \begin{equation}
        \frac{1}{m} \sum_{j=1}^m g\big( \| \P_{\Omega} \bPsi \x_j\|_2^2 \big) \ge g(1 - \delta). \label{eq:RIP_training_avg}
   \end{equation}
   Then for any signal $\xnew$ with $\|\xnew\|_2 = 1$ such that $\frac{1}{m} \sum_{j=1}^m \|\P_{\Omega^c}\bPsi(\xnew - \x_j)\|_2^2 \le \epsilon$, we have
    \begin{equation}
        \| \P_{\Omega} \bPsi \xnew\|_2^2 \ge 1 - \big(\sqrt{\delta} + \sqrt{\epsilon}\big)^2. \label{eq:RIP_new_avg}
   \end{equation}
\end{thm}
\begin{proof}
    Substituting  \eqref{eq:pythag} into \eqref{eq:RIP_training_avg}, we obtain $\frac{1}{m} \sum_{j=1}^m g(1 - \| \P_{\Omega^c} \bPsi \x_j\|_2^2) \ge g(1 - \delta)$.  Recalling that $g$ is concave by assumption, applying Jensen's inequality on the left-hand side gives $g\big( 1 - \frac{1}{m} \sum_{j=1}^m \| \P_{\Omega^c} \bPsi \x_j\|_2^2\big) \ge g(1 - \delta)$.  Since $g$ is also monotonically increasing by assumption, it follows that $\frac{1}{m} \sum_{j=1}^m \| \P_{\Omega^c} \bPsi \x_j\|_2^2 \le \delta$.  
    
    Next, since the average of squares is at least as large as the square of the average, the previous condition implies $\frac{1}{m} \sum_{j=1}^m \| \P_{\Omega^c} \bPsi \x_j\|_2 \le \sqrt{\delta}$, and the assumption $\frac{1}{m} \sum_{j=1}^m \|\P_{\Omega^c}\bPsi(\xnew - \x_j)\|_2^2 \le \epsilon$ similarly implies $\frac{1}{m} \sum_{j=1}^{m} \|\P_{\Omega^c}\bPsi(\xnew - \x_j)\|_2 \le \sqrt{\epsilon}$.  We thus obtain from the triangle inequality that
    \begin{align}
        &\| \P_{\Omega^c} \bPsi \xnew\|_2 \nonumber \\
            &\quad\le \frac{1}{m} \sum_{j=1}^m \Big( \| \P_{\Omega^c} \bPsi \x_j\|_2 + \| \P_{\Omega^c} \bPsi (\xnew - \x_j)\|_2 \Big) \\
            &\quad\le \sqrt{\delta} + \sqrt{\epsilon}.
    \end{align}
    Taking the square and applying \eqref{eq:pythag}, we obtain \eqref{eq:RIP_new_avg}.
\end{proof}

The case $f=\favg$ corresponds to $f = \fgen$ with $g(\alpha) = \alpha$, and in this case, the condition in \eqref{eq:RIP_training_avg} takes the particularly simple form
\begin{equation}
    \frac{1}{m} \sum_{j=1}^m \| \P_{\Omega} \bPsi \x_j\|_2^2  \ge 1 - \delta, 
\end{equation}
requiring that at least a fraction $1-\delta$ of the training signal energy be captured on average.

We now turn to the worst-case criterion.

\begin{thm} \label{thm:gb_min}
    \emph{(Deterministic generalization bound for $f = \fmin$)} Fix $\delta>0$ and $\epsilon > 0$, and suppose that for a set of training signals $\x_1,\dotsc,\x_m$ with $\|\x_j\|_2 = 1$, we have a sampling set $\Omega$ such that 
    \begin{equation}
        \min_{j=1,\dotsc,m} \| \P_{\Omega} \bPsi \x_j\|_2^2 \ge 1 - \delta. \label{eq:RIP_training}
   \end{equation}
   Then for any signal $\xnew$ with $\|\xnew\|_2 = 1$ such that $\|\P_{\Omega^c}\bPsi(\xnew - \x_j)\|_2^2 \le \epsilon$ for some $j \in \{1,\dotsc,m\}$, we have
    \begin{equation}
        \| \P_{\Omega} \bPsi \xnew\|_2^2 \ge 1 - \big(\sqrt{\delta} + \sqrt{\epsilon}\big)^2. \label{eq:RIP_new}
   \end{equation}
\end{thm}
\begin{proof}
    It follows from \eqref{eq:pythag} and \eqref{eq:RIP_training} that $\| \P_{\Omega^c} \bPsi \x_j\|_2^2 \le \delta$ for all $j$.  Hence, letting $j$ be an index such that $\|\P_{\Omega^c}\bPsi(\xnew - \x_j)\|_2^2 \le \epsilon$, we obtain from the triangle inequality that
    \begin{align}
        \| \P_{\Omega^c} \bPsi \xnew\|_2 
            &\le \| \P_{\Omega^c} \bPsi \x_j\|_2 + \| \P_{\Omega^c} \bPsi (\xnew - \x_j)\|_2 \\
            &\le \sqrt{\delta} + \sqrt{\epsilon}.
   \end{align}
    Taking the square and applying \eqref{eq:pythag}, we obtain \eqref{eq:RIP_new}.
\end{proof}

We note that a sufficient condition for $\|\P_{\Omega^c}\bPsi(\xnew - \x_j)\|_2^2 \le \epsilon$ is that $\|\xnew - \x_j\|_2^2 \le \epsilon$.  However, this is certainly not necessary; for example, when $\xnew = -\x_j$, the latter is large, whereas the former is small provided that the index set $\Omega$ captures most of the signal energy. 

We note that Theorems \ref{thm:gb_avg} and \ref{thm:gb_min} are rather different despite appearing to be similar. In Theorem \ref{thm:gb_min}, the definition of $\delta$ corresponds to capturing the energy in \emph{all} training signals, and the definition of $\epsilon$ corresponds to the new signal being close to \emph{some} training signal (i.e., the minimum distance).  In contrast, in Theorem \ref{thm:gb_avg}, both $\delta$ and $\epsilon$ are defined with respect to the corresponding average.

%Finally, we can choose $\epsilon$ in  to be the minimum distance among all training signals whereas the $\epsilon$ in Theorem  is fixed. 
 \vspace{-3mm}
\subsection{Statistical Approach} \label{sec:STATISTICAL}

Thus far, we have treated all of our signals as being deterministic.  We now turn to a statistical approach, in which the training signals $\x_1,\dotsc,\x_m$ and the test signal $\x$ are independently drawn from a common probability distribution $\PP$ on $\C^p$.  We assume that this distribution is \emph{unknown}, and thus cannot be exploited as prior knowledge for the design of $\Omega$.  We also assume that $\|\x\|_2 = 1$ almost surely for $\x \sim \PP$; otherwise, we can simply replace $\x$ by $\frac{ \x }{ \|\x\|_2 }$ throughout.

In this setting, there is precise notion of a ``best'' subsampling set: The set $\Omega^* \in \calA$ that captures the highest proportion of the signal energy \emph{on average} is given by
\begin{equation}
    \Omega^* = \argmax_{\Omega \in \calA} \EE\big[ \|\P_{\Omega} \bPsi \x\|_2^2 \big]. \label{eq:omega_opt}
\end{equation}
It is worth noting that, under our statistical model, one cannot improve on this choice by moving to randomized subsampling strategies.  To see this,  we write for any randomized $\Omega$
\begin{equation}
    \EE\big[ \|\P_{\Omega} \bPsi \x\|_2^2 \big] = \EE\Big[\EE\big[ \|\P_{\Omega} \bPsi \x\|_2^2 \,\big|\, \Omega \big]\Big].
\end{equation}
Then, by the mean value theorem, there exists a deterministic set such that the argument to the outer expectation has a value at least as high as the average.

Since the distribution $\PP$ of $\x$ is unknown, we cannot expect to solve \eqref{eq:omega_opt} exactly.  However, given $m$ training samples, we can approximate the true average by the empirical average: %, yielding the estimate
\begin{align}
    \hat{\Omega} 
    & = \hat{\Omega}(\x_1,\dotsc,\x_m) \nonumber \\
    & = \argmax_{\Omega \in \calA} \frac{1}{m} \sum_{j=1}^m \|\P_{\Omega} \bPsi \x_j\|_2^2. \label{eq:omega_hat_erm}
\end{align}
This idea is known as \emph{empirical risk minimization} in statistical learning theory, and in the present setting, we see that it yields precisely the optimization problem in \eqref{eq:opt_gen} with $f = \favg$.

We are now interested in determining how the average energy captured by $\hat{\Omega}$ compares to the optimal choice $\Omega^*$.  To this end, we introduce some definitions.  We let $\EE_{\x}[\cdot]$ denote averaging with respect to $\x$ alone, so that for any function $h(\x,\x_1,\dotsc,\x_m)$, $\EE_{\x}[ h(\x,\x_1,\dotsc,\x_m) ]$ is a random variable depending on the training signals.  With this notation, we define the random variable
\begin{align}
    \Delta_n
        &= \Delta_n(\x_1,\dots,\x_m) \\
        &:= \EE\big[ \|\P_{\Omega^*} \bPsi \x\|_2^2 \big] - \EE_{\x}\big[ \|\P_{\hat{\Omega}} \bPsi \x\|_2^2 \big] \label{eq:delta_n}
\end{align}
representing the gap to optimality as a function of the training signals.  In the following theorem, we bound this gap \emph{independently of the distribution $\PP$}.

\begin{thm} \label{thm:erm}
    \emph{(Statistical generalization bound for $f = \favg$)}
    Under the above statistical model, for any $\eta > 0$, we have with probability at least $1-\eta$ that
    \begin{equation}
        \Delta_n \le \sqrt{ \frac{2}{m}\bigg( \log|\calA| + \log\frac{2}{\eta} \bigg) }. \label{eq:erm_bound}
    \end{equation}
\end{thm}
\begin{proof}
    For brevity, we define $\gamma_{\Omega}(\x) := \|\P_{\Omega}\bPsi\x\|_2^2$.  Moreover, we define the empirical average
    \begin{equation}
        \EEhat_m[\gamma_{\Omega}(\cdot)] := \frac{1}{m}\sum_{i=1}^m \gamma_{\Omega}(\x_j), \label{eq:empirical_distr}
    \end{equation}
    which is a random variable depending on $\x_1,\dotsc,\x_m$.
    
    With these definitions, we have the following:
    \begin{align}
        \Delta_n &= \EE[\gamma_{\Omega^*}(\x)] - \EE_{\x}[\gamma_{\hat{\Omega}}(\x)] \\
            &= \big(\EE[\gamma_{\Omega^*}(\x)] - \EEhat_m[\gamma_{\Omega^*}(\cdot)]\big)   \nonumber \\ 
            &\qquad + \big(\EEhat_m[\gamma_{\Omega^*}(\cdot)] - \EEhat_m[\gamma_{\hat{\Omega}}(\cdot)] \big) \nonumber \\ 
            &\qquad + \big(\EEhat_m[\gamma_{\hat{\Omega}}(\cdot)] - \EE_{\x}[\gamma_{\hat{\Omega}}(\x)]\big)\\
            &\le \big|\EEhat_m[\gamma_{\Omega^*}(\cdot)] - \EE[\gamma_{\Omega^*}(\x)]\big| \nonumber \\ &\qquad +\big|\EEhat_m[\gamma_{\hat{\Omega}}(\cdot)] - \EE_{\x}[\gamma_{\hat{\Omega}}(\x)]\big| \label{eq:Dn_bound3} \\
            &\le 2\max_{\Omega \in \calA} \big|\EEhat_m[\gamma_{\Omega}(\cdot)] - \EE[\gamma_{\Omega}(\x)]\big|, \label{eq:Dn_bound4}
    \end{align}
    where \eqref{eq:Dn_bound3} follows since $\EEhat_m[\gamma_{\Omega^*}(\cdot)] \le \EEhat_m[\gamma_{\hat{\Omega}}(\cdot)]$ by the definition of $\hat{\Omega}$ in \eqref{eq:omega_hat_erm}.
    
    By the assumption that $\|\x\|_2 = 1$, we have $\gamma_{\Omega}(\x) \in [0,1]$.  Moreover, using the definition in \eqref{eq:empirical_distr} and the fact that both $\x$ and $\{\x_j\}_{j=1}^m$ have distribution $\PP$, the average of $\EEhat_m[\gamma_{\Omega}(\cdot)]$ is given by $\EE[\gamma_{\Omega}(\x)]$.  It thus follows from Hoeffding's inequality \cite[Sec.~2.6]{boucheron2013conc} that
    \begin{equation}
        \PP\Big[ \big|\EEhat_m[\gamma_{\Omega}(\cdot)] - \EE[\gamma_{\Omega}(\x)]\big| > t \Big] \le 2e^{-2mt^2}
    \end{equation}
    for any set $\Omega$ and constant $t > 0$.  Thus, by the union bound, we have
    \begin{equation}
        \PP\bigg[ \max_{\Omega\in\calA} \big|\EEhat_m[\gamma_{\Omega}(\cdot)] - \EE[\gamma_{\Omega}(\x)]\big| > t \bigg] \le 2|\calA|e^{-2mt^2}.
    \end{equation}
     The proof is concluded by setting $t = \sqrt{\frac{1}{2m}\log(\frac{2}{\eta} |\calA|) }$ and substituting the condition of the event into \eqref{eq:Dn_bound4}.
\end{proof}

In the case that $\calA = \{\Omega\,:\,|\Omega| = n\}$, we have $|\calA| = {p \choose n}$, and hence, Theorem \ref{thm:erm} reveals that the performance of $\hat{\Omega}$ can be made arbitrarily close to that of $\Omega^*$ using $m = O(n \log \frac{p}{n})$ training signals, for any distribution $\PP$. Our numerical findings in Section \ref{sec:NUMERICAL} suggest that this bound may be pessimistic in some specific scenarios, as we achieve good results even when $m$ is smaller than $n$.  Nevertheless, it is reassuring that the optimal performance can always be approached when the number of training signals is large enough. 

Clearly, considering a smaller set $\calA$ yields an improved bound in \eqref{eq:erm_bound}; in particular, this occurs for the multi-level sampling and rooted connected tree structures in accordance with the cardinalities given in  Section \ref{sec: example constraints}.  On the other hand, using a smaller set can also worsen the performance of $\Omega^*$ itself, due to a smaller maximization set in \eqref{eq:omega_opt}.  Thus, there is a trade-off between the performance of the best choice within the class considered, and the extent to which the bound guarantees that the corresponding performance is approached using few training samples.

\begin{rem}
    {\em Although we focused on $f=\favg$ for clarity, the analysis remains valid with $f=\fgen$ provided that $g$ is bounded in $[0,1]$ (e.g., $g(\alpha) = 1-(1-\alpha)^2$).  Specifically, we modify \eqref{eq:omega_opt}--\eqref{eq:delta_n} by replacing the squared-norms $\|\cdot\|_2^2$ by $g(\|\cdot\|_2^2)$, and similarly in the definition of $\gamma_{\Omega}(\x)$ in the proof.}
\end{rem}

%% file: experiments.tex
%!TEX root = JSTSP_OPT_SAM.tex
\vspace{-3mm}
\section{Numerical Experiments} \label{sec:NUMERICAL}
In this section, we present the results of numerical experiments that illustrate the effectiveness of the learning-based compressive subsampling approach. For a given number of measurements, our consistent observation is that our approach matches or improves the quality of signal recovery over the randomized variable-density based sampling approaches. A likely reason for the improvements is that we directly optimize the sampling indices, as opposed to only optimizing auxiliary parameters that are related to those indices.

We also identify several scenarios in which our learning-based subsampling in tandem with the simple \emph{linear} decoder \eqref{eq:linear_dec} outperforms randomized sampling techniques used in conjunction with the \emph{non-linear} decoder \eqref{eq: BP}. Moreover, we find in all of our examples that the improvements of the non-linear decoders over the linear one are quite marginal even for the randomized subsampling indices. 

This is not conclusive evidence against the use of non-linear decoders in these problems, since it may be possible to devise more sophisticated techniques to optimize $\Omega$ specifically for these decoders.  Nevertheless, our observations support the practitioners' historical preference of simple decoders.  

Throughout this section, the experiments are done using the algorithms proposed in Section \ref{sec:STRATEGIES}: Sorting for $\favg$, the greedy algorithm for $\fgen$, and the Saturate algorithm for $\fmin$.

\input{exp_kenya.tex}
\input{exp_ieeg.tex}

\input{exp_mri.tex}

%% file: exp_kenya.tex
%!TEX root = JSTSP_OPT_SAM.tex
\subsection{Kenya \& ImageNet Data sets}
We created an image data set in $16$-bit \texttt{tiff} format of $187$  Kenya images of resolution $2048 \times 2048$ from one of the authors' personal collection, without any compression. Given the high-resolution images, we used MATLAB's  \texttt{imresize} function to create $1024 \times 1024$, $512 \times 512$, and $256 \times 256$ pixel images. We split the data set into a training set of the first $137$ images and a test set with the remaining $50$ images. Here are some examples from this data set: % \\  [1mm]
%\begin{figure}[!h]
%\centering
%\vspace{1mm}
%\begin{center}
% \vspace{-1mm}
\begin{tabular}{ccccc}
\hspace{-1mm}
\includegraphics[width=0.185\columnwidth]{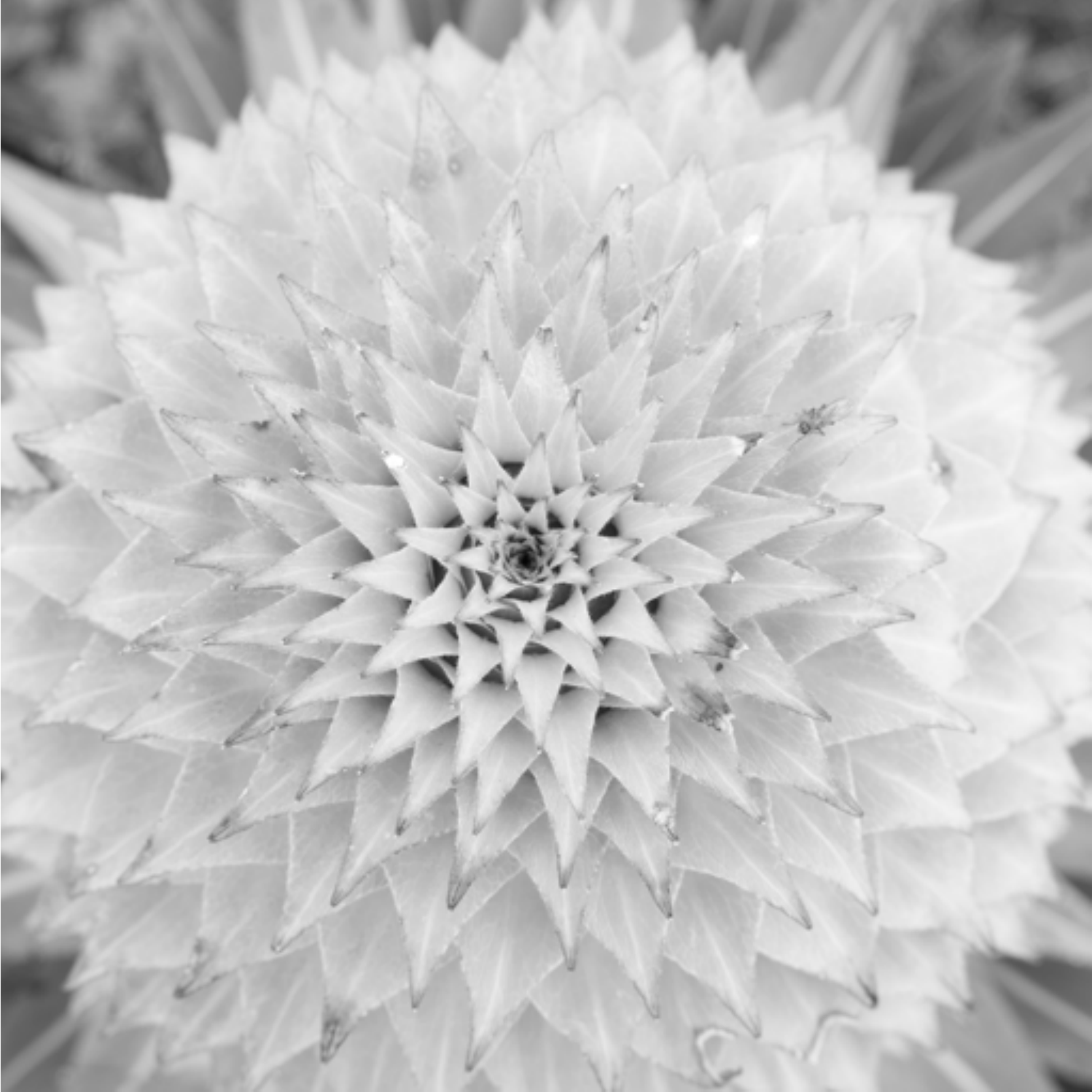} &
\hspace{-4mm}\includegraphics[width=0.185\columnwidth]{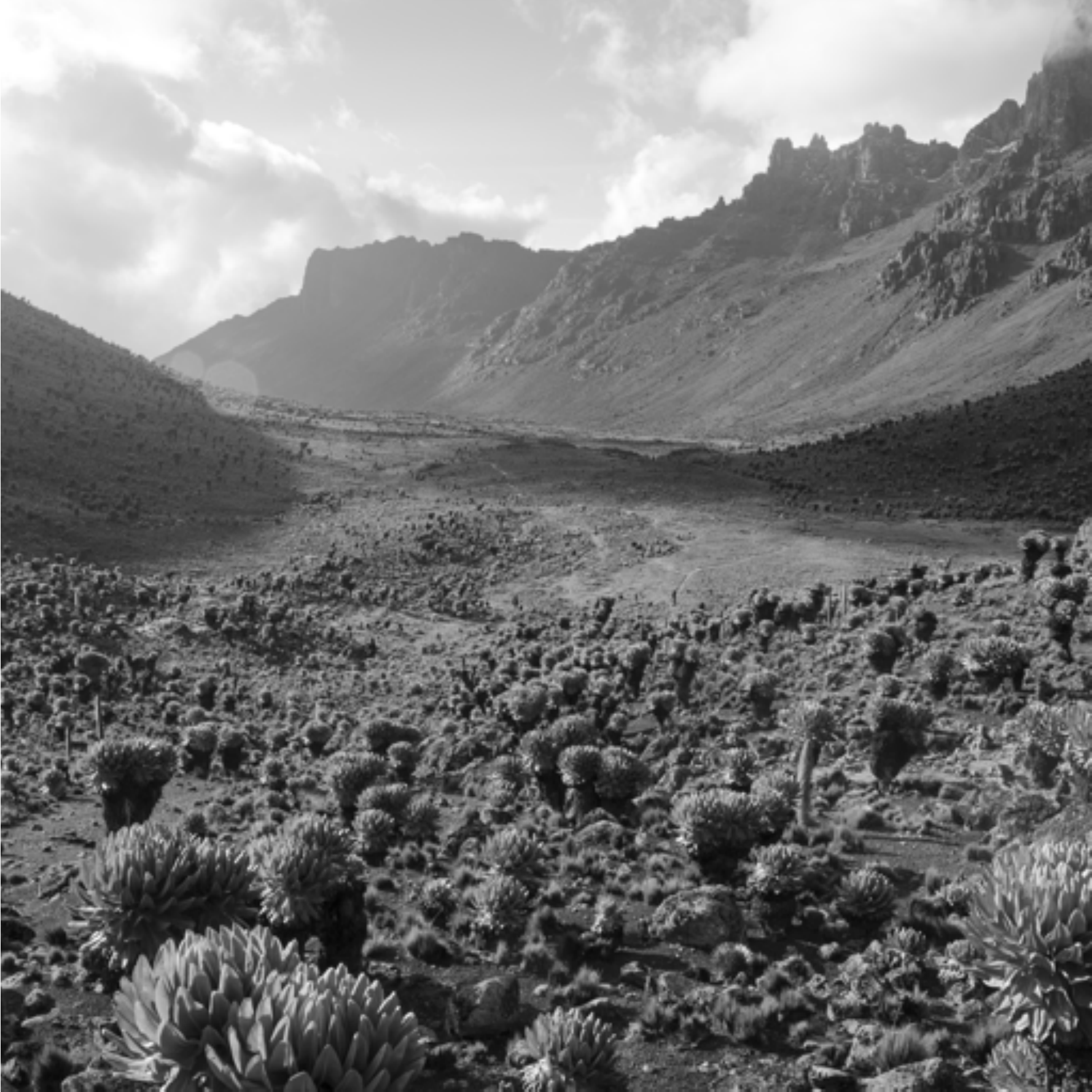} &
\hspace{-4mm}\includegraphics[width=0.185\columnwidth]{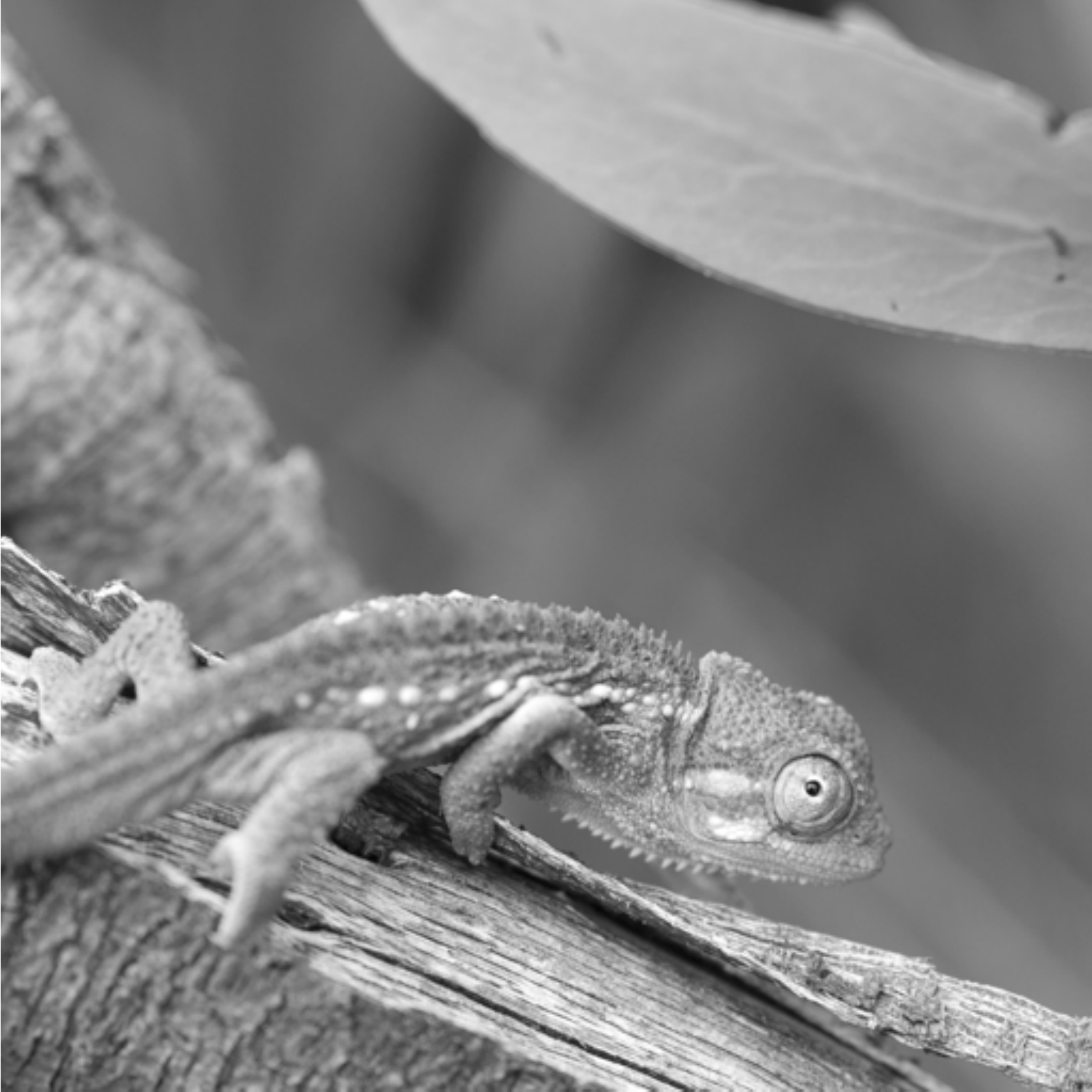} &
\hspace{-4mm}\includegraphics[width=0.185\columnwidth]{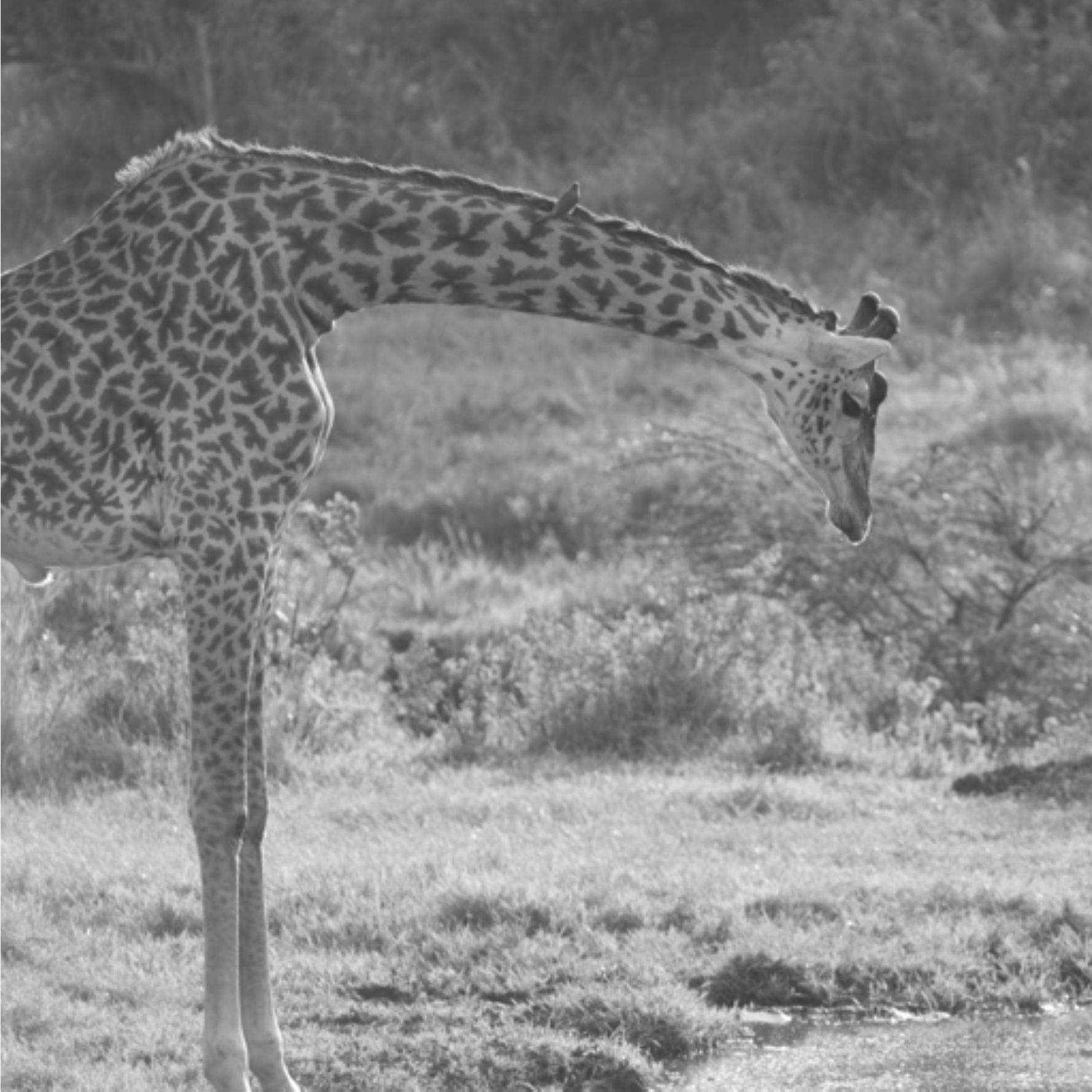} &
\hspace{-4mm}\includegraphics[width=0.185\columnwidth]{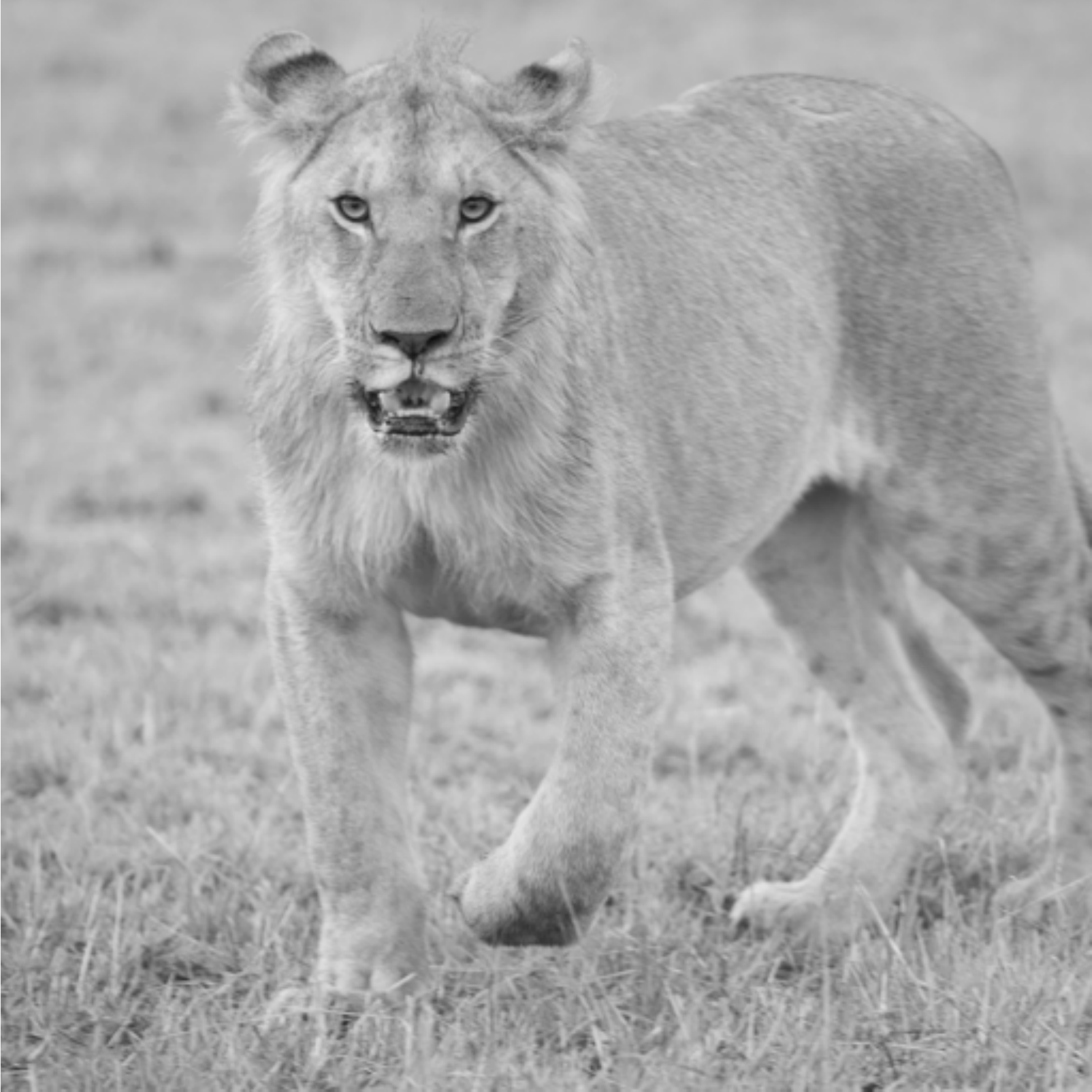}
\end{tabular}
%\end{center}
%\end{figure}

% COMPRESSION RATE SCALING
\begin{figure}
\centering
\begin{tabular}{ccc}
\footnotesize{Hadamard} & \footnotesize{DCT} & \footnotesize{Wavelets} \\
\hspace{-2.5mm}\includegraphics[width=0.32\columnwidth]{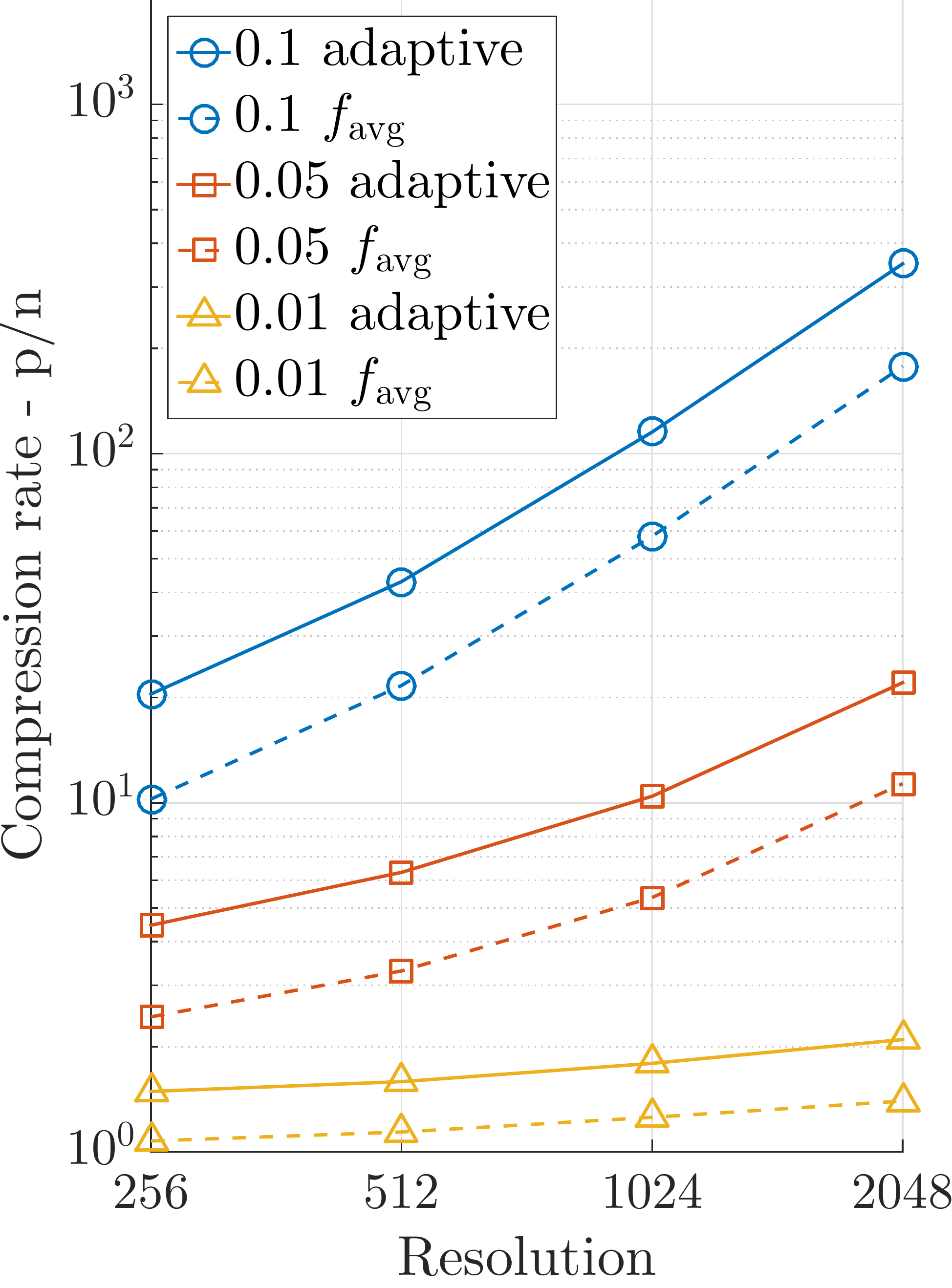} &
\hspace{-2.5mm}\includegraphics[width=0.32\columnwidth]{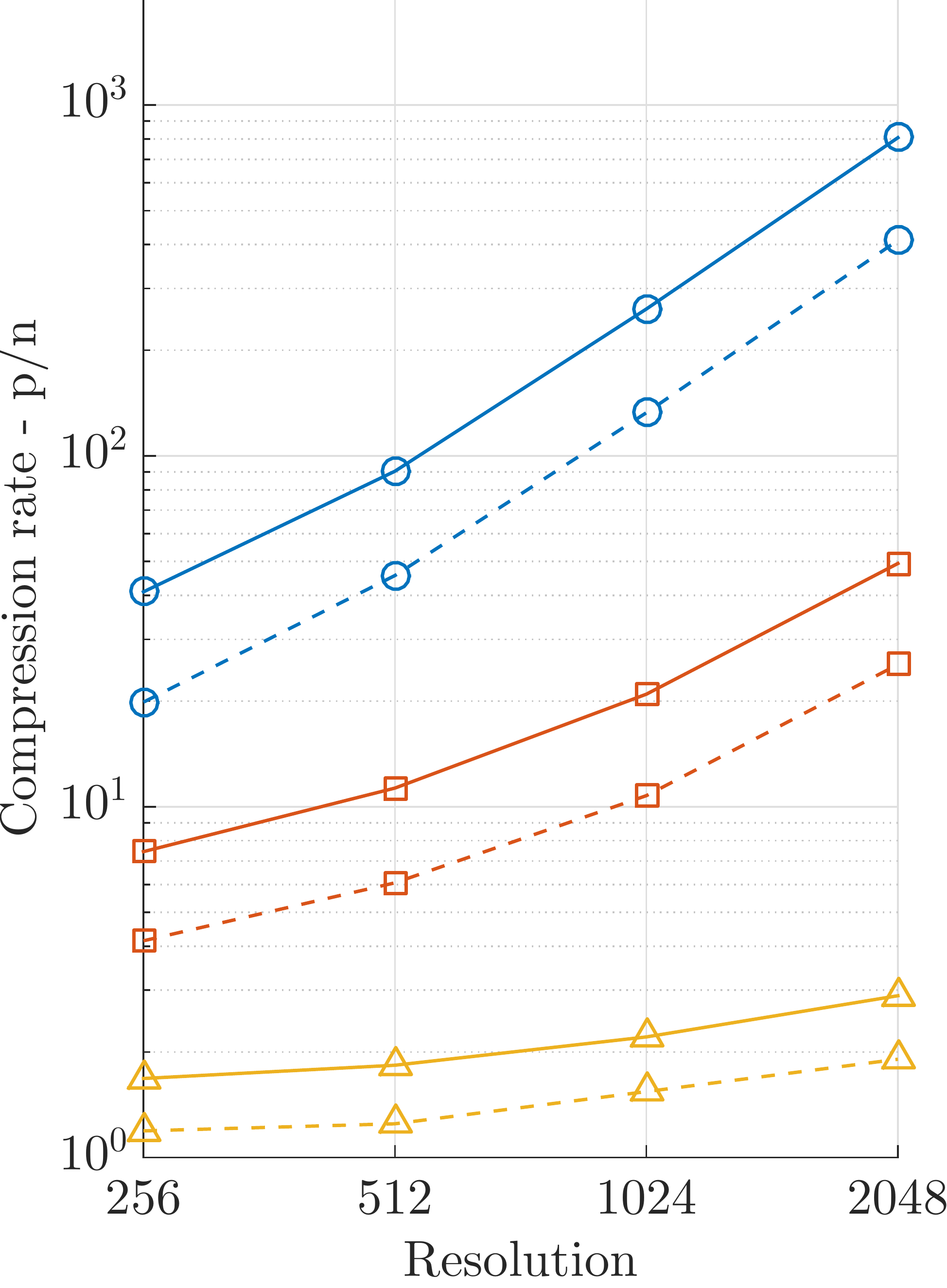} &
\hspace{-2.5mm}\includegraphics[width=0.32\columnwidth]{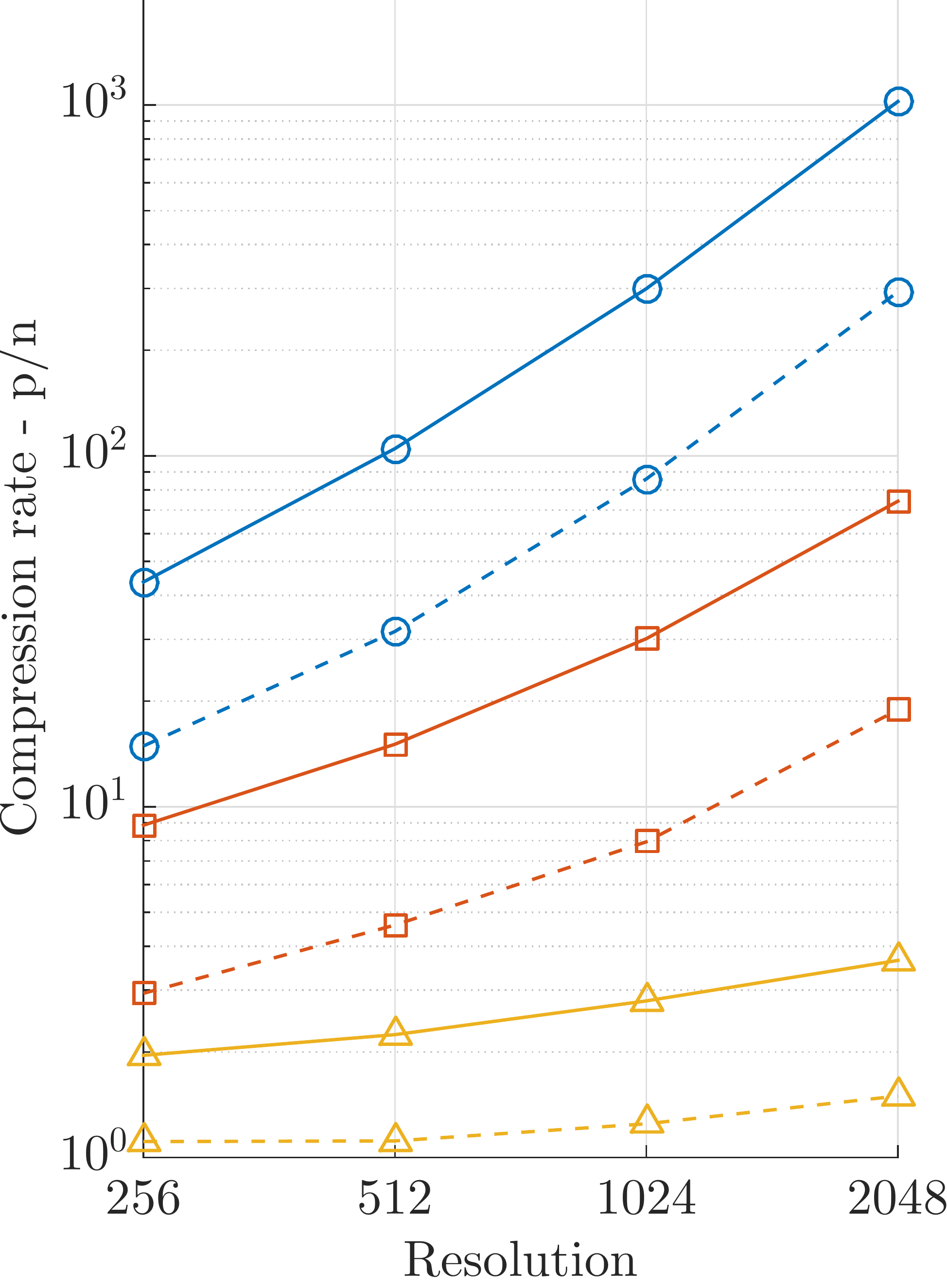}
\end{tabular}
\begin{tabular}{cc}
\scriptsize{256 $\times$ 256} & \scriptsize{512 $\times$ 512} \\
\includegraphics[width=0.46\columnwidth]{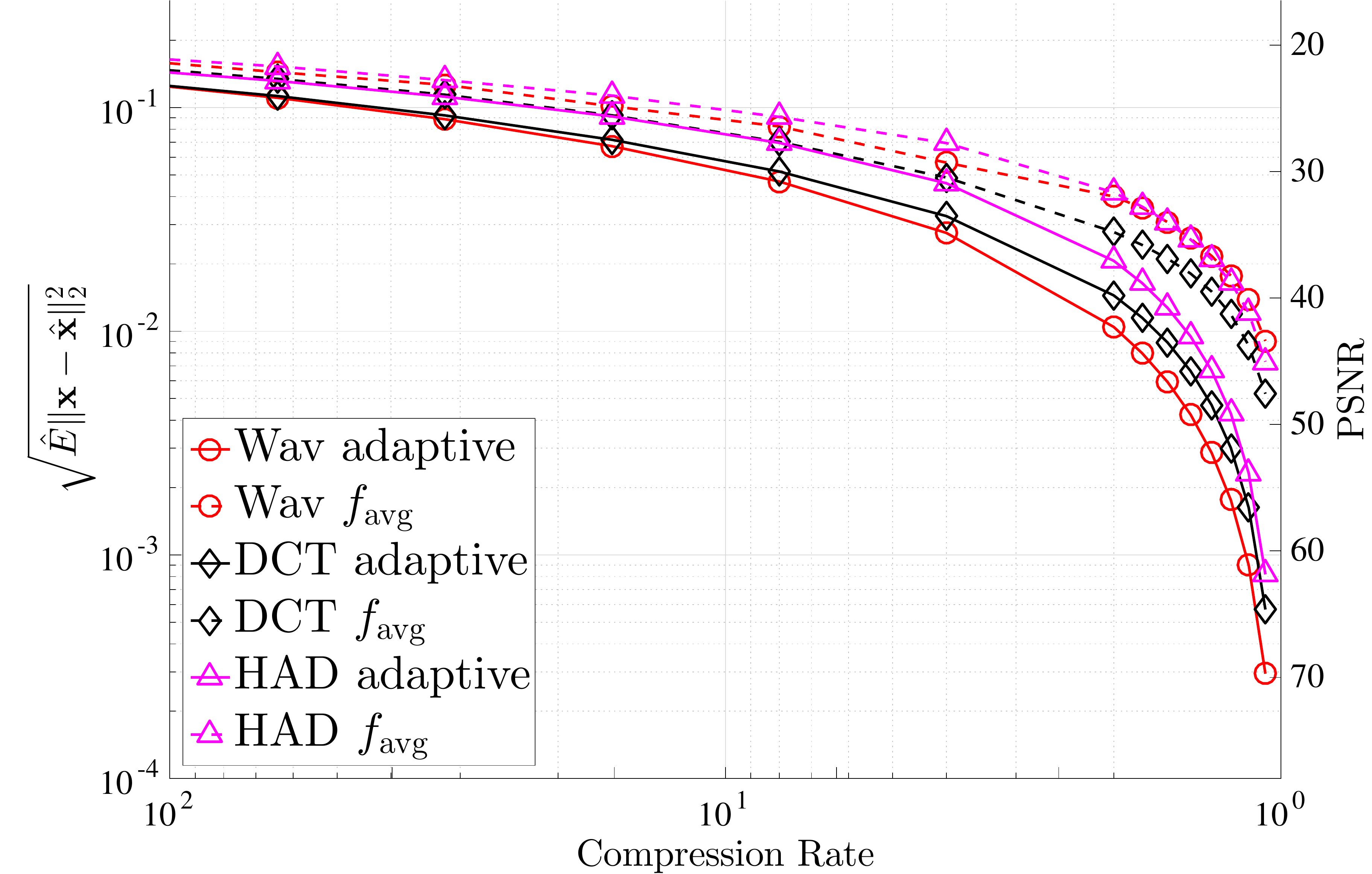} &
\includegraphics[width=0.46\columnwidth]{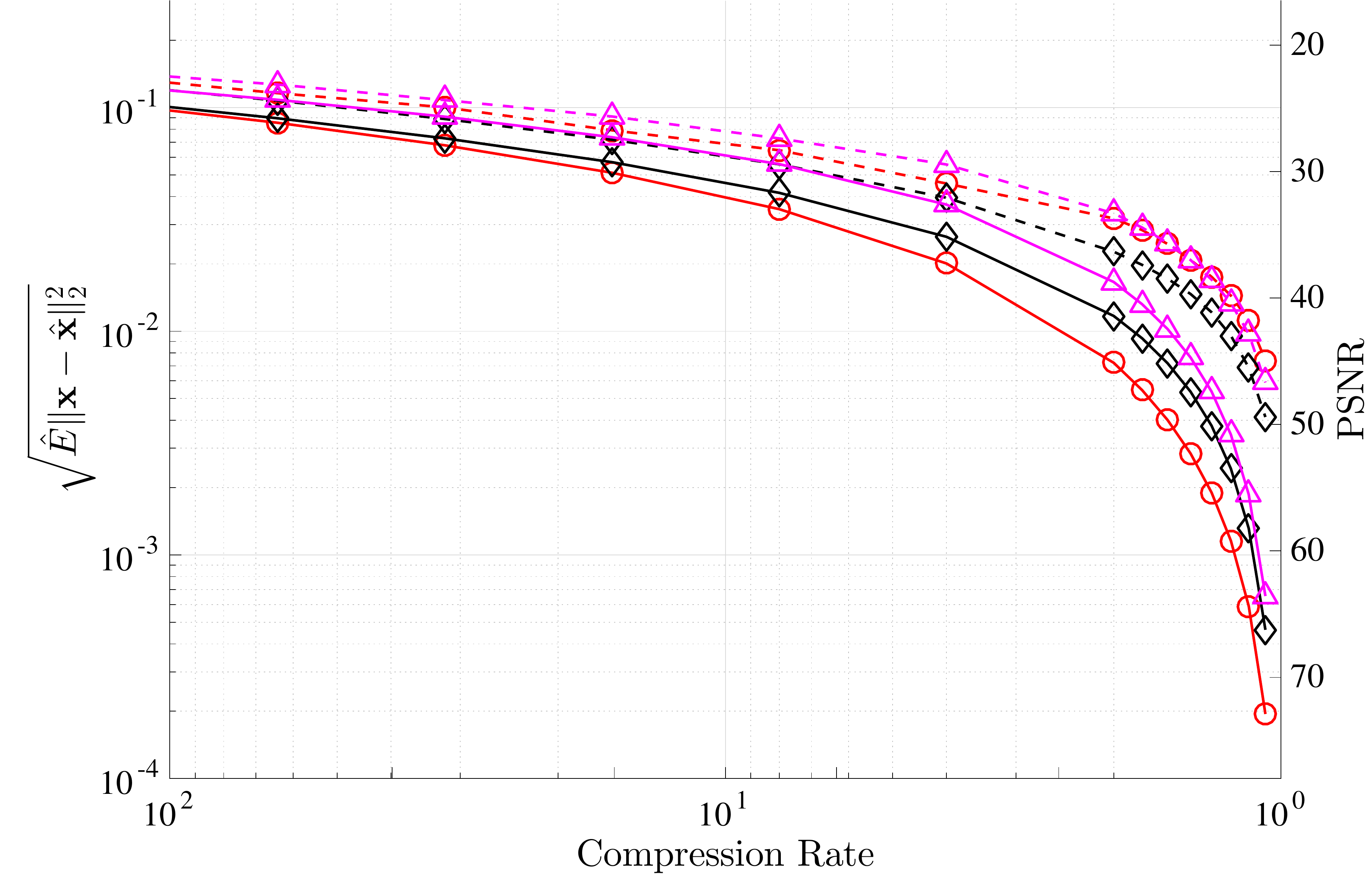} \\
\scriptsize{1024 $\times$ 1024} & \scriptsize{2048 $\times$ 2048} \\
\includegraphics[width=0.46\columnwidth]{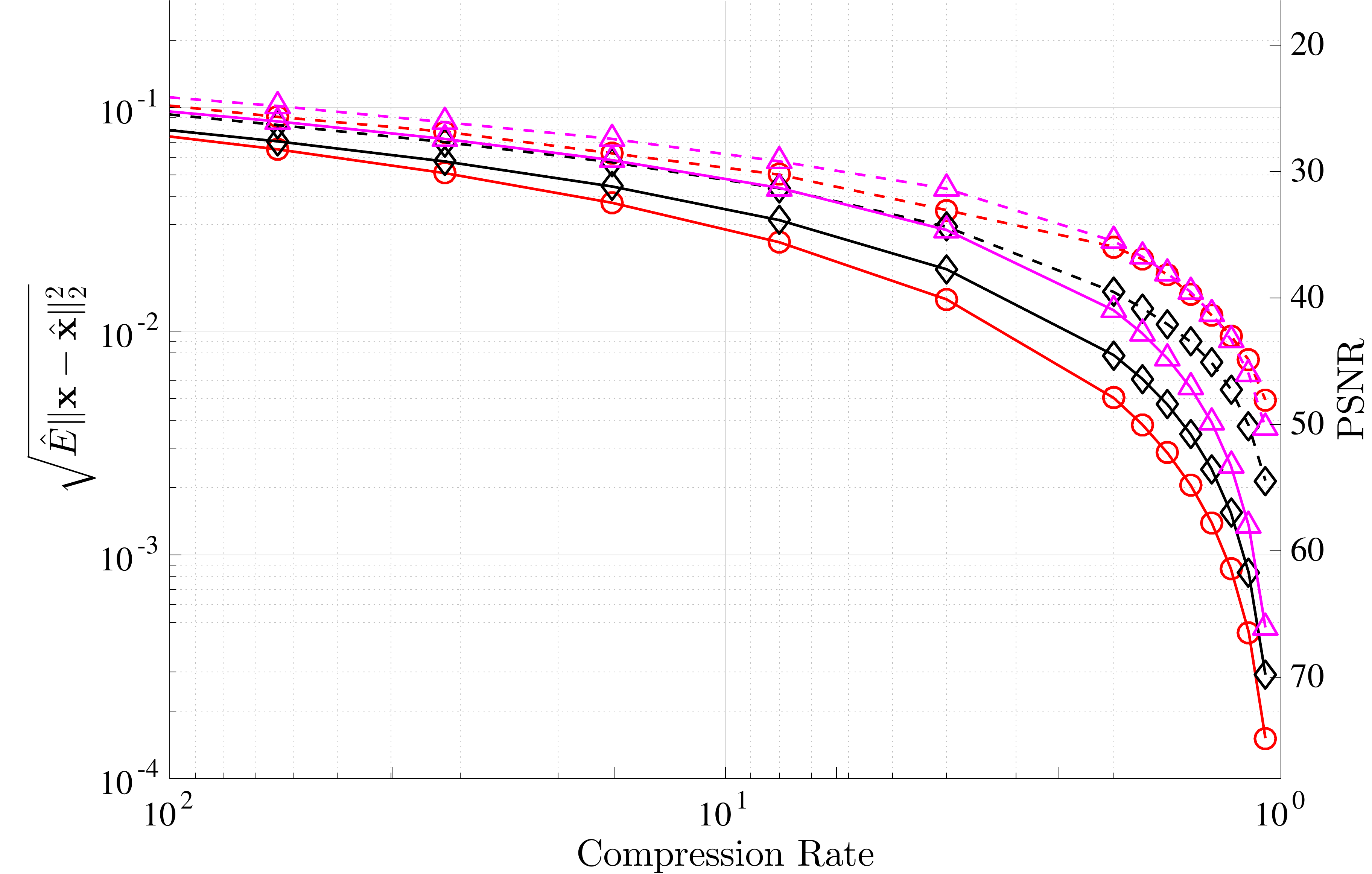} &
\includegraphics[width=0.46\columnwidth]{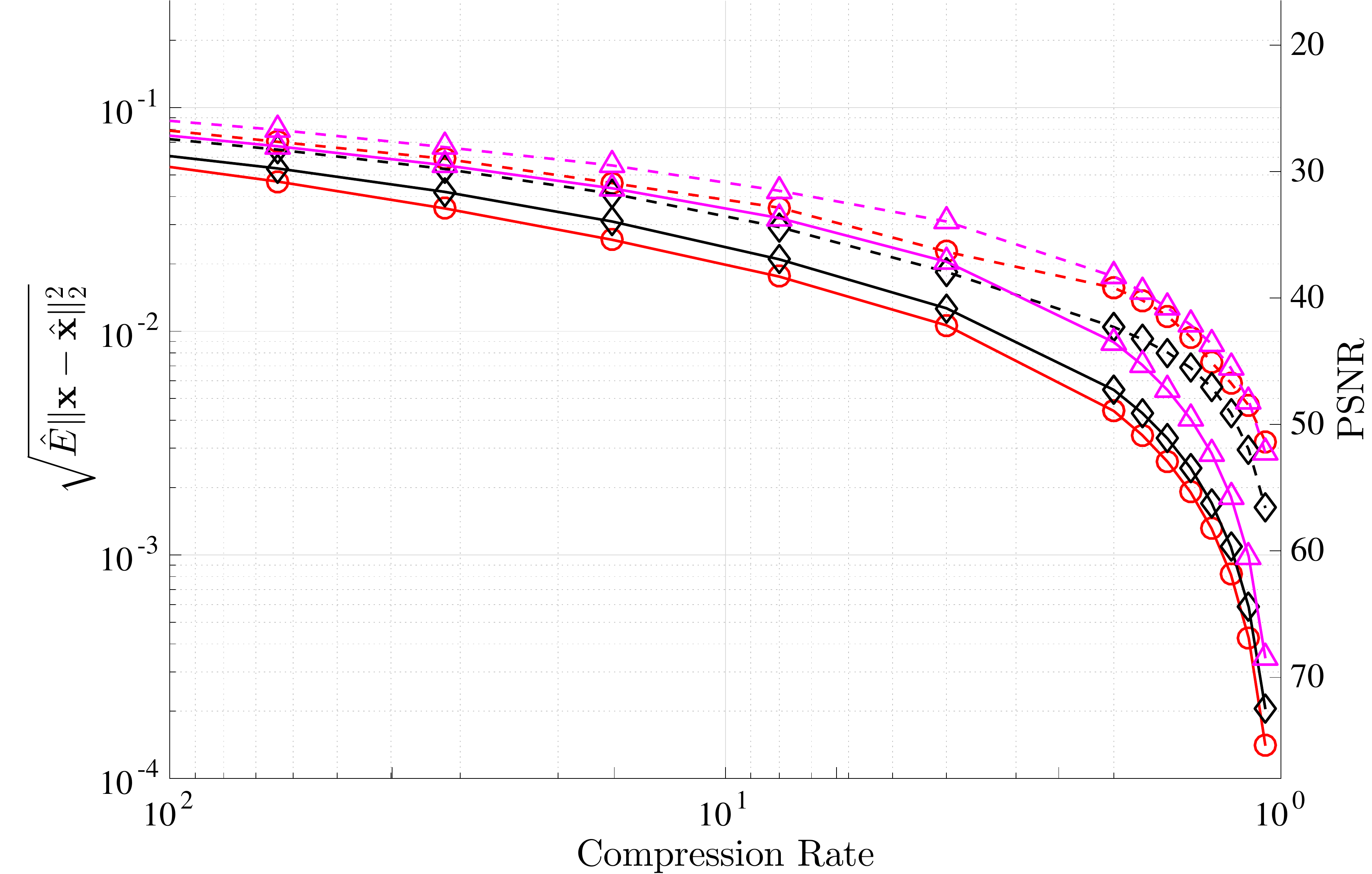}
\end{tabular}
\begin{tabular}{ccc}
\footnotesize{Hadamard} & \footnotesize{DCT} & \footnotesize{Wavelets} \\
\includegraphics[width=0.25\columnwidth]{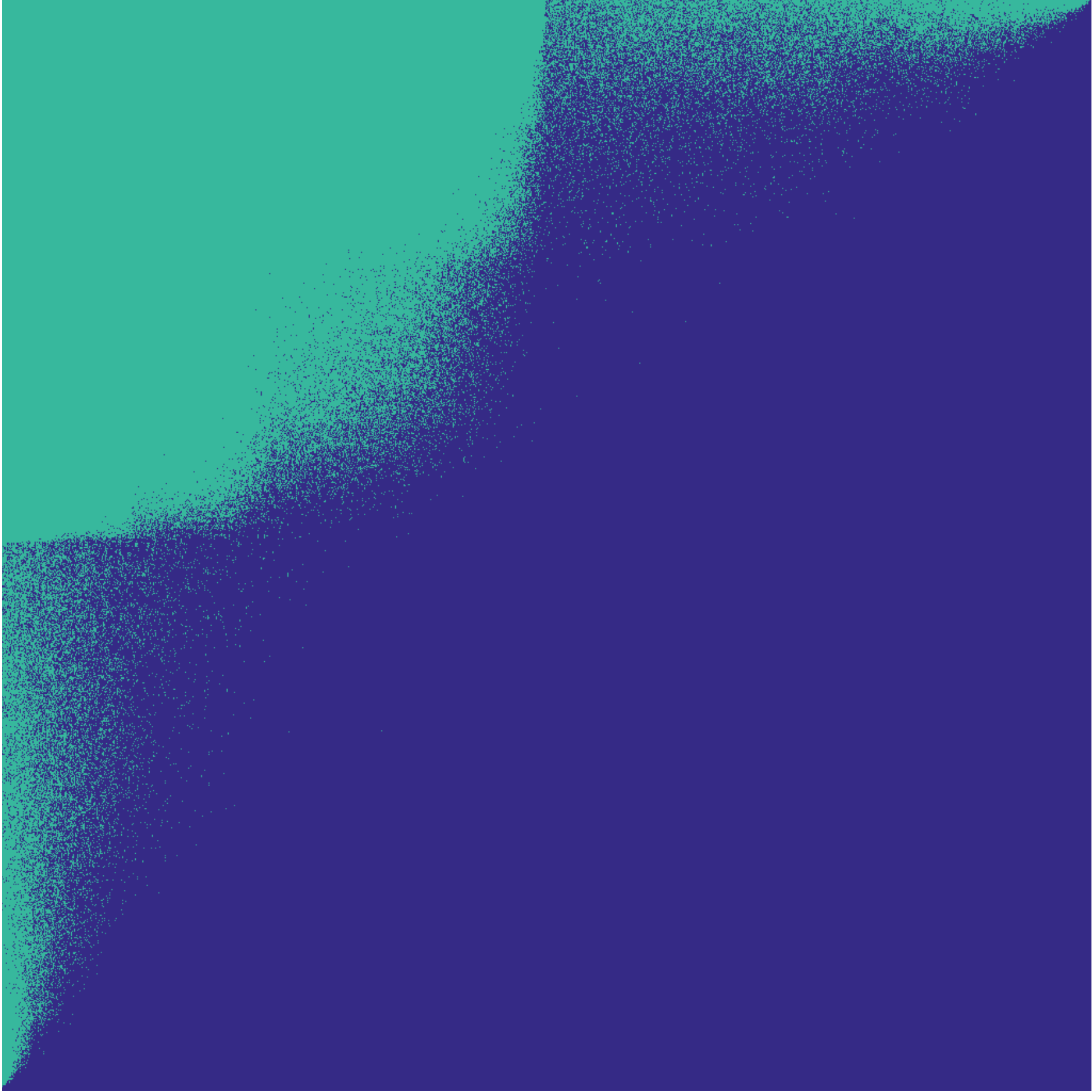} \hspace{1mm} &\hspace{1mm}
\includegraphics[width=0.25\columnwidth]{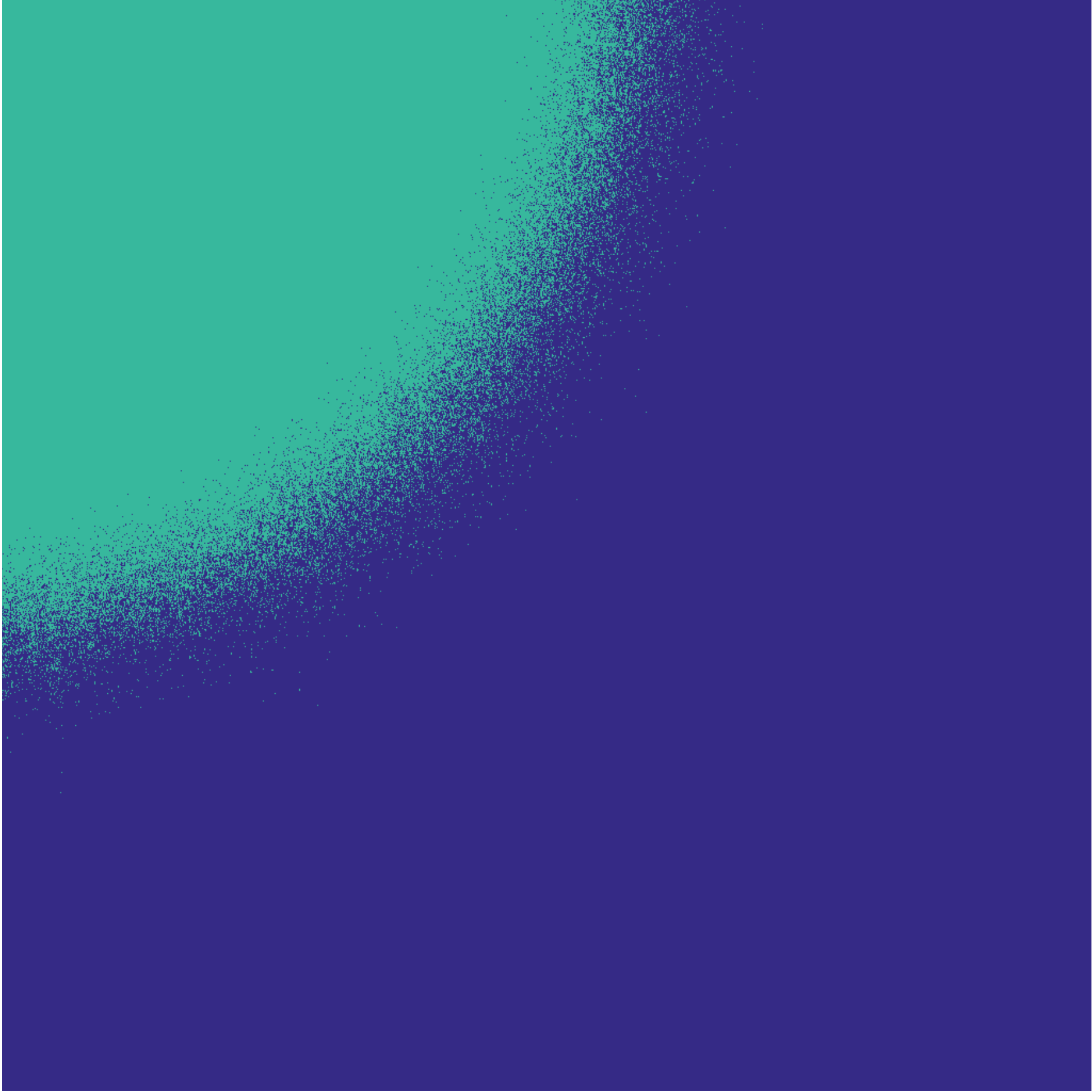} \hspace{1mm} &\hspace{1mm} 
\includegraphics[width=0.25\columnwidth]{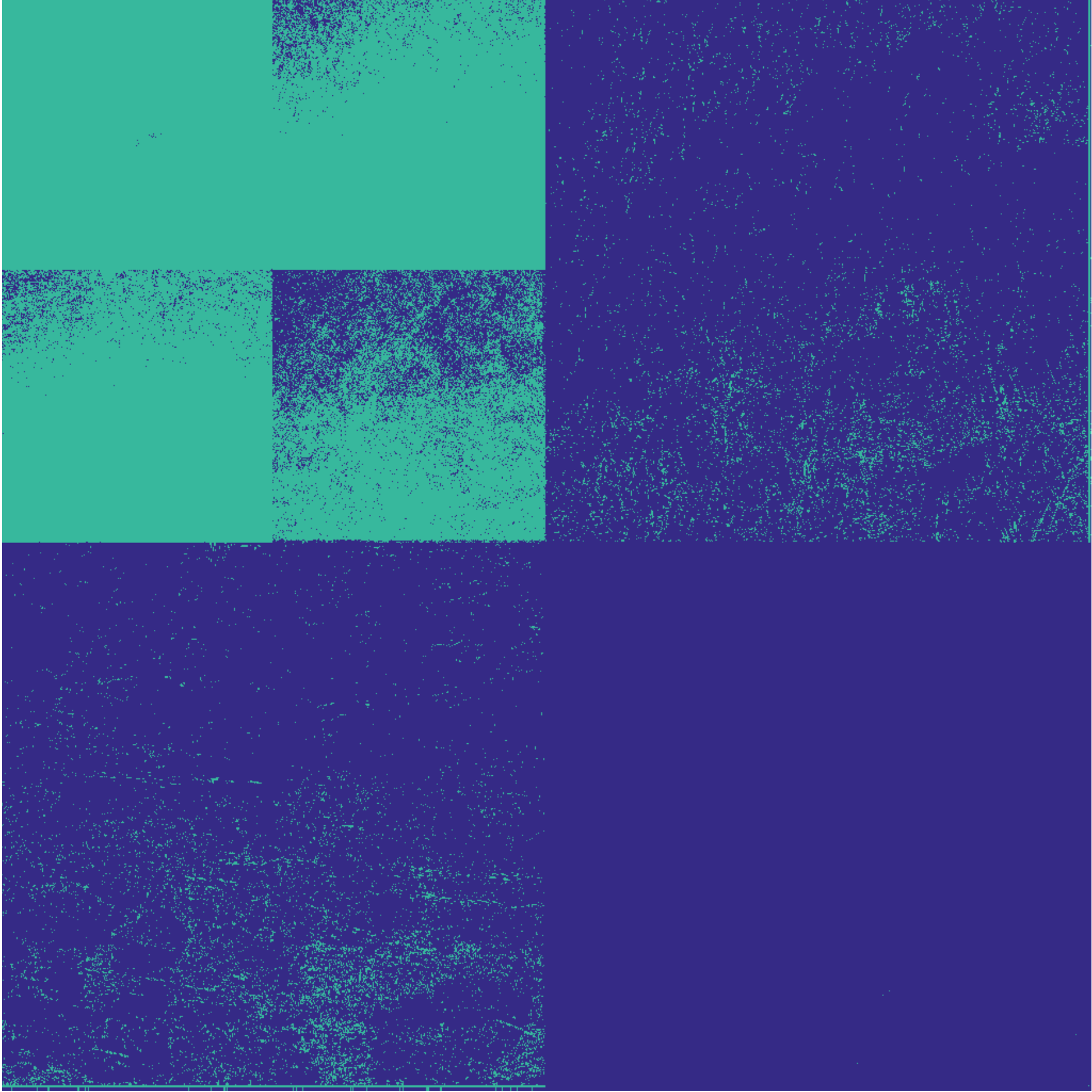}
\end{tabular}
\caption{ (Top) Maximum compression factor to achieve a given $\ell_2$-error performance. (Middle) Average estimation error in conjunction with average PSNR values. (Bottom) Selected indices at 25\% compression with $f_\text{avg}$. }\label{fig:kenya_images_compression_factor_scaling}
\end{figure}

Figure \ref{fig:kenya_images_compression_factor_scaling} illustrates the compression performance of our index selection using the Hadamard, discrete cosine (DCT), and wavelet (Daub-4) transforms. In the figure, we compute the square root of the empirical average of $\|\x - \hat{\x}\|_2^2$  over the test set for a range of $n$, and also determine the best adaptive compression (i.e., the best \emph{image-dependent} $n$-term approximation in the measurement basis)  as the baseline, to reach an error level of $0.1$, $0.05$ and $0.01$.  Here we normalize so that the total energy in each image is $1$.
% We  set up a baseline by considering adaptive best $n$-term approximations of the signals. 

The learned indices for Hadamard-subsampling and DCT-subsampling get within a factor of 2 of their corresponding adaptive compression rates on the test data. In contrast, the learned indices for wavelets are within a factor of 4 of their adaptive compression rates. While the adaptive DCT and wavelet compressions are close, the learned DCT-subsampling obtains compression rates within a factor of 2 of the adaptive wavelet compression. 

Overall, the learned DCT indices obtain the best compression performance on the test data. Moreover, to quadruple the resolution at an error of $0.1$, which corresponds to $25$dB peak signal-to-noise ratio (PSNR), we simply need to double the number of samples. While we observe a similar behavior for the average error level of $0.05$, we need to commensurately increase the number of samples with resolution for the error level of $0.01$, which corresponds to high levels of PSNR values, as shown in  Figure \ref{fig:kenya_images_compression_factor_scaling} (Middle). 

Figure \ref{fig:kenya_images_compression_factor_scaling} (Bottom) shows the learned subsampling patterns. The learned wavelet samples exhibit the expected parent-child correlation on its decomposition hierarchy. Intriguingly, while the learned DCT samples show circular symmetry, the learned Hadamard samples exhibit an unexpected shape which is typically not predicted by randomized approaches. 

% INDICES
\begin{figure}
%\centering
\begin{minipage}{0.7\columnwidth}
\includegraphics[width=\columnwidth]{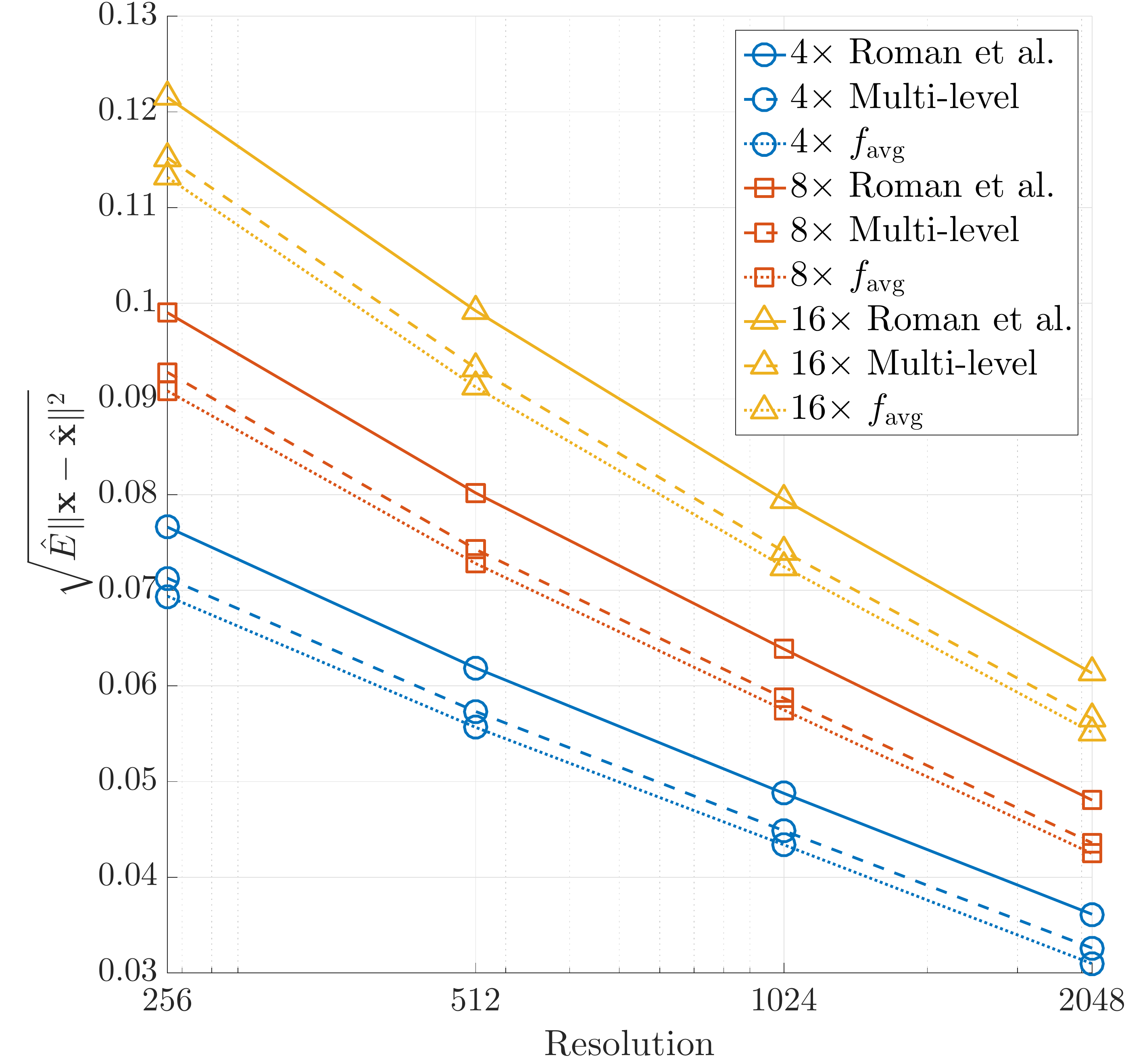}
\end{minipage}
\begin{minipage}{0.25\columnwidth}
\begin{tabular}{c}
\includegraphics[width=1\columnwidth]{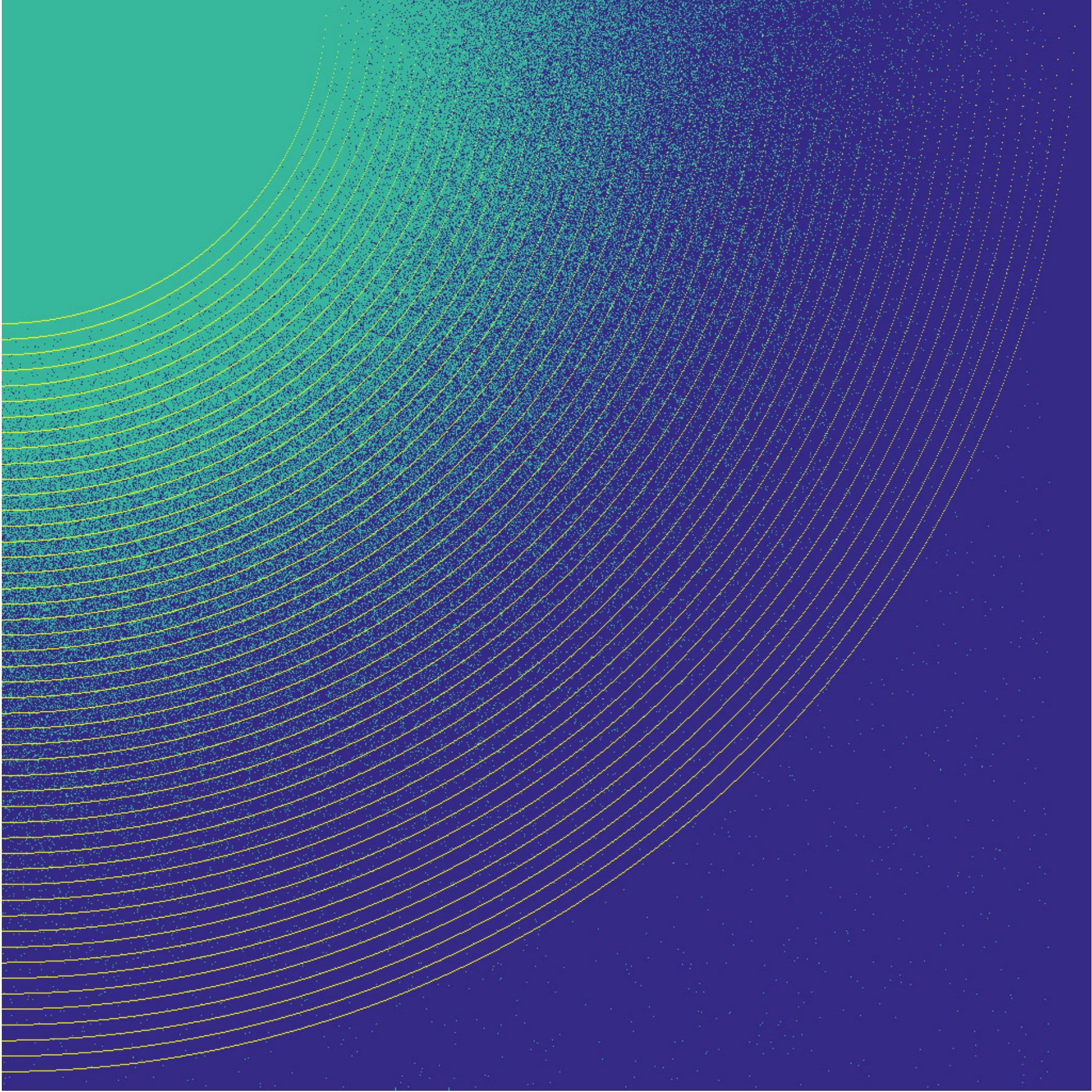} \\
\footnotesize{Roman et al.} \\ 
\includegraphics[width=1\columnwidth]{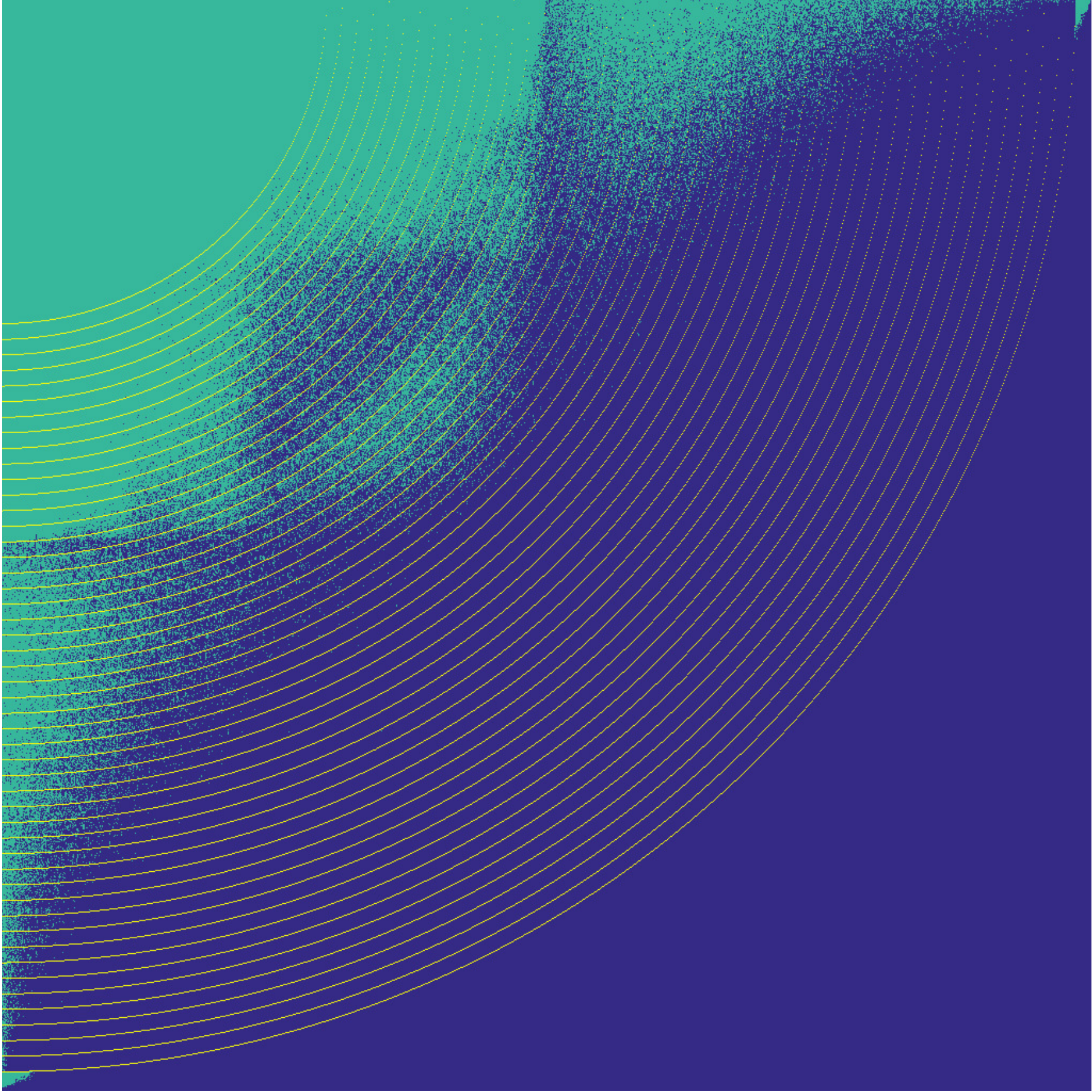} \\
\footnotesize{$f_{\text{avg}}$ with multi-level} \\ 
%\includegraphics[width=1\columnwidth]{FIGURES/kenya_images_train_1024_4x_had_indices} \\
%\footnotesize{$f_{\text{avg}}$} \\ 
\end{tabular} 
\end{minipage} 
\caption{(Left) Recovery performance with the linear decoder. (Right) Tuned randomized variable-density pattern [Top] vs.\ Learned indices with multi-level constraints [Bottom]. The overlayed circles are the non-overlapping  multi-level partitions.}\label{fig:kenya_images_1024_had_patterns} 
\end{figure}

We made use of indices provided to us by the first author of \cite{roman2014asymptotic} for Hadamard sampling, which are based on the randomized approach therein but with tuning done for the goal of effectively compressing natural images.  See Figure \ref{fig:kenya_images_1024_had_patterns} (Right), where a 25\% compression rate is used.  Using these indices and the corresponding non-overlapping partitions (also shown in the figure), we counted the number of samples in each partition and created a corresponding set of constraints (\emph{cf.}, the multi-level sampling defined in Section \ref{sec:STRATEGIES}).  We then applied our learning-based approach using the resulting constraint set $\calA$.  Surprisingly, the shape of the learned pattern differs from the circularly symmetric version and outperforms the randomized indices when we use the linear decoder; see Figure \ref{fig:kenya_images_1024_had_patterns} (Left).   

% ERROR SCALING
\begin{figure}
\centering
\begin{tabular}{ccc}
\footnotesize{4$\times$ compression} & \footnotesize{8$\times$ compression} & \footnotesize{16$\times$ compression} \\
\hspace{-.5cm}
\includegraphics[width=0.33\columnwidth]{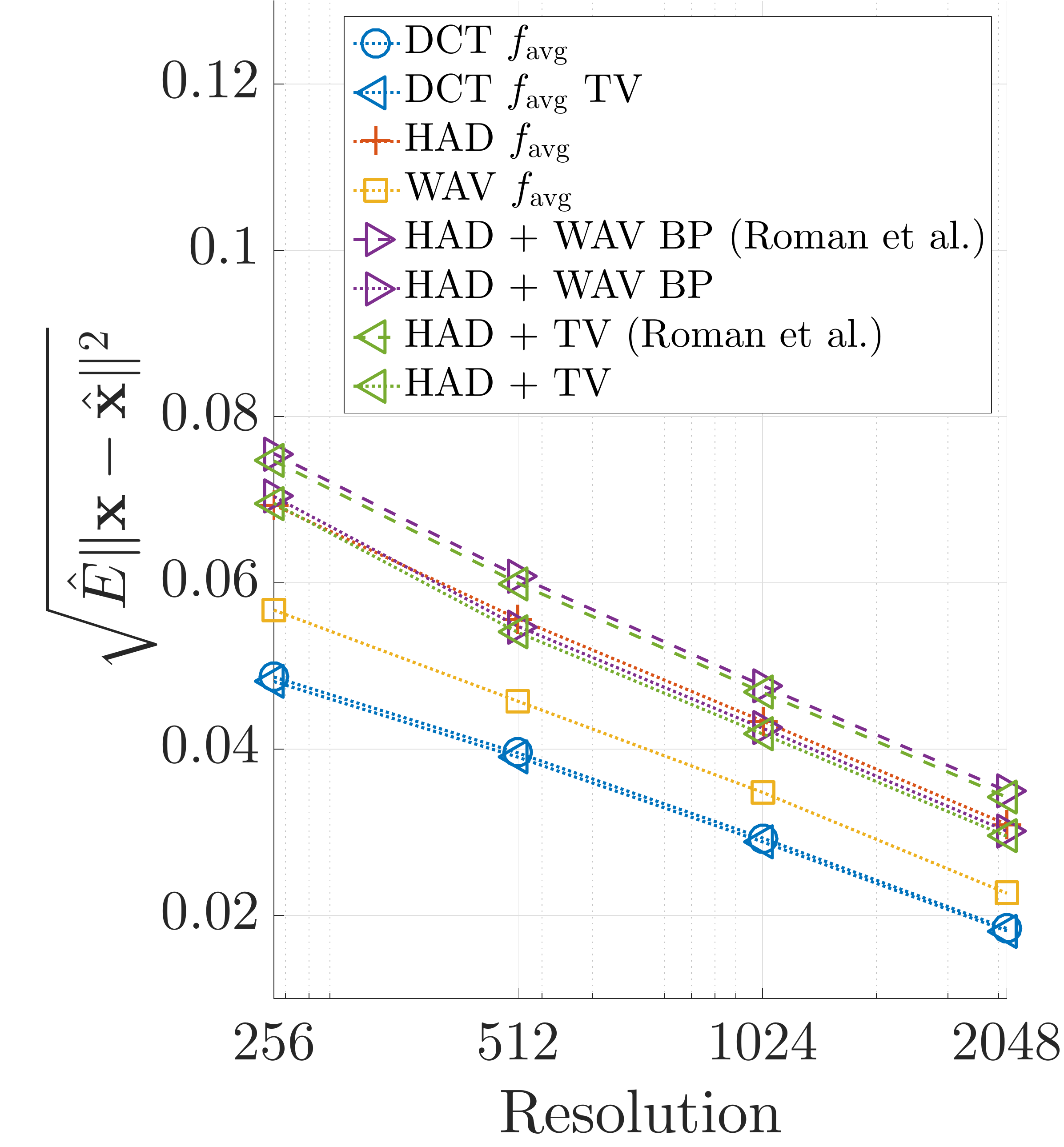} &
\hspace{-.5cm}
\includegraphics[width=0.33\columnwidth]{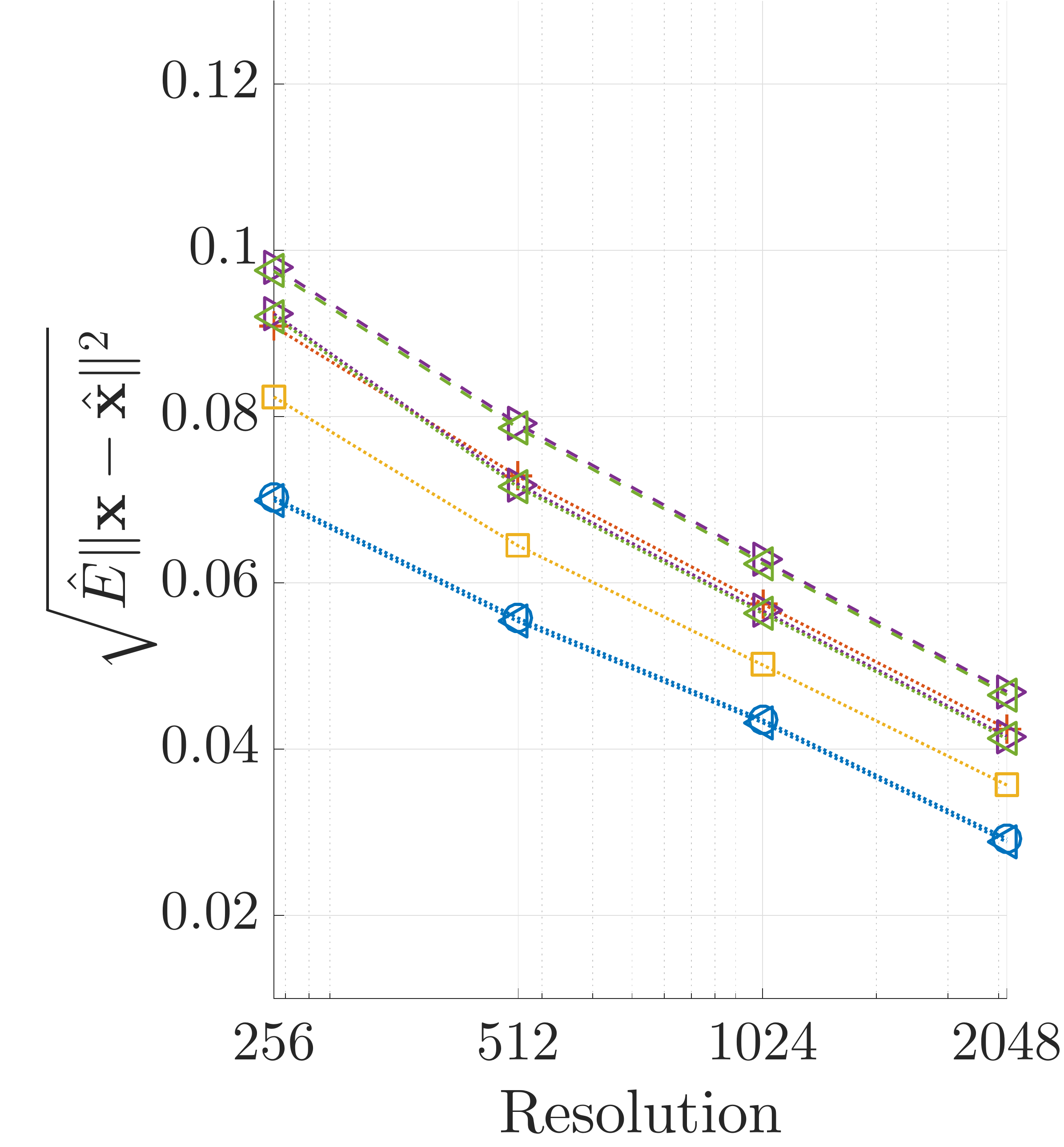} &
\hspace{-.5cm}
\includegraphics[width=0.33\columnwidth]{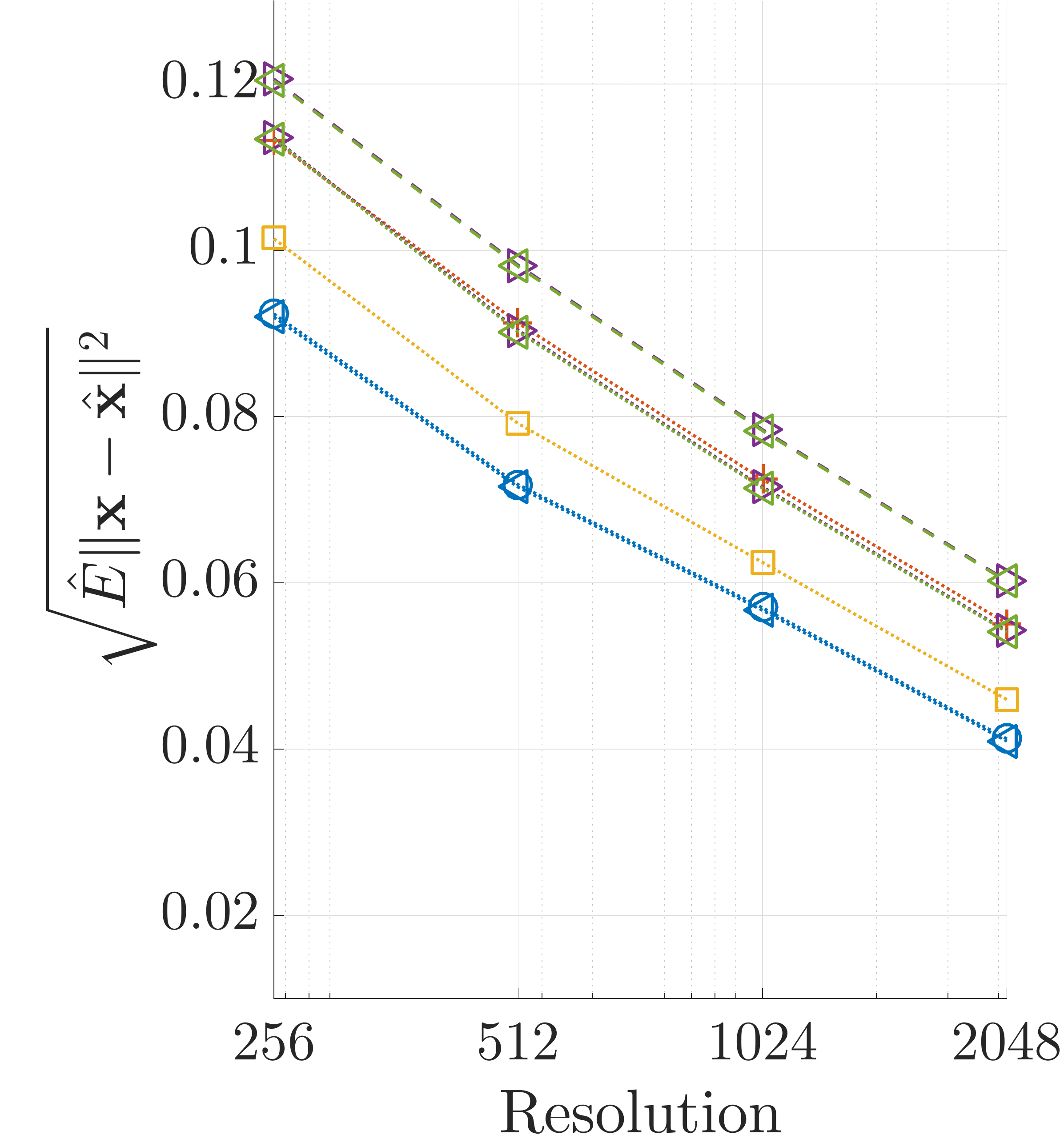}
\end{tabular}
\caption{The performance of the nonlinear decoder vs.\ the linear decoder.  }\label{fig:kenya_images_accuracy_compression}
\end{figure}

% ERROR SCALING - DCT ONLY
\begin{figure}
\centering
\begin{tabular}{ccc}
\footnotesize{4$\times$ compression} & \footnotesize{8$\times$ compression} & \footnotesize{16$\times$ compression} \\
\hspace{-.5cm}
\includegraphics[width=0.33\columnwidth]{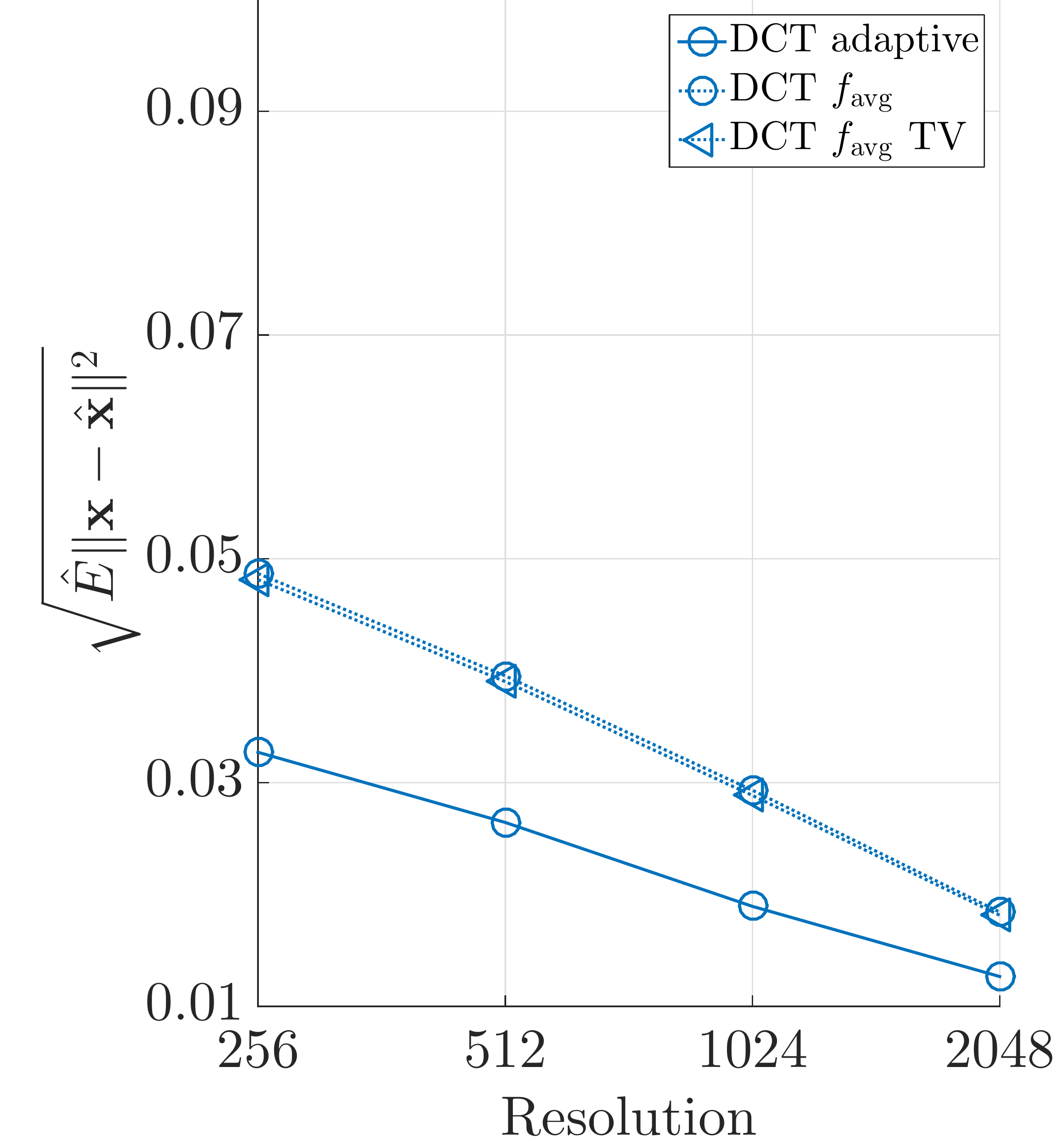} &
\hspace{-.5cm}
\includegraphics[width=0.33\columnwidth]{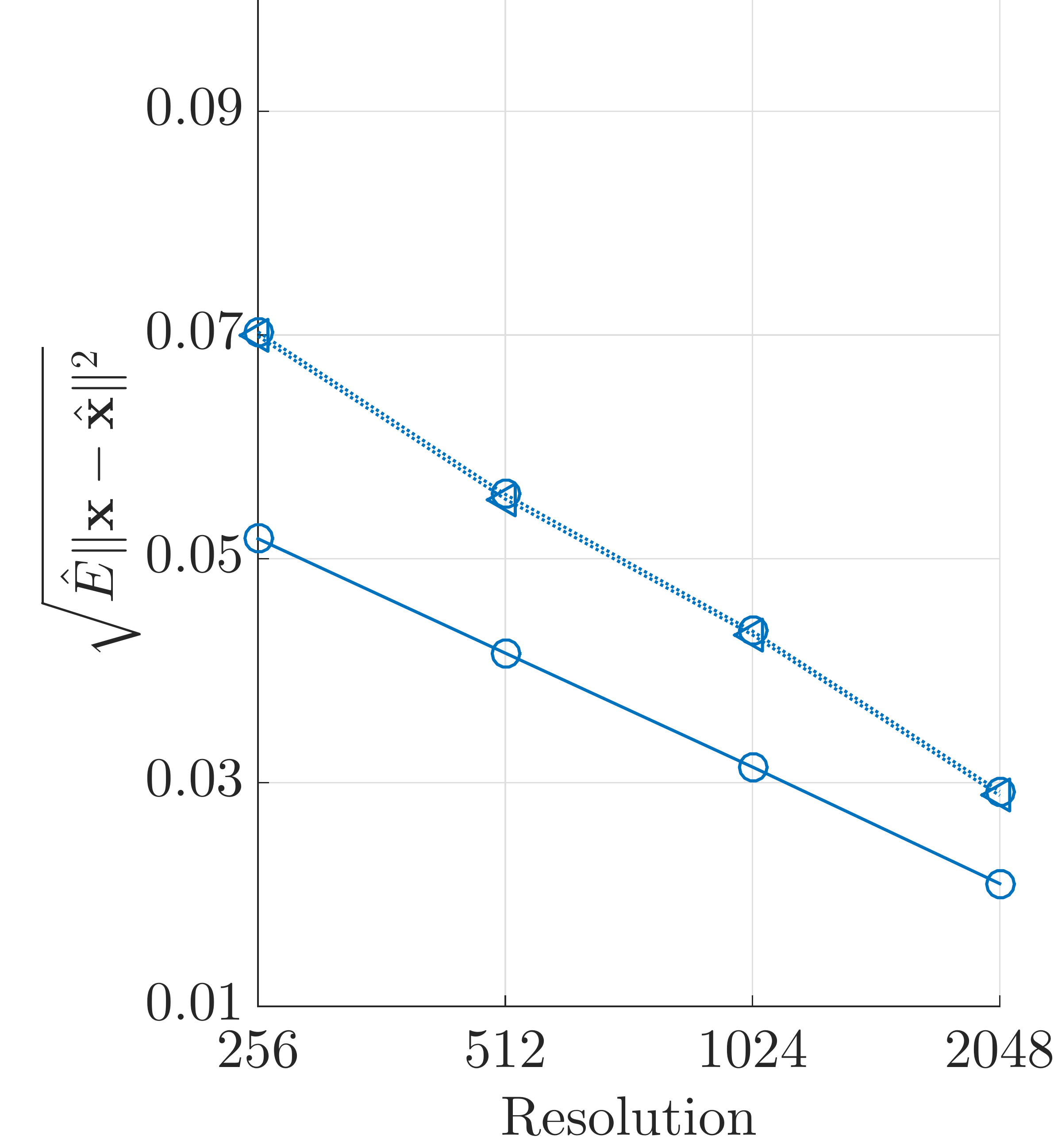} &
\hspace{-.5cm}
\includegraphics[width=0.33\columnwidth]{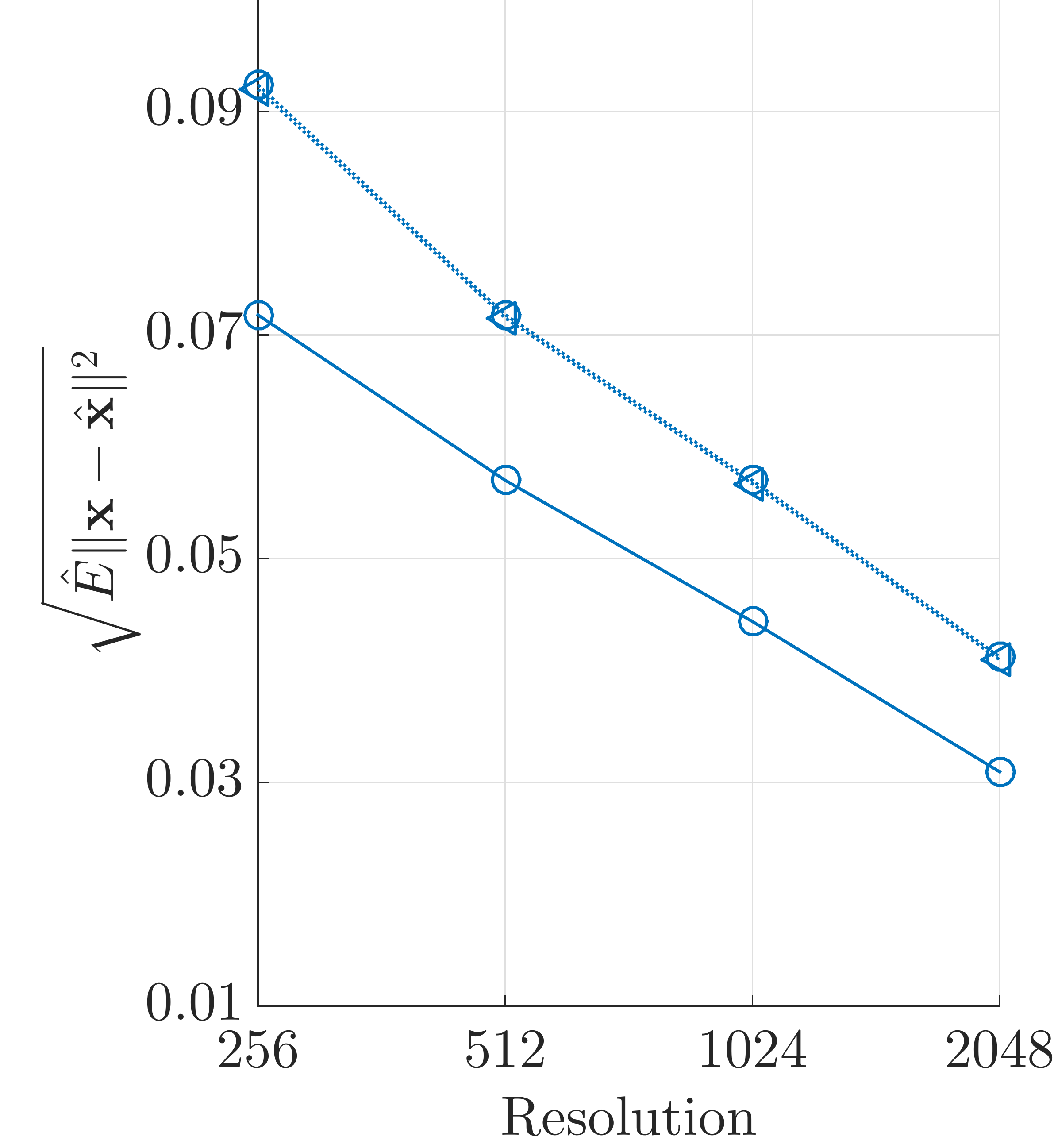}
\end{tabular}
\caption{The performance of the nonlinear decoder vs.\ the linear decoder for DCT measurements only.}\label{fig:kenya_images_accuracy_compression_dct_only}
\end{figure}

The theory associated with the approach of \cite{roman2014asymptotic} requires the basis pursuit (BP) decoder \eqref{eq: BP},  with the sparsity basis $\Phi$ being the wavelet decomposition. Figure \ref{fig:kenya_images_accuracy_compression} illustrates that the overall improvement of this decoder over the linear one is limited, and similarly for an additional non-linear decoder based on minimizing the total variation (TV). 

In summary, for this data set, replacing the random index set coming from the randomized approach with the learned indices leads to a noticeable improvement, whereas the improvements from switching to non-linear decoders are marginal. It should be noted that these decoders incur heavy computational costs; some running times are summarized in Table \ref{tab: times} based on a state-of-the-art primal-dual decomposition method  \cite{trandinh2014primaldual}.
% illustrates the heavy computational burden required for these marginal improvements. 

\begin{table}[!h]
\caption{Kenya images: $\ell_2$-errors vs.\ running times}\label{tab: times}
\centering
\begin{tiny}
\begin{tabular}{|c|c|lll|}	
\hline
\multirow{2}{*}{Resolution} 	& \multirow{2}{*}{Recovery} 		& \multicolumn{3}{c|}{Sampling rate} 		\\ \cline{3-5}
						& 							& $6.25\%$	& $12.50\%$	& $25\%$ \\ \hline
						
\multirow{3}{*}{256} 			& BP 						& $0.102$ / $6$s 		& $0.083$ / $6$s 		& $0.063$ / $6$s 	\\
						& TV 						& $0.102$ / $27$s 		& $0.082$ / $22$s 		& $0.062$ / $20$s \\
						& Adjoint						& $0.103$ / $0.01$s 		& $0.084$ / $0.01$s 		& $0.064$ / $0.01$s 	\\ \hline
\multirow{3}{*}{512}			& BP 						& $0.080$ / $23$s		& $0.063$ / $22$s 		& $0.048$ / $22$s \\
						& TV							& $0.080$ / $151$s		& $0.063$ / $162$s 		& $0.047$ / $153$s \\
						& Adjoint						& $0.081$ / $0.03$s 		& $0.064$ / $0.03$s 		& $0.049$ / $0.02$s 	\\ \hline
\multirow{3}{*}{1024} 		& BP							& $0.062$ / $85$s		& $0.049$ / $85$s 		& $0.036$ / $93$s \\
						& TV							& $0.062$ / $340$s 		& $0.049$ / $614$s 		& $0.036$ / $65$s \\
						& Adjoint						& $0.063$ / $0.08$s 		& $0.050$ / $0.08$s 		& $0.037$ / $0.09$s 	\\ \hline
\multirow{3}{*}{2048}			& BP							& $0.047$ / $381$s 		& $0.036$ / $366$s 		& $0.026$ / $333$s \\
						& TV							& $0.047$ / $1561$s 	& $0.036$ / $2501$s 	& $0.025$ / $2560$s \\
						& Adjoint						& $0.048$ / $0.26$s 		& $0.037$ / $0.29$s 		& $0.027$ / $0.28$s 	\\ \hline
\end{tabular}

\end{tiny}
\end{table}

\begin{figure}
\centering
\begin{footnotesize}
 ImageNet ILSVRC2010 data set details
 \begin{tabular}{|c|c|c|c|}
\hline
Resolution 	& Train 		& Test		& $\frac{m}{p}$	\\ \hline
256 	$\times$ 256		& $1,035,802$ 	& $124,235$ 	& $15.8$ 		\\
512	$\times$ 512		& $52,736$ 	& $8,556$ 	& $0.2$		\\
1024 $\times$ 1024		& $10,445$ 	& $1,474$ 	& $0.01$ 		\\ \hline
\end{tabular}
\end{footnotesize}
\begin{tabular}{ccc}
\scriptsize{256 $\times$ 256} & \scriptsize{512 $\times$ 512} & \scriptsize{1024 $\times$ 1024} \\
\hspace{-.5cm}
\includegraphics[width=0.35\columnwidth]{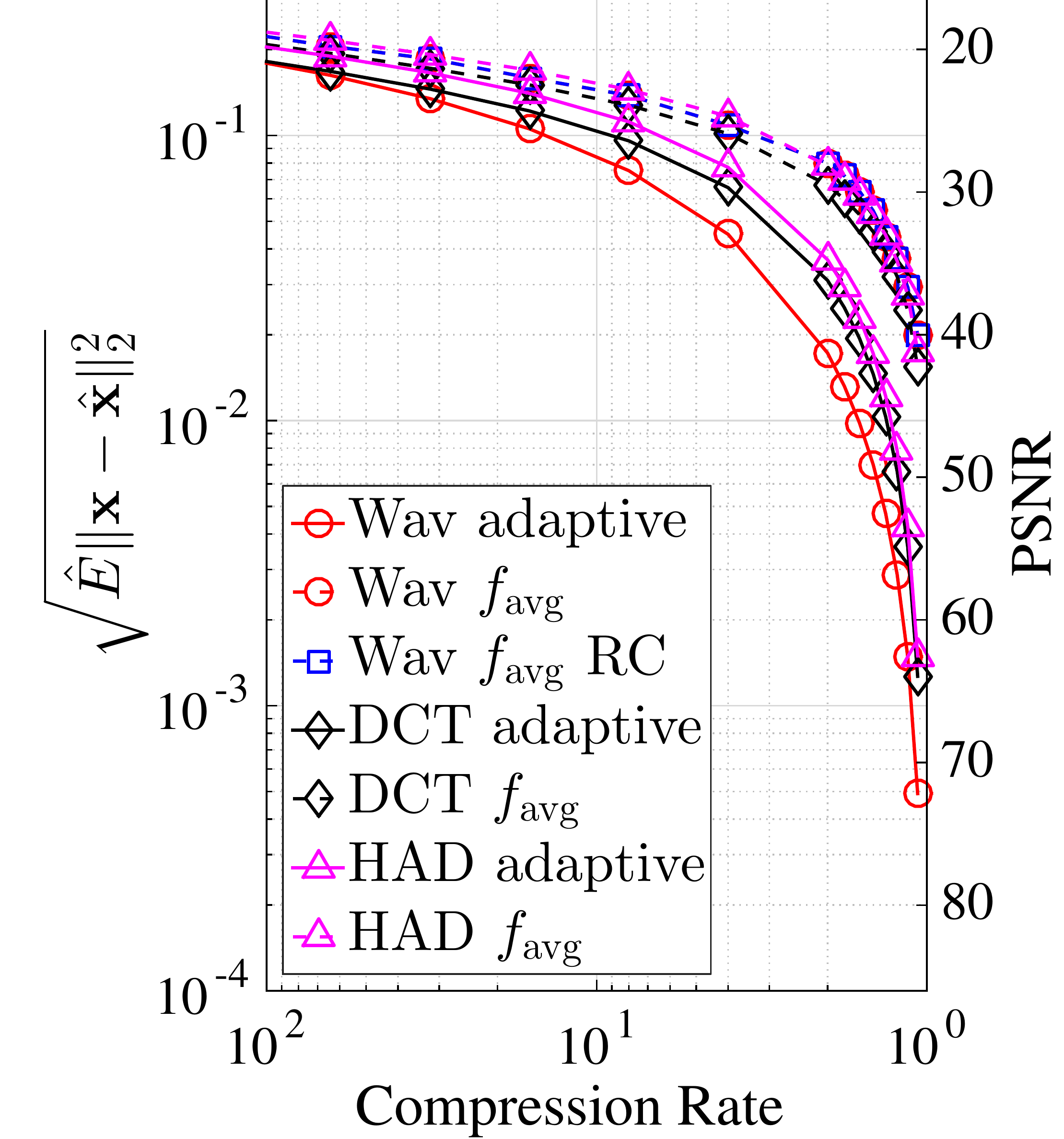} &
\hspace{-.5cm}
\includegraphics[width=0.35\columnwidth]{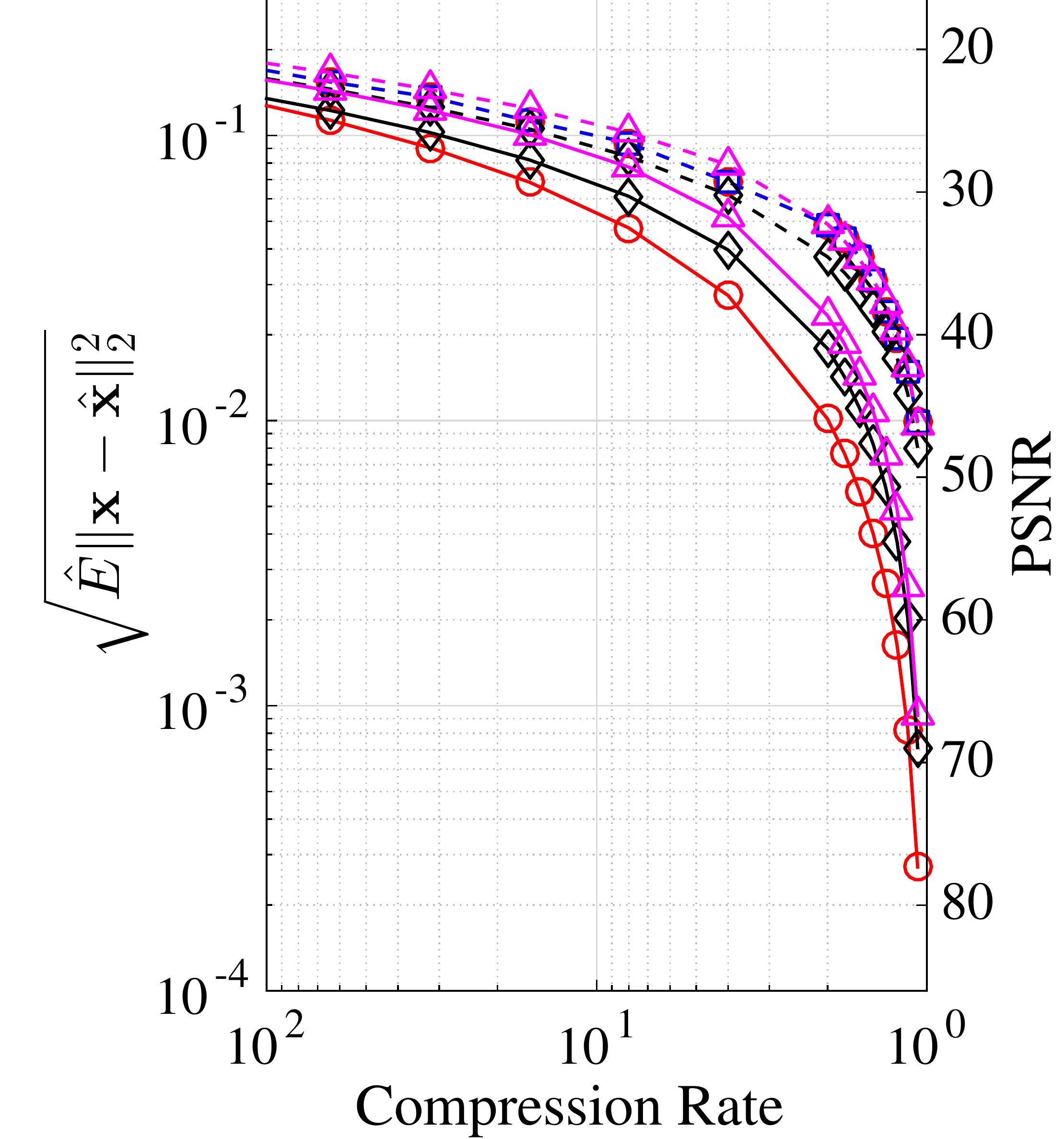} &
\hspace{-.5cm}
\includegraphics[width=0.35\columnwidth]{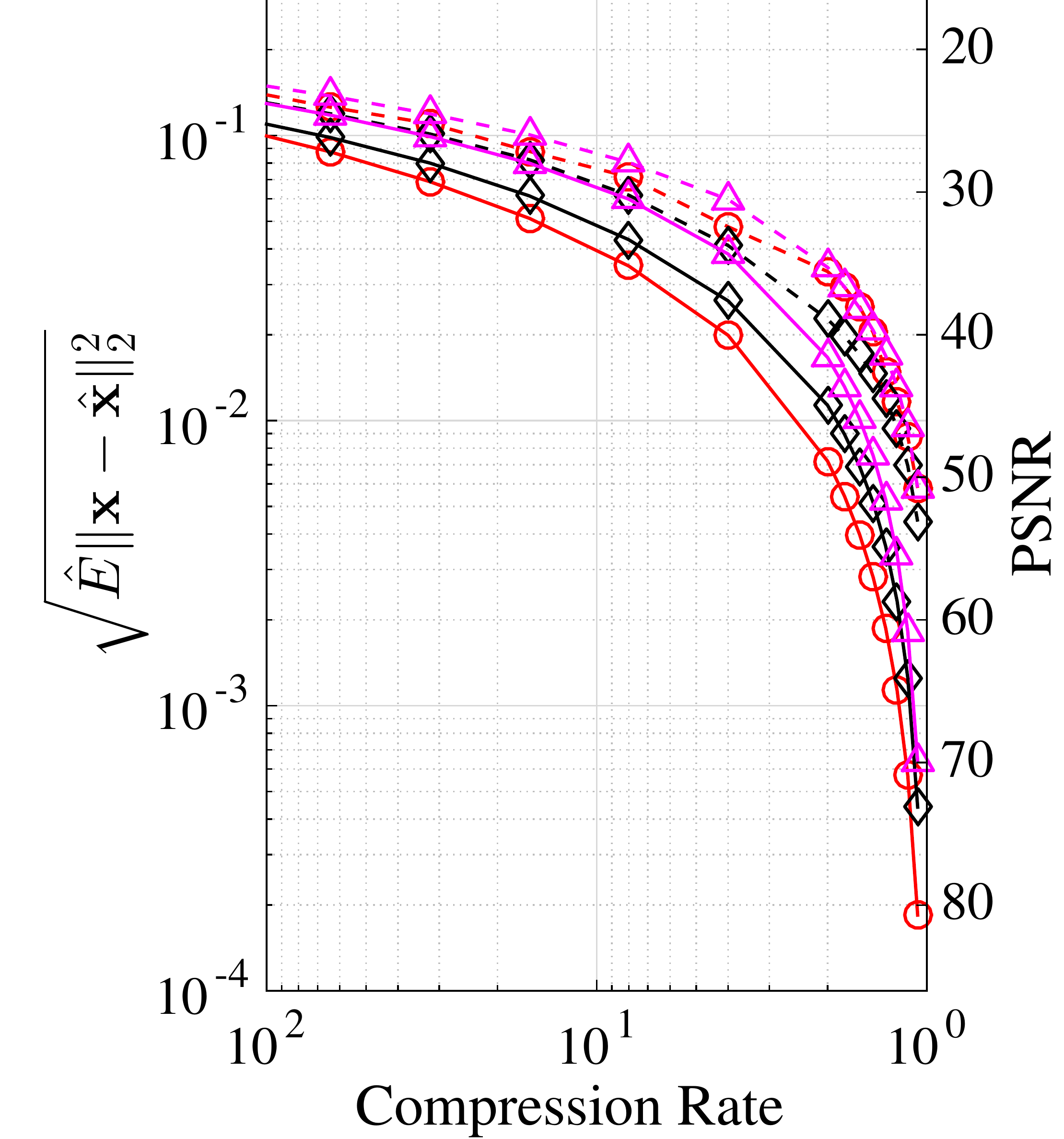} \\
\end{tabular}
\begin{tabular}{ccc}
\footnotesize{Hadamard} & \footnotesize{DCT} & \footnotesize{Wavelets} \\
\includegraphics[width=0.3\columnwidth]{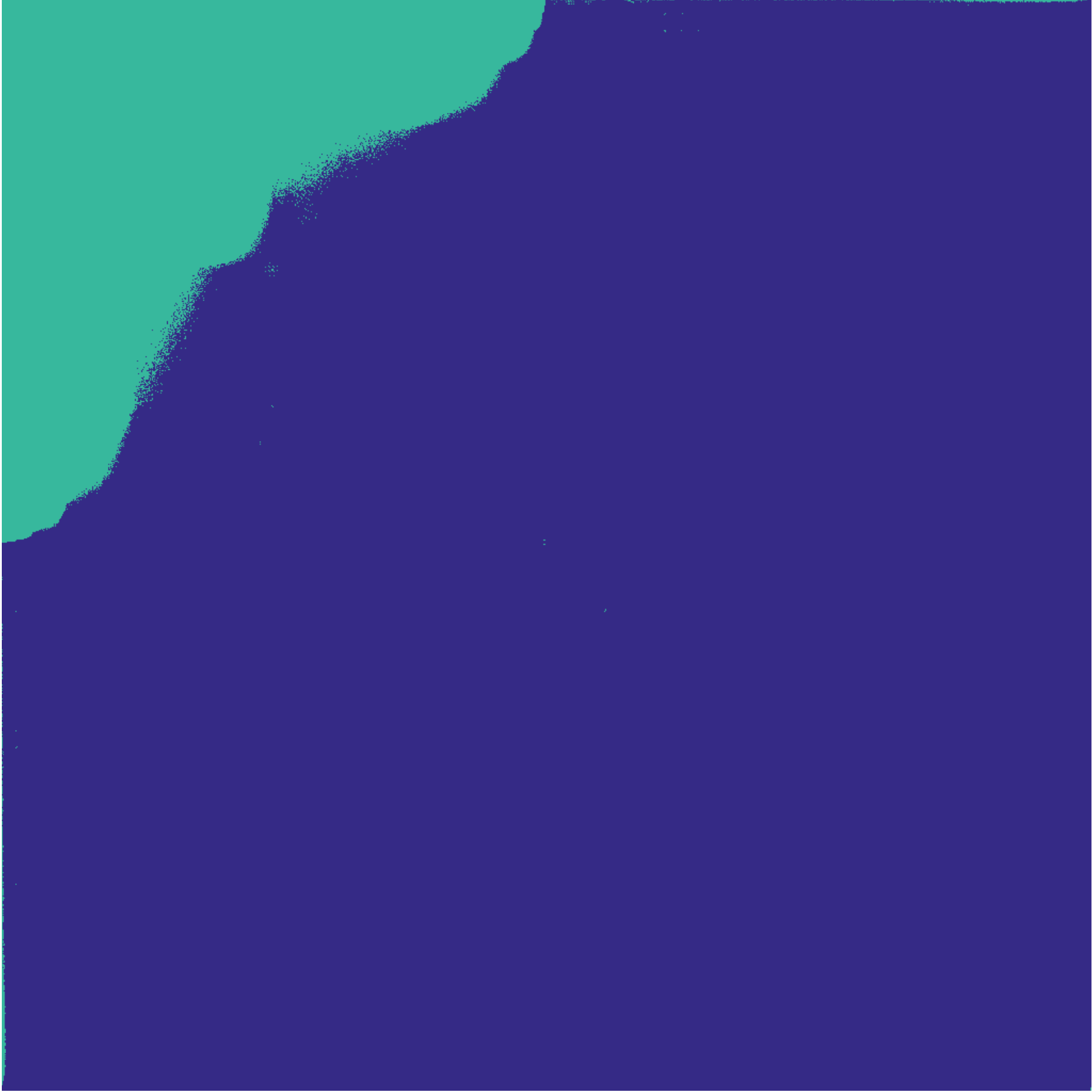} &
\includegraphics[width=0.3\columnwidth]{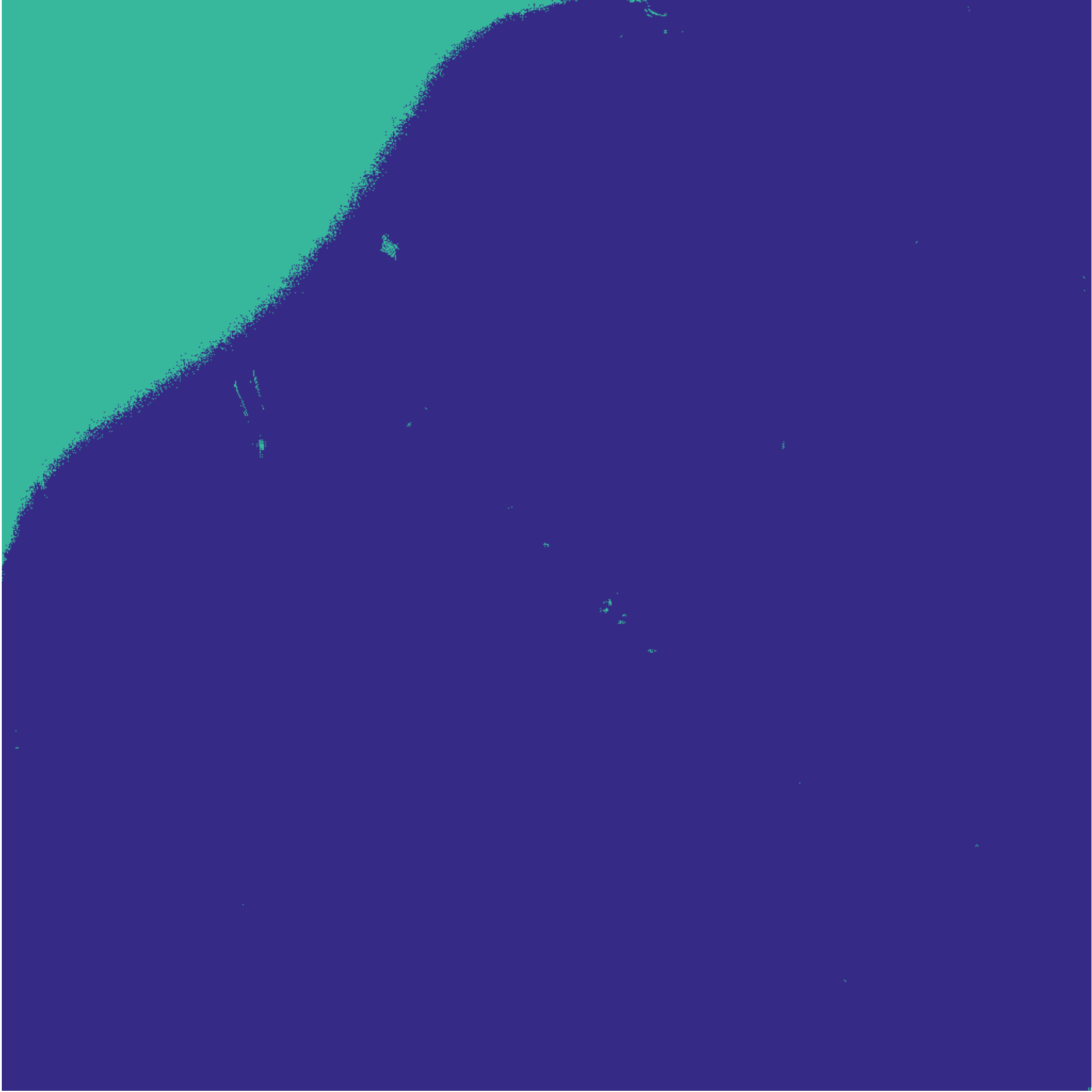} &
\includegraphics[width=0.3\columnwidth]{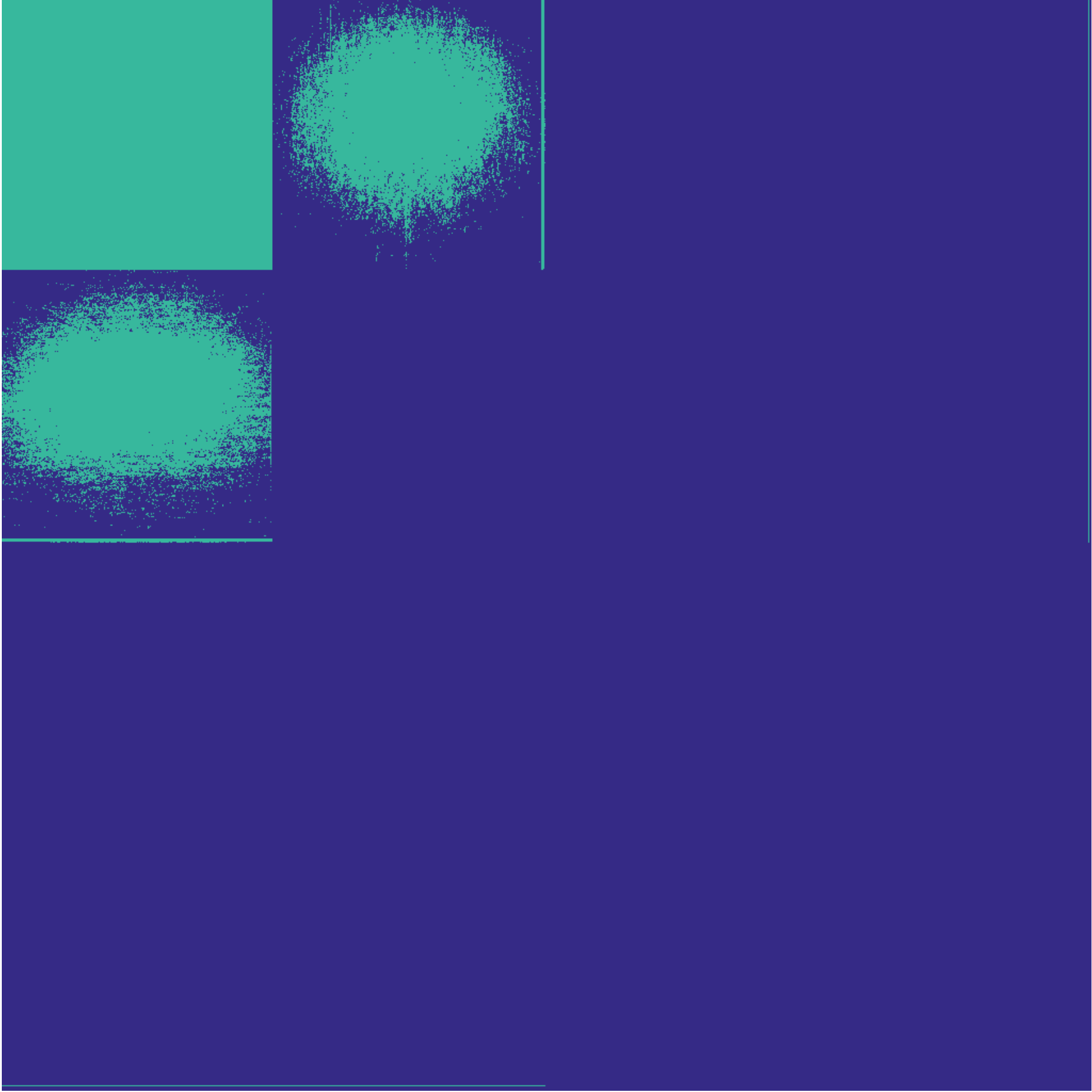}
\end{tabular}

\caption{ (Top) The test and training split sizes on the ImageNet data set. (Middle) Average estimation error in conjunction with average PSNR values. (Bottom) Selected indices at 12.5\% compression with $f_\text{avg}$. }\label{fig: imagenet}
\end{figure}

As the Kenya data set size is quite limited, we performed similar experiments on a much larger data set called ImageNet.\footnote{Available at \url{http://image-net.org}} In particular, we consider the 2010 ImageNet Large Scale Visual Recognition Challenge data set. This data set consists of approximately 1.4 million images of various resolutions; see Figure \ref{fig: imagenet} (top) for our training and test splits. Since the data is already JPEG compressed with 8 bits, we crop the existing high-resolution images in the central part of the image in order to obtain low-resolution ones.

While the images are already compressed, Figure \ref{fig: imagenet} (Middle) illustrates that the qualitative behavior we have seen so far does not change: Wavelets achieve the best adaptive compression and the learned DCT subsampling performs the best in the test data. We also observe that the learned indices for the rooted-connected (RC) wavelet tree model generalize marginally better than the learned wavelet sampling. Finally, Figure \ref{fig: imagenet} (Bottom) shows that the distribution of the sampling patterns for Hadamard and DCT  do not exhibit circularly symmetric shapes. Moreover, the learned wavelet indices appear to concentrate on the spatial center of the images. 
%
%\begin{table}[!t]
%\centering
%\caption{ImageNet ILSVRC2010 data set details}\label{tab: imnet}
%\begin{tabular}{|c|c|c|c|}
%\hline
%Resolution 	& Train 		& Test		& $\frac{m}{p}$	\\ \hline
%256 	$\times$ 256		& $1,035,802$ 	& $124,235$ 	& $15.8$ 		\\
%512	$\times$ 512		& $52,736$ 	& $8,556$ 	& $0.2$		\\
%1024 $\times$ 1024		& $10,445$ 	& $1,474$ 	& $0.01$ 		\\ \hline
%\end{tabular}
%\end{table}

% ERROR VS SAMPLE SIZE

%% file: exp_ieeg.tex
\subsection{iEEG Data set}
This subsection focuses on an intracranial electro-encephalography (iEEG) data set from the \texttt{iEEG.org} portal.
%The data is freely available for researchers to study the onset of seizures. 

Currently, iEEG is an invasive procedure that permits the recording of neuro-electrical signals via an array of electrodes placed directly on the brain surface.
%The patient needs to be constantly monitored since a small portion of his scalp is left open for the wires to pass through, limiting the recording time to few days at most.
A very active research area concerns the design of implantable wireless devices that do not require the patient to be tethered to an external recorder.
%would the patient to continue his normal life while his brain signals are being recorded.
%These devices need to be ultra-small and consume as little power as possible. 
Efficient compression schemes based on compressive sensing have been recently proposed in order to reduce transmission power consumption \cite{laska2007theory, chen2012design, shoaran2014compact}; however, they may use a large chip area \cite{chen2012design} or not be able to compress beyond the number of channels \cite{shoaran2014compact}.

We show the effectiveness of our approach on the micro-electrode recordings from the \texttt{I001\_P034\_D01} data set.
We consider only the recordings annotated as seizures, with the exception of seizure $7$, which is corrupted.  Moreover, we remove channel 1, which is inactive, and channel 7, which erroneously records the AC signal powering the system instead of the neuronal signal.
We define signal windows of $p = 1024$ samples and use seizures $1,2,3,4,5,6,8$ for learning the indices of the subsampled Hadamard transform for compression, for a total of $5.8$ million samples over $5$ channels.
We test the reconstruction performance of the chosen indices on channels $1,2,3, 5$ and $6$ on the last seizure.
For the randomized variable-density sampling approach, we consider non-linear reconstruction via wavelets and a tree-structure promoting norm, shown in \cite{baldassarre2015structured} to yield the best performance on this type of data.  For our learning-based approach, we again use the simple linear decoder given in \eqref{eq:linear_dec}.

\begin{figure}
\centering
\begin{minipage}{0.45\columnwidth}
%\begin{tabular}{cc}
%\includegraphics[width=\columnwidth]{FIGURES/ieeg_data_I001-P034-D01_seizure_16_clean_hadamard_errors.pdf}
%\includegraphics[width=\columnwidth]{FIGURES/ieeg_I001_P034_D01_micro_test_hadamard_errors.pdf}
\includegraphics[width=\columnwidth]{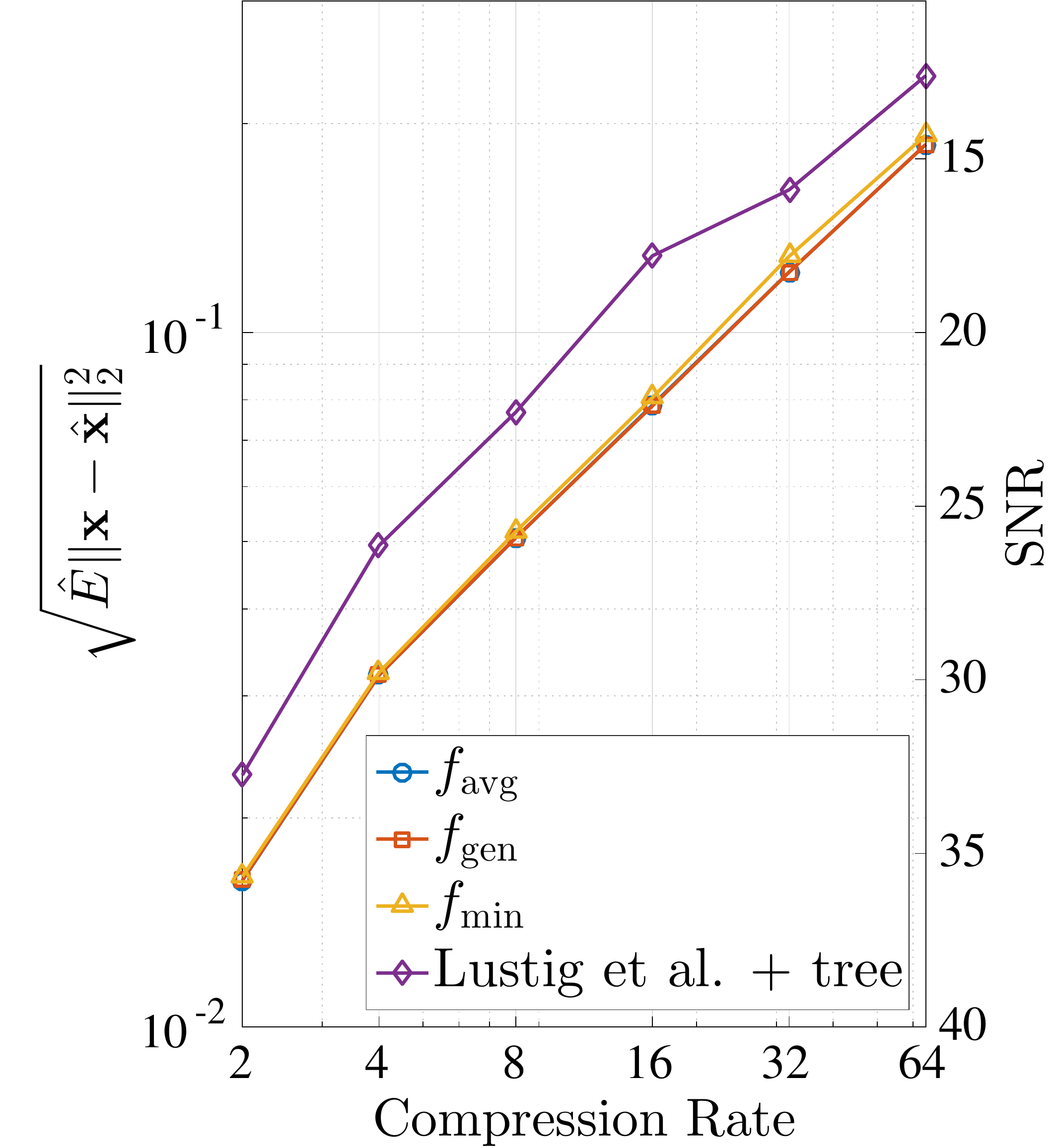}
\end{minipage}
\begin{minipage}{0.45\columnwidth}
	\begin{tabular}{c}
%		$16 \times$\\
%		\includegraphics[width=\columnwidth]{FIGURES/ieeg_data_I001-P034-D01_seizure_16_clean_compression_16_sample_121_reconstruction.pdf} \\
		\includegraphics[width=\columnwidth]{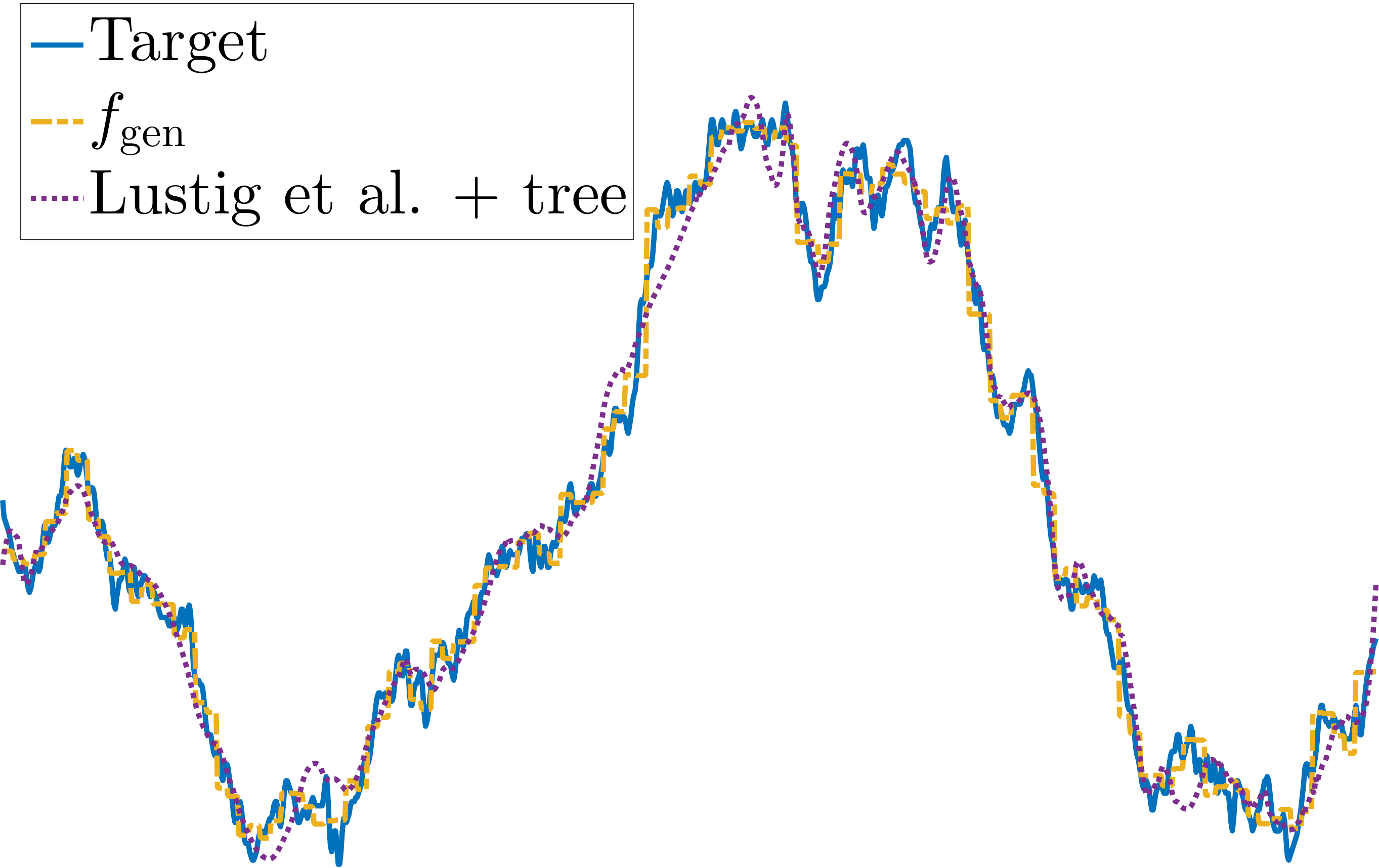} \\
%		$32 \times$\\
%		\includegraphics[width=\columnwidth]{FIGURES/ieeg_data_I001-P034-D01_seizure_16_clean_compression_32_sample_121_reconstruction.pdf} 
		\includegraphics[width=\columnwidth]{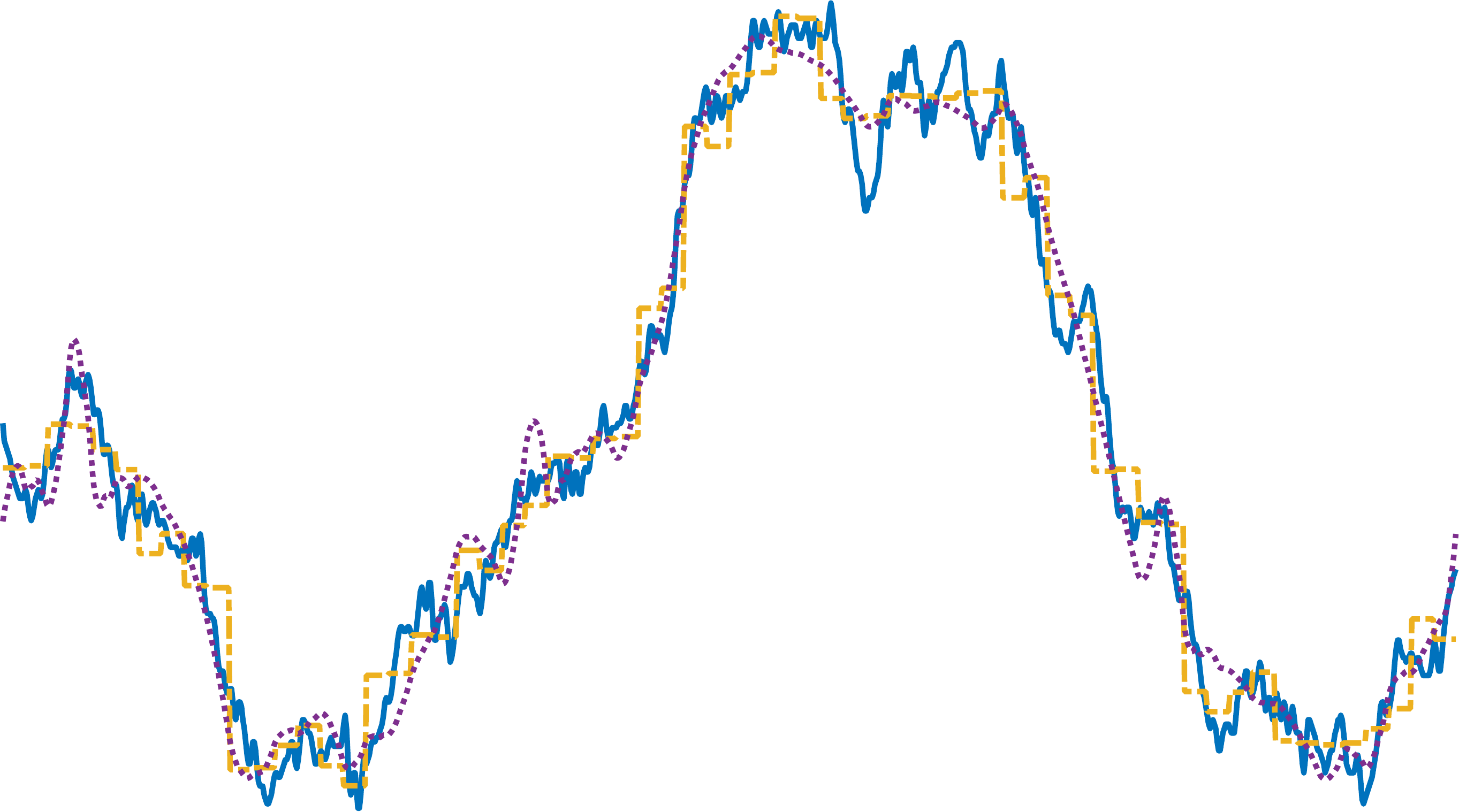} 
	\end{tabular}
%\end{tabular}
\end{minipage}
\caption{iEEG data: Recovery performance (Left) and recovery examples (right): $16 \times$ compression (Top) and $32 \times$ compression (Bottom).} \label{fig: iEEG}
\end{figure}

Figure \ref{fig: iEEG} (Left) illustrates that the learning-based approach outperforms the randomized variable-density approach.
Note that here we use the function $\fgen$ with $g(\alpha) = 1 - (1-\alpha)^2$, as introduced in Section \ref{sec:F_GEN}.  
We use Hadamard sampling, since it is easy to implement in digital hardware \cite{chen2012design}. 
The density function of \cite{lustig2007sparse} is parametrized by the radius of fully sampled region, $r$, and the polynomial degree, $d$. 
We choose the parameters' values that yield the lowest reconstruction error on the training set for each compression rate in the ranges $r = 
\frac{4}{2p}, \frac{5}{2p}, \ldots, \frac{16}{2p}$ and $d = 1, 2, \ldots, 75$, taking the best realization over 20 random draws in each case.

% We then use these values to draw $20$ different realizations for each compression rate on each signal of the test set and report the average reconstruction error.
%
% The resulting optimized parameters were as follows for the compressions ratios 2, 4, 8, 16, 32, and 64 respectively: ($d$ values) 6, 9, 8, 22, 33, and 70; ($r$ values multiplied by $2p$) 16, 5, 11, 12, 14, and 6.

%, and take the realization with the best performance on the training set. 

%For the randomized approach, we have manually tuned\footnote{The best parameters were $d = 15$ and $r = 0.03$ for compression rates up to $16\times$, $d = 45$ and $r = 0.03$ for $32\times$, and $d = 70$ and $r = 0.015$ for $64\times$ compression, see Section \ref{sec:exp_mri} for more details about these parameters.} the density function of \cite{lustig2007sparse} in order to obtain the best training performance and drew $100$ different realizations for each compression rate. 
Figure \ref{fig: iEEG} (Left) shows example reconstructions at $16\times$ and $32\times$ compressions. 
A likely reason for the reduced error due to our approach is that we do not assume any shape for the distribution of the indices. 

In Table \ref{tab:ieeg_rips}, we present the values of $\favg$, $\fgen$ and $\fmin$ obtained by our learning-based algorithms for various compression ratios with $p=1024$.  Note that we normalize the signals to have unit energy, and hence $1$ is the highest possible value of each objective. As expected, the procedure trained for a given objective always gives the best value of that objective on the training data.  However, it is sometimes the case that the indices obtained for optimizing $\favg$ and $\fgen$ yield a better value of $\fmin$ on the \emph{test} set compared to the indices used for optimizing $\fmin$ itself; for example, see the final column corresponding to $4\times$ or $32\times$ compression.

\begin{table}
\centering
\tiny
\caption{\label{tab:ieeg_rips} Objective values obtained for various recovery criteria on the iEEG data set with $p=1024$. CR stands for compression rate.}
\begin{tabular}{|*{8}{c|}}
\hline
\multirow{3}{*}{CR} 			& \multirow{3}{*}{Criterion}	& \multicolumn{6}{c|}{Metric} \\ \cline{3-8}
 							& 					& \multicolumn{2}{c|}{$f_\mathrm{avg}$} 	& \multicolumn{2}{c|}{$f_\mathrm{gen}$}	& \multicolumn{2}{c|}{$f_\mathrm{min}$} \\ \cline{3-8}
 							& 					& Train 	& Test	& Train 	& Test	& Train 	& Test	\\ \hline
\multirow{3}{*}{2$\times$}		& $f_\mathrm{avg}$		& 0.9980  & 0.9975 & 0.999998 & 0.999997 & 0.9955 & 0.9945	\\ \cline{2-8}
 							& $f_\mathrm{gen}$  	& 0.9980  & 0.9975 & 0.999998 & 0.999997 & 0.9955 & 0.9945	\\ \cline{2-8}
 							& $f_\mathrm{min}$		& 0.9977 & 0.9972 & 0.999997 & 0.999996 & 0.9963 & 0.9943 	\\ \cline{1-8}
\multirow{3}{*}{4$\times$}		& $f_\mathrm{avg}$		& 0.9919 & 0.9902 & 0.99997  & 0.99995  & 0.9843 & \bf 0.9804 \\ \cline{2-8}
 							& $f_\mathrm{gen}$  	& 0.9919 & 0.9901 & 0.99997  & 0.99995  & 0.9843 & \bf 0.9805	\\ \cline{2-8}
 							& $f_\mathrm{min}$		& 0.9913 & 0.9896 & 0.99996  & 0.99994  & 0.9862 & \bf 0.9797	\\ \cline{1-8}
\multirow{3}{*}{8$\times$}		& $f_\mathrm{avg}$		& 0.9790  & 0.9756 & 0.9998   & 0.9997   & 0.9534 & 0.9539 	\\ \cline{2-8}
 							& $f_\mathrm{gen}$  	& 0.9789 & 0.9757 & 0.9998   & 0.9997   & 0.9555 & 0.9541	\\ \cline{2-8}
 							& $f_\mathrm{min}$		& 0.9747 & 0.9724 & 0.9997   & 0.9996   & 0.9614 & 0.9504	\\ \cline{1-8}
\multirow{3}{*}{16$\times$}		& $f_\mathrm{avg}$		& 0.9466 & 0.9407 & 0.9987   & 0.9982   & 0.8501   & 0.8063	\\ \cline{2-8}
 							& $f_\mathrm{gen}$  	& 0.9460  & 0.9422 & 0.9987   & 0.9983   & 0.8633  & 0.8338	\\ \cline{2-8}
 							& $f_\mathrm{min}$		& 0.9372 & 0.9330  & 0.9983   & 0.9977   & 0.8737  & 0.8149	\\ \cline{1-8}
\multirow{3}{*}{32$\times$}		& $f_\mathrm{avg}$		& 0.8642 & 0.8618 & 0.9919    & 0.9906   & 0.6734  & 0.6921	\\ \cline{2-8}
 							& $f_\mathrm{gen}$  	& 	0.8642 & 0.8618 & 0.9922    & 0.9906   & 0.6731  & 0.6920 \\ \cline{2-8}
 							& $f_\mathrm{min}$		& 	0.8254 & 0.8299 & 0.9873   & 0.9861   & 0.6970  & 0.6522 \\ \cline{1-8}
\multirow{3}{*}{64$\times$}		& $f_\mathrm{avg}$		& 0.6643 & 0.6883 & 0.9554   & 0.9539   & 0.4549  & 0.2939	\\ \cline{2-8}
 							& $f_\mathrm{gen}$  	& 0.6643 & 0.6883 & 0.9554   & 0.9539   & 0.4550  & 0.2944 	\\ \cline{2-8}
 							& $f_\mathrm{min}$		& 0.6296 & 0.6639 & 0.9462   & 0.9471    & 0.4815  & 0.2946 \\ \hline
												
\end{tabular}

\end{table}

%% file: exp_mri.tex
%!TEX root = JSTSP_OPT_SAM.tex
\subsection{MRI}
\label{sec:exp_mri}
Our final example considers a classical MRI application in medical imaging. For our illustrations, we use a data set consisting of 3-dimensional volumes of knee images, fully scanned on a GE clinical 3T scanner.\footnote{Available at \url{http://mridata.org/fullysampled}} We focus our attention on a commonly-considered sampling technique based on subsampling the $k$-space in the $x$ and $y$ directions, while fully sampling the $z$ direction.  Note that we do not consider certain physical constraints such as those addressed in \cite{boyer2014algorithm}; our focus is on using the data to illustrate our learning-based approach, rather than demonstrating direct real-world applicability.

We pick the first half of the patients in the given data set for training and test our results on the remaining 10 patients. For the nonlinear decoder, we use basis pursuit (BP) combined with complex dual-tree wavelets, since this has been shown to give superior performance compared to other basis/solver combinations such as those involving total variation (TV) minimization \cite{majumdar2012choice}. Also following \cite{majumdar2012choice}, in this work we are not concerned with the denoising aspect, but instead we are only comparing our reconstructions to a fully sampled noisy image which constitutes a ground truth that is used to compute the error and PSNR. To solve the BP algorithm, we made use of NESTA \cite{becker2011nesta}.  
% to solve the basis pursuit algorithm as it allows to have a redundant sparsity transform which is implemented using ... (refer) with 6 stages and ... filter. \textbf{[TODO]} % to make it reproducable add that reference and filter

As a baseline for the subsampling map, we use the variable-density functions proposed in \cite{lustig2007sparse} and \cite{roman2014asymptotic}, which are determined by various parameters.  Specifically, the former has a radius $r$ within which full sampling is done and a polynomial degree $d$, and the latter has a radius $r$ within which full sampling is done, a number of levels $N$, and two exponential parameters $a$ and $b$.\footnote{We use slightly different notation to \cite{roman2014asymptotic} to avoid clashes with our notation.}  We note that letting $N$ be large and $r$ be small in \cite{roman2014asymptotic}, we recover a very similar sampling pattern to that proposed in \cite{wang2010variable}, which has only the parameters $a$ and $b$.

For each of \cite{lustig2007sparse} and \cite{roman2014asymptotic}, we do a sweep over the parameters, generate $20$ random subsampling patterns for each set of parameters, and finally choose the pattern with the best average PSNR on $100$ fixed and randomly selected training images.  For \cite{lustig2007sparse}, we sweep over $r \in \{0, 0.025, \ldots, 0.5\}$ and $d \in \{0, 0.25, \ldots, 10\}$.  For \cite{roman2014asymptotic}, we let the number of levels $N=100$ be fixed and large, sweep over $r \in \{0, 0.025, \ldots, 0.5\}$ and $a \in \{0.05,0.1,\dotsc,8,9,\dotsc,25\}$, and compute $b$ via a binary search in order to ensure that the total number of samples is exactly $n$.  We note that the parameter sweeps for the non-linear decoder are highly time consuming, taking considerably longer than our combinatorial optimization problems for the linear decoder.

%which is parametrized by the radius of fully sampled region, $r$, and the polynomial degree, $d$. Note that this density function assumes circular symmetry in the $k$-space.  We do a parameter sweep over , drawing $20$ subsampling patterns from the corresponding variable-density distribution in each case, and finally choosing the pattern with the highest average PSNR on the training data. 
%The best values were the following: for $6.25\%$ sampling rate, $r = 0.15$ and $d = 7$; for $12.5\%$ sampling rate, $r = 0.25$ and $d = 5.5$, and for $25\%$ sampling rate, $r = 0.35$ and $d = 3.5$.
%For the test data, we fix the indices of the best performing random index set from this procedure, as suggested by \cite{lustig2007sparse}. 

%\begin{figure}
 \begin{figure*}[!ht]
\centering
\begin{tabular}{ccccccc}
\includegraphics[width=.24\columnwidth]{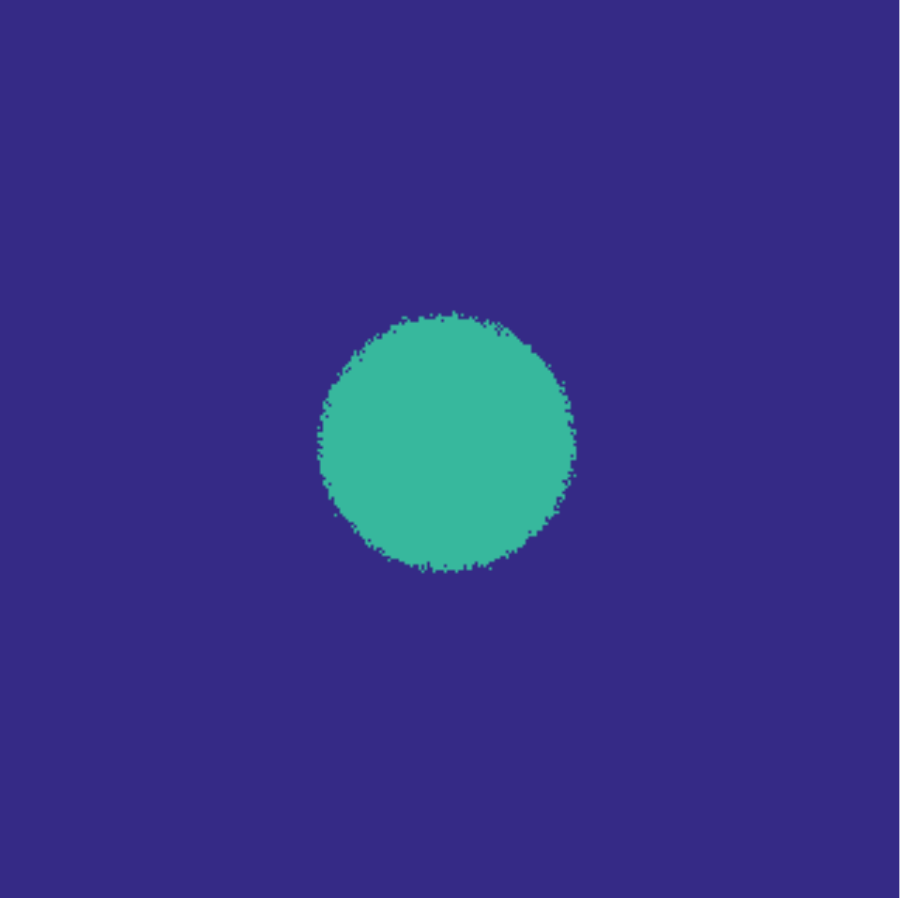} &
\hspace{-4mm}\includegraphics[width=.24\columnwidth]{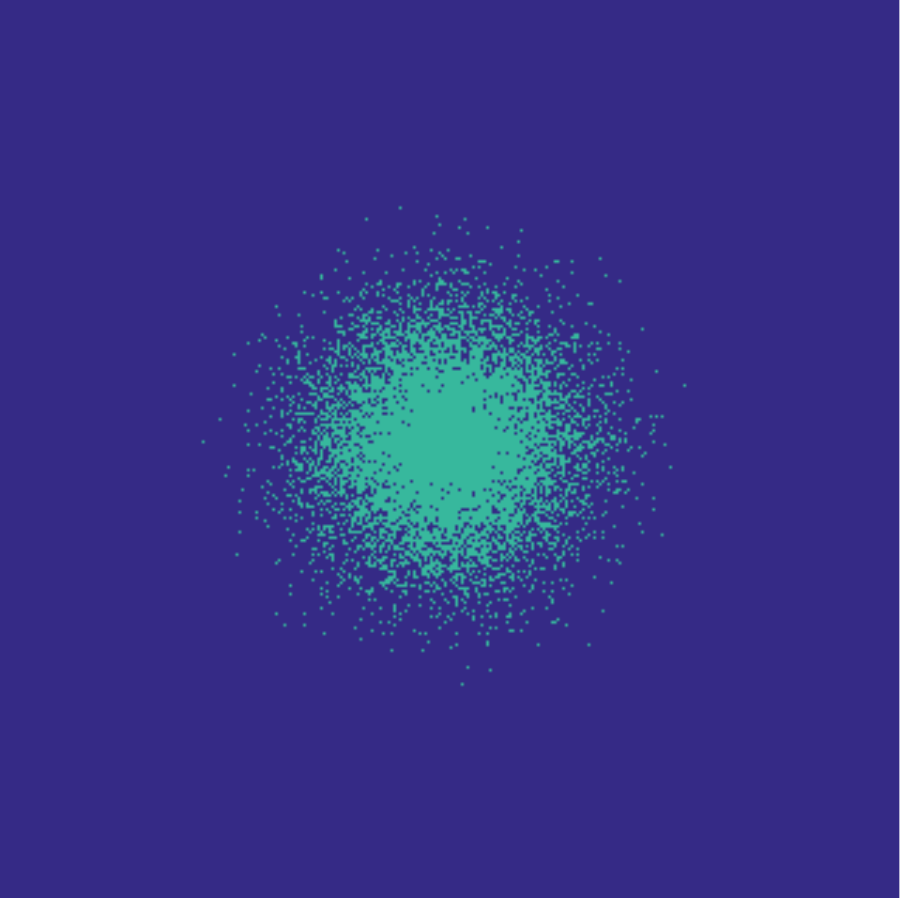} &
\hspace{-4mm}\includegraphics[width=.24\columnwidth]{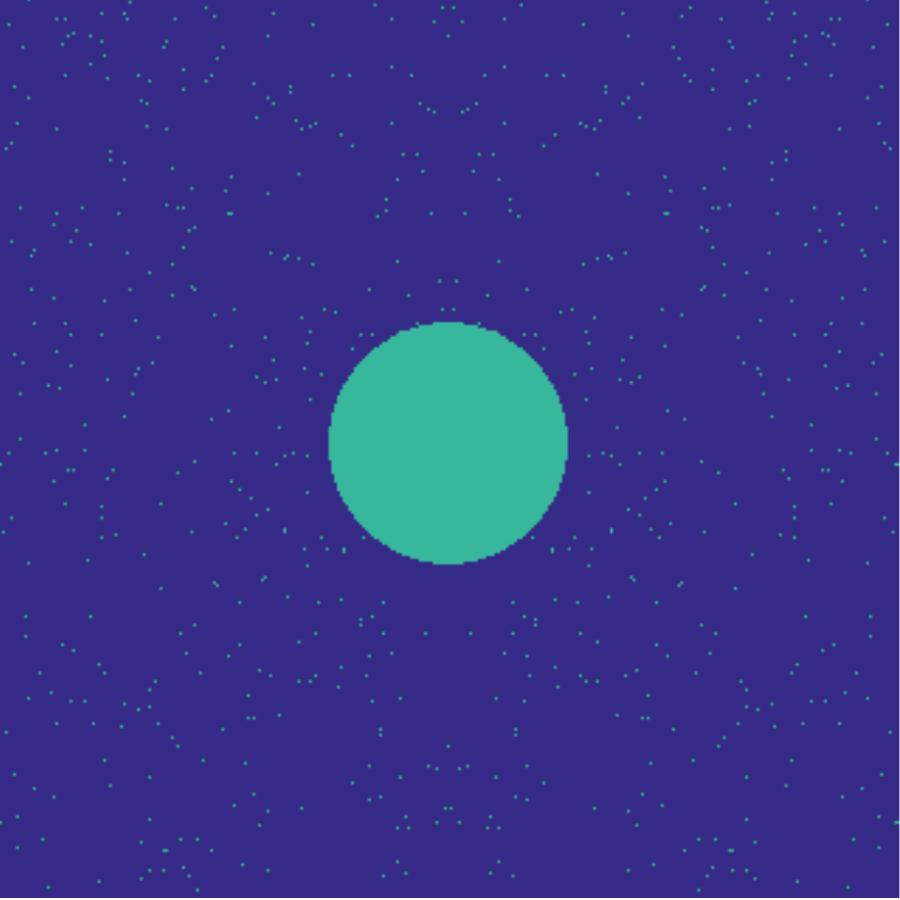} &
\hspace{-4mm}\includegraphics[width=.24\columnwidth]{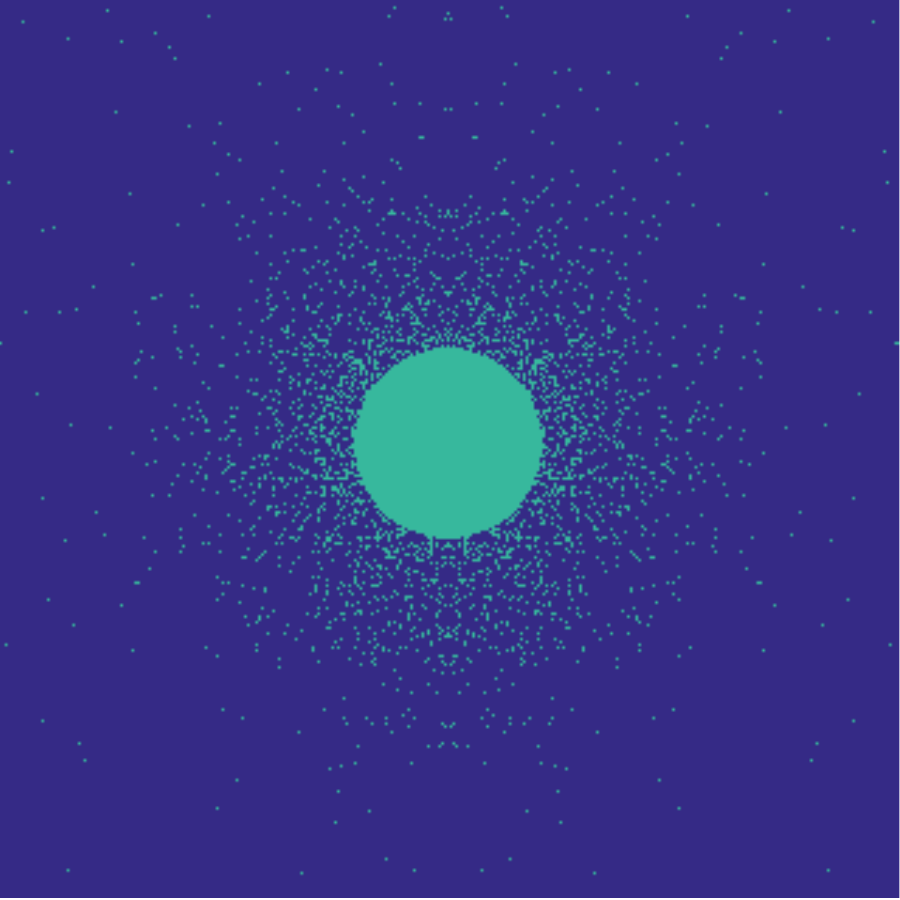} &
\hspace{-4mm}\includegraphics[width=.24\columnwidth]{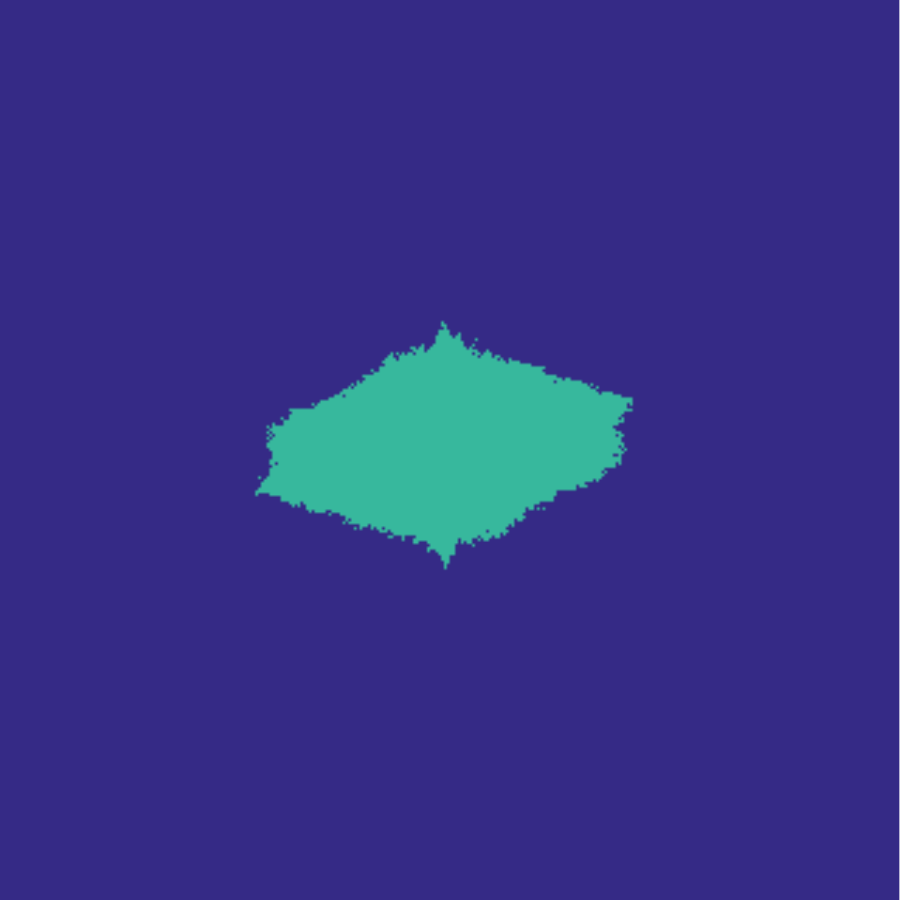} &
\hspace{-4mm}\includegraphics[width=.24\columnwidth]{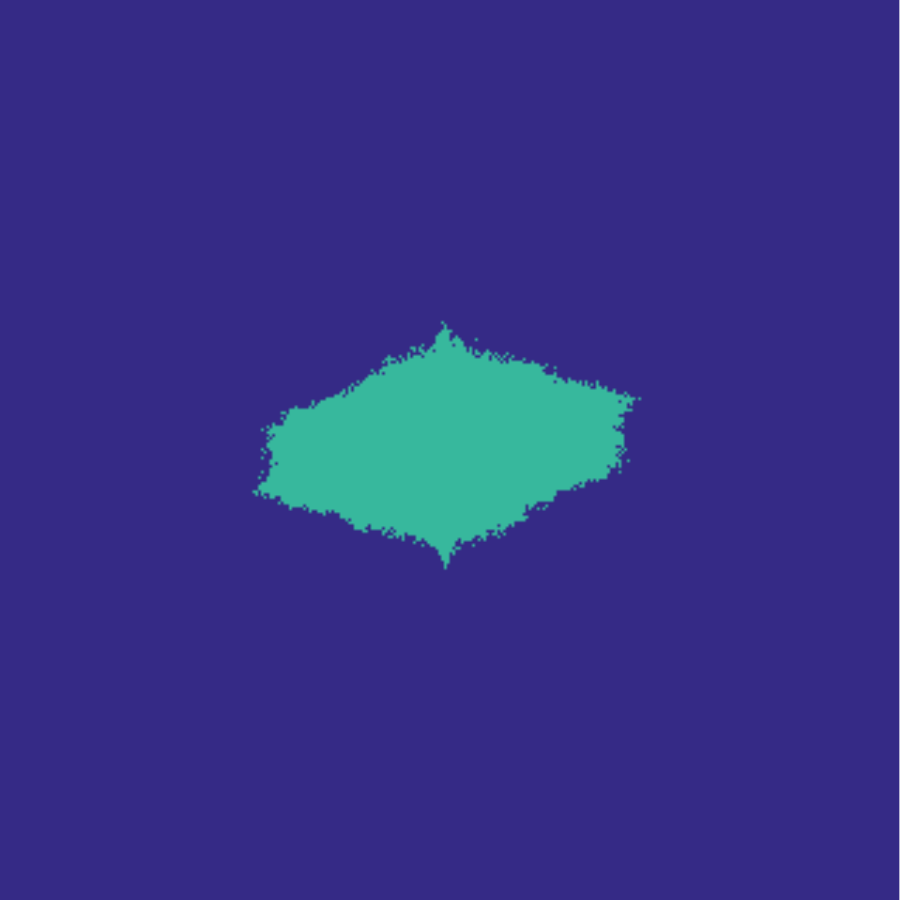} &
\hspace{-4mm}\includegraphics[width=.24\columnwidth]{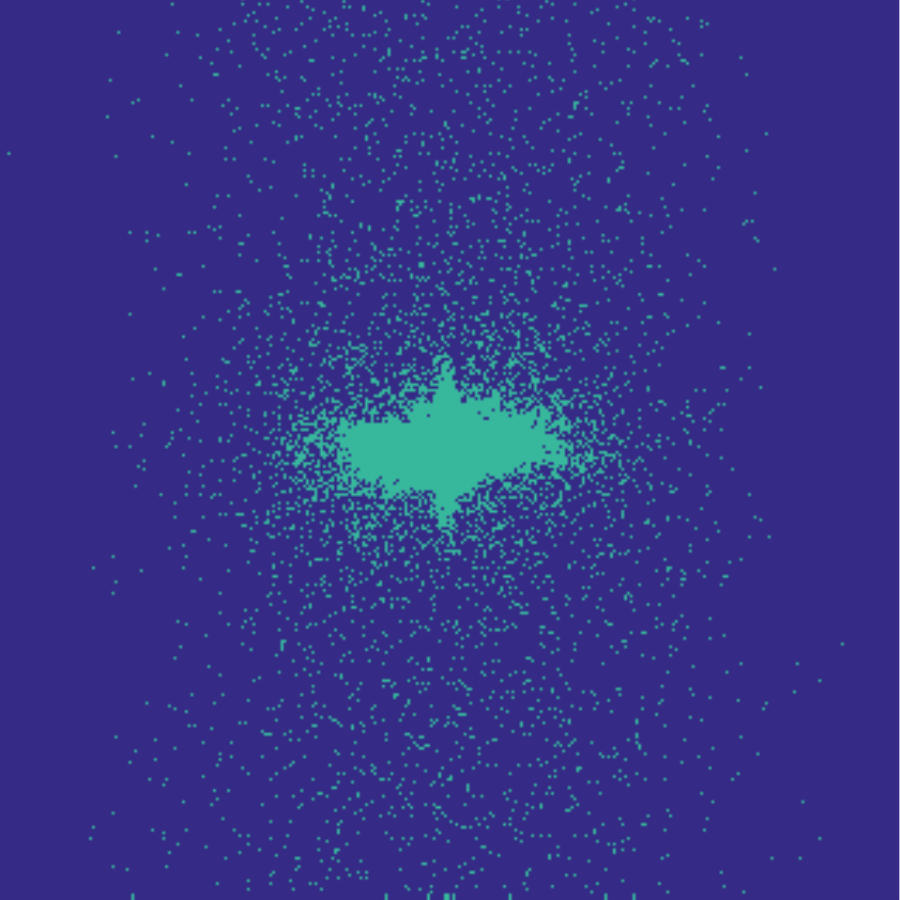} \\[-1mm]
\includegraphics[width=.24\columnwidth]{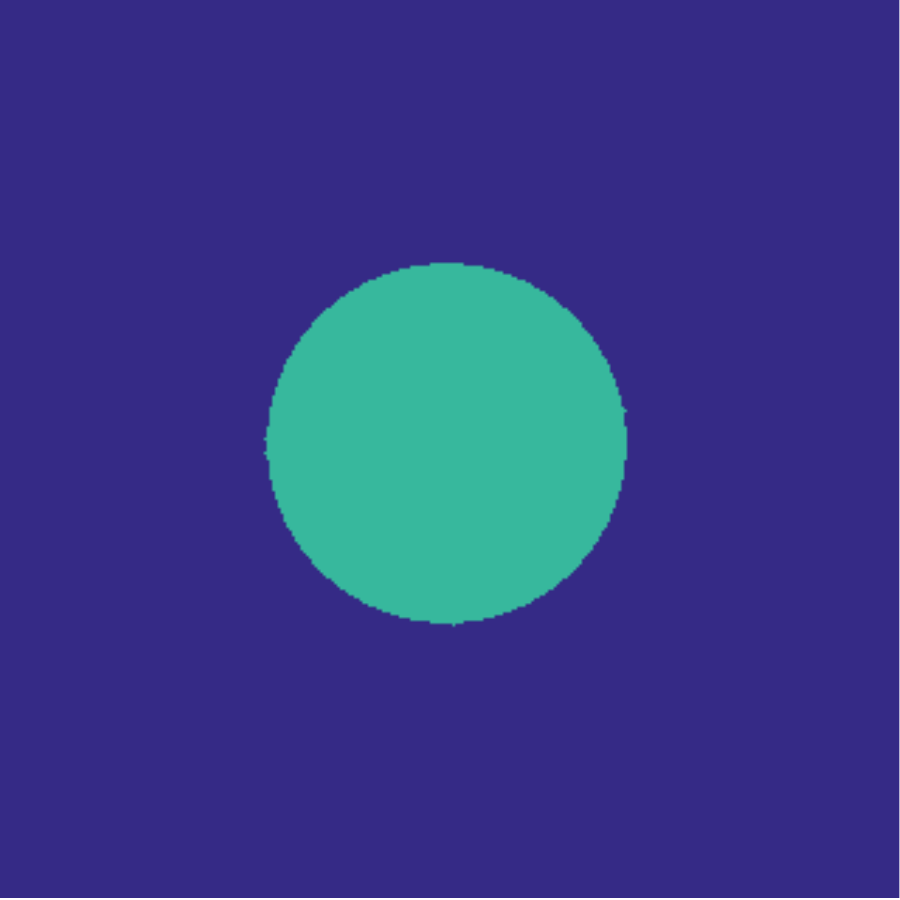} &
\hspace{-4mm}\includegraphics[width=.24\columnwidth]{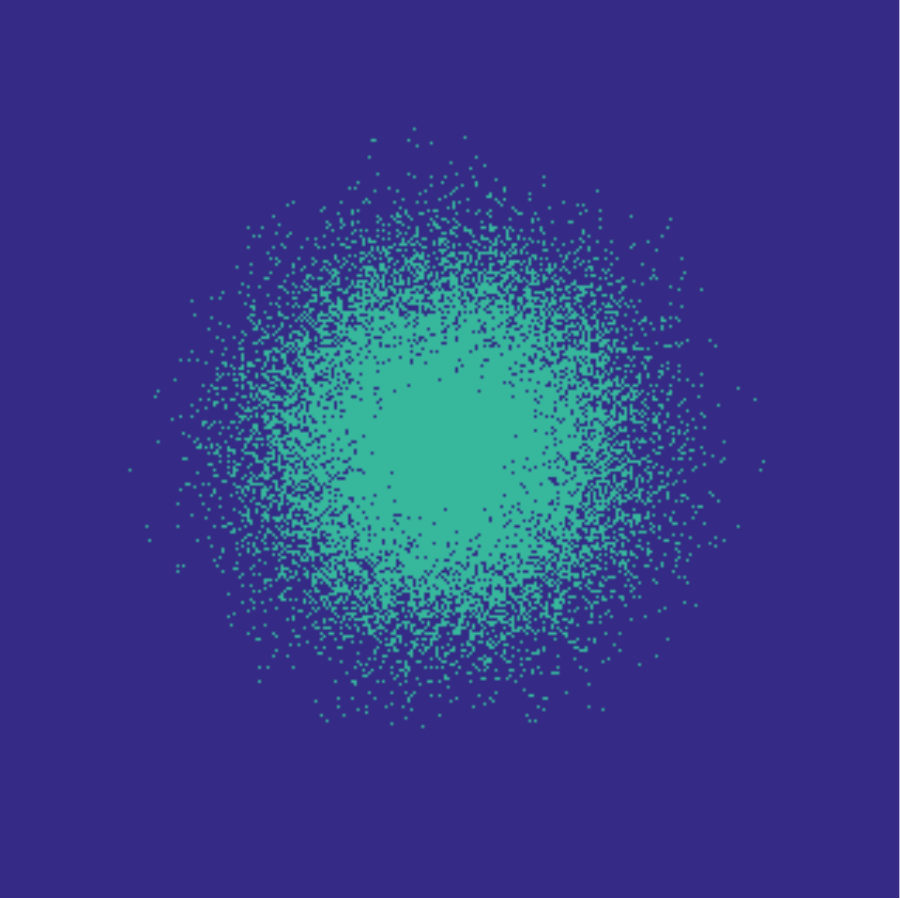} &
\hspace{-4mm}\includegraphics[width=.24\columnwidth]{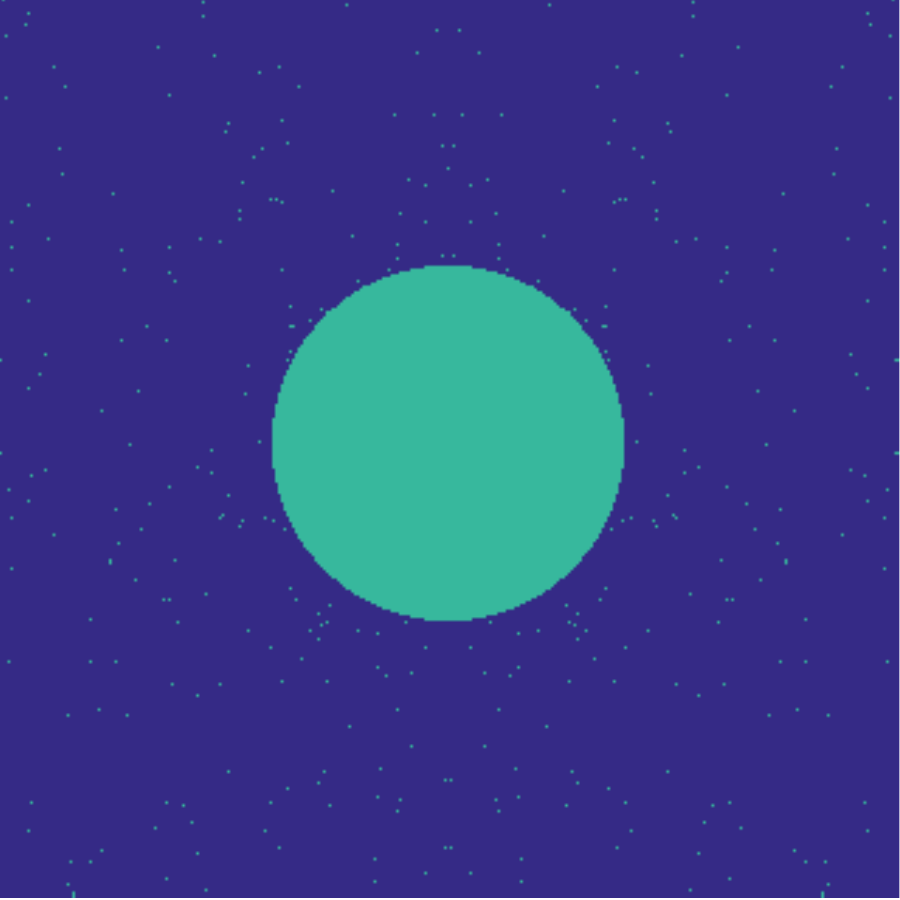} &
\hspace{-4mm}\includegraphics[width=.24\columnwidth]{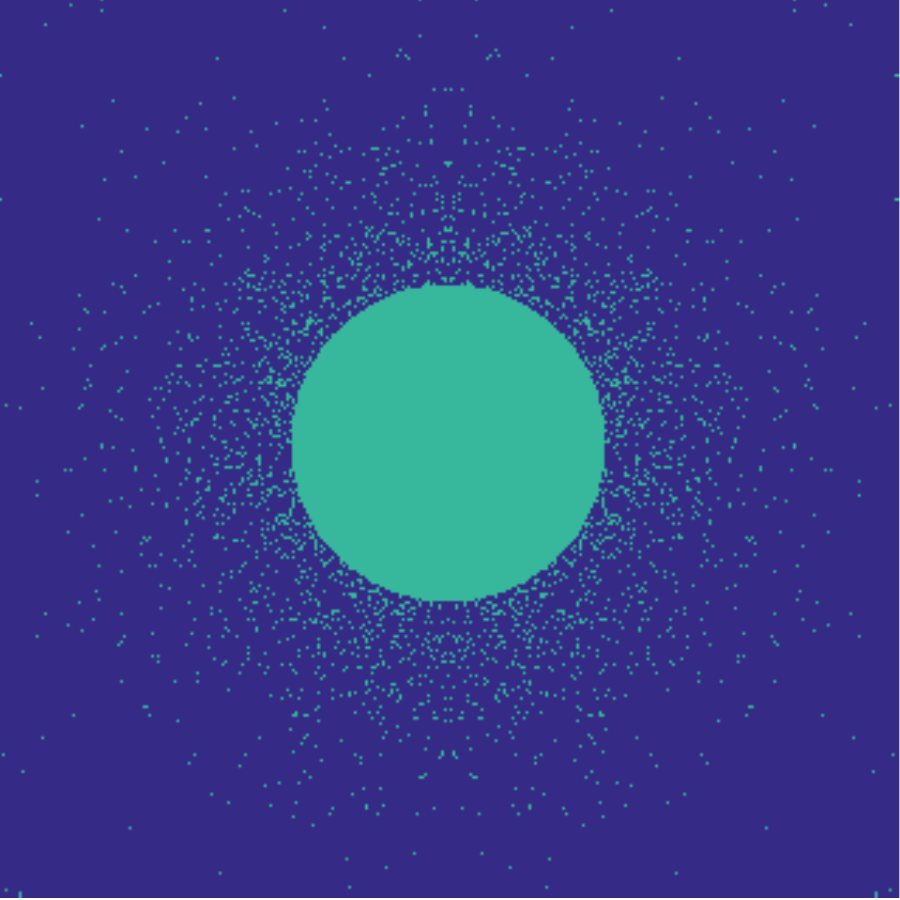} &
\hspace{-4mm}\includegraphics[width=.24\columnwidth]{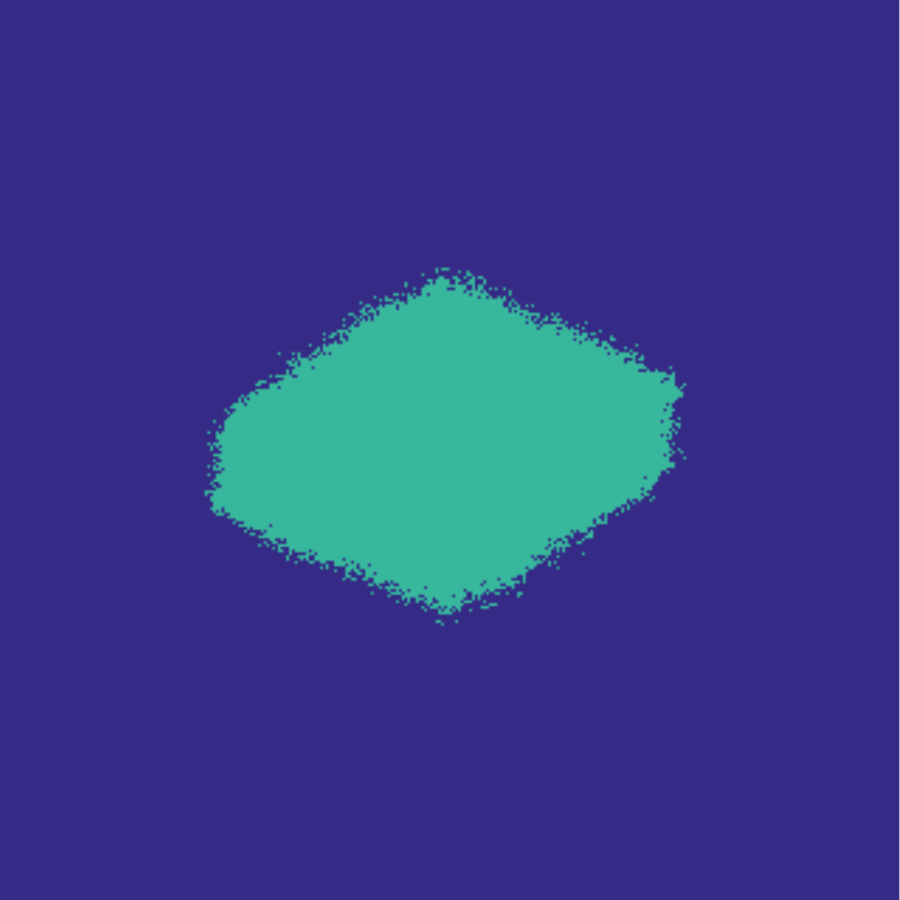} &
\hspace{-4mm}\includegraphics[width=.24\columnwidth]{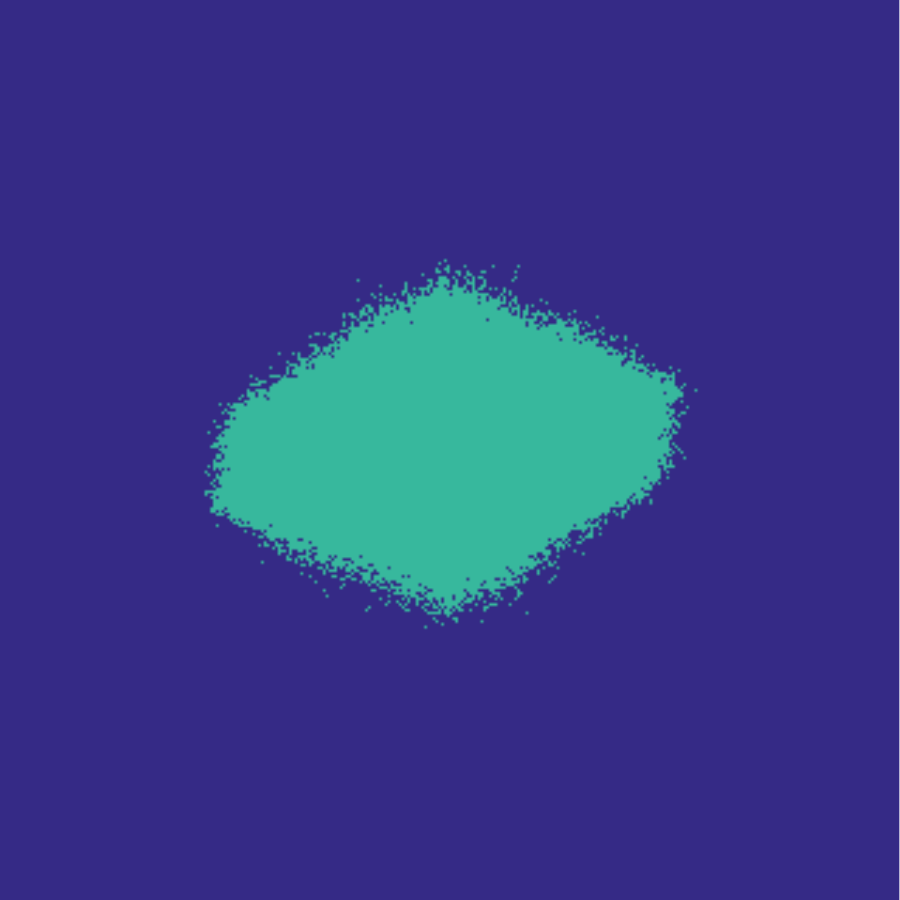} &
\hspace{-4mm}\includegraphics[width=.24\columnwidth]{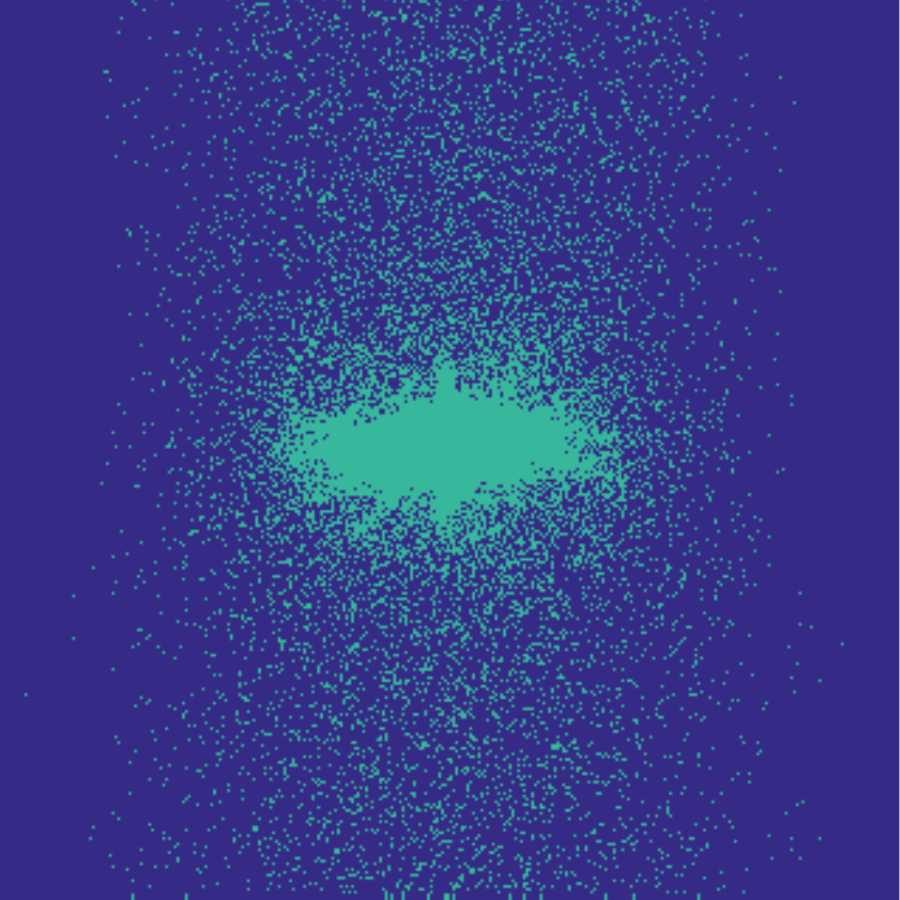} \\[-1mm]
\includegraphics[width=.24\columnwidth]{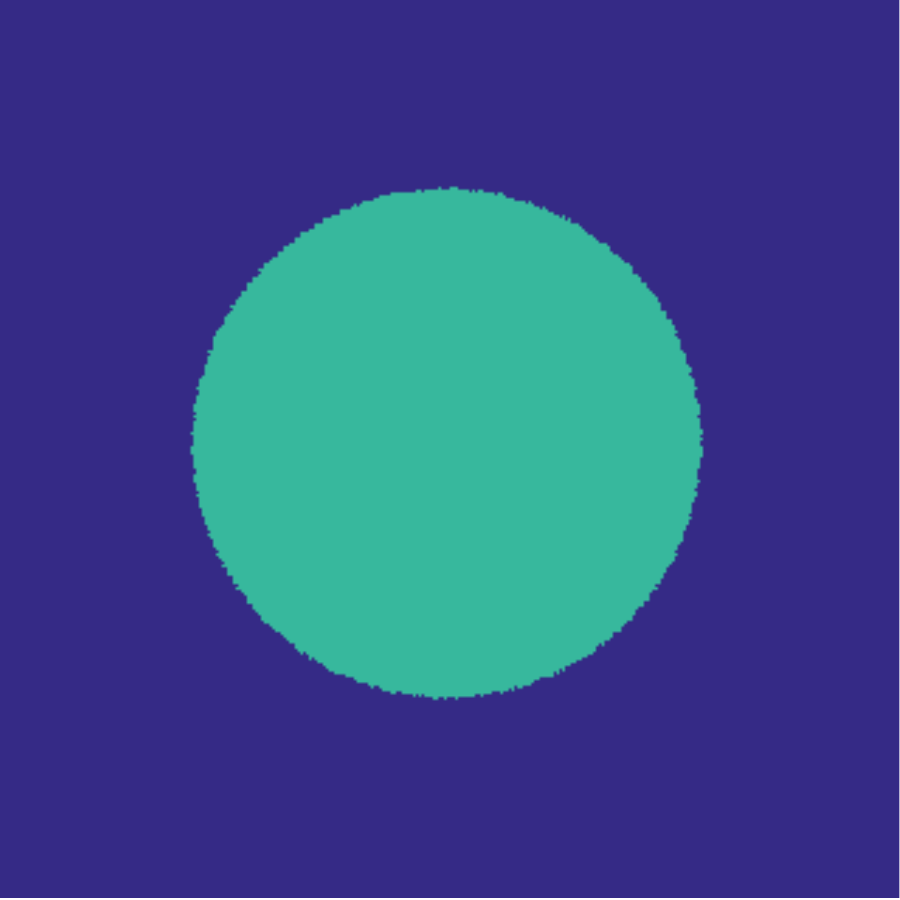} &
\hspace{-4mm}\includegraphics[width=.24\columnwidth]{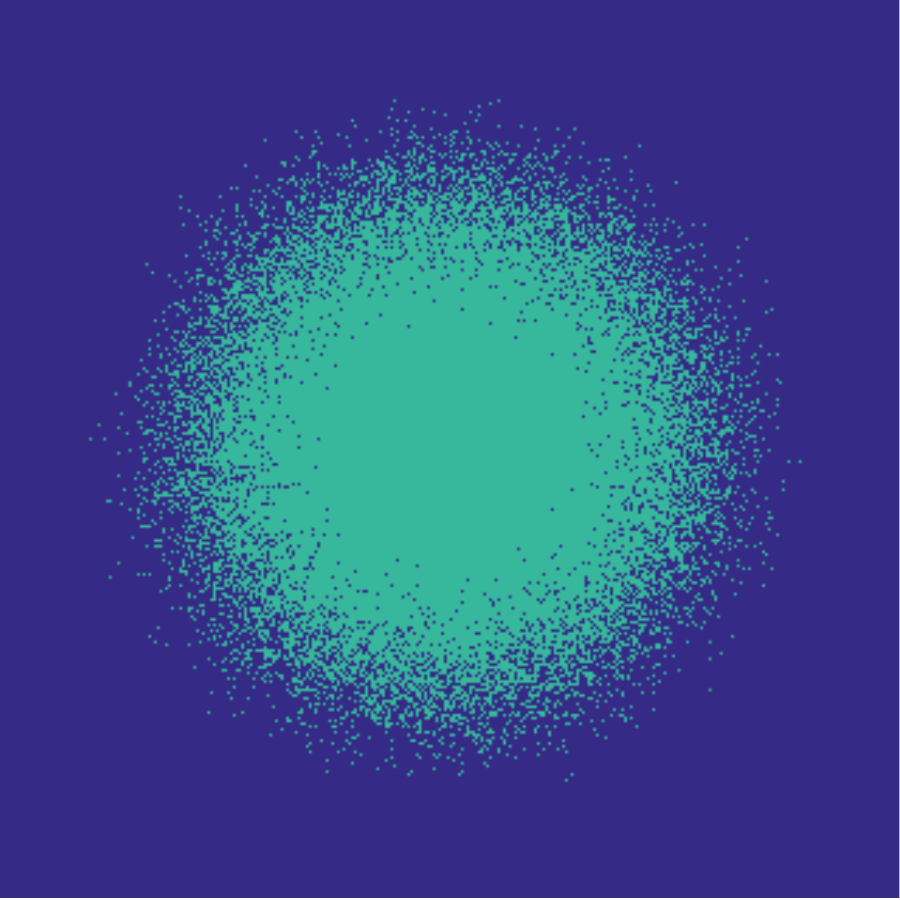} &
\hspace{-4mm}\includegraphics[width=.24\columnwidth]{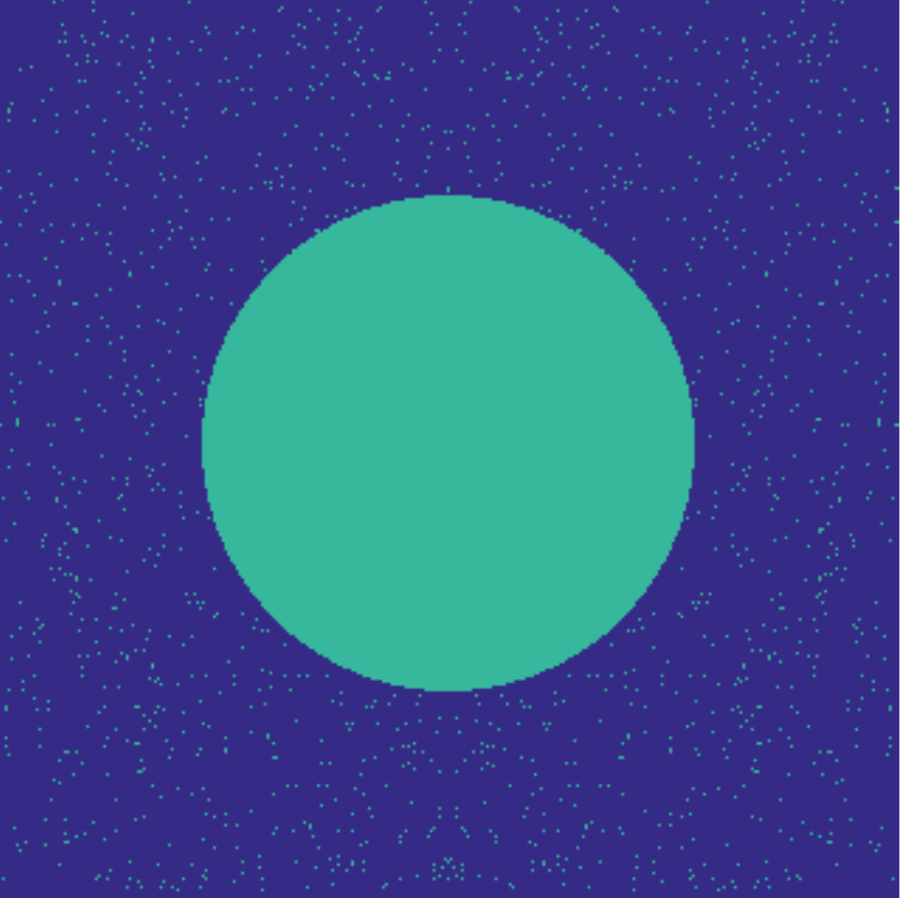} &
\hspace{-4mm}\includegraphics[width=.24\columnwidth]{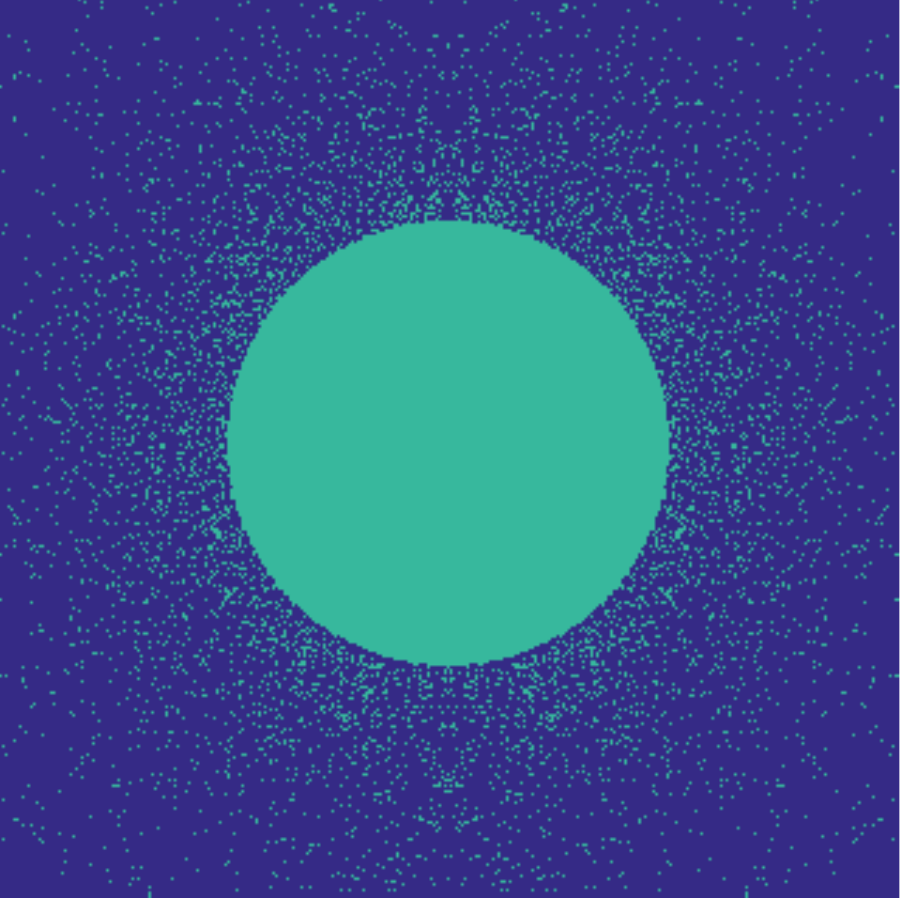} &
\hspace{-4mm}\includegraphics[width=.24\columnwidth]{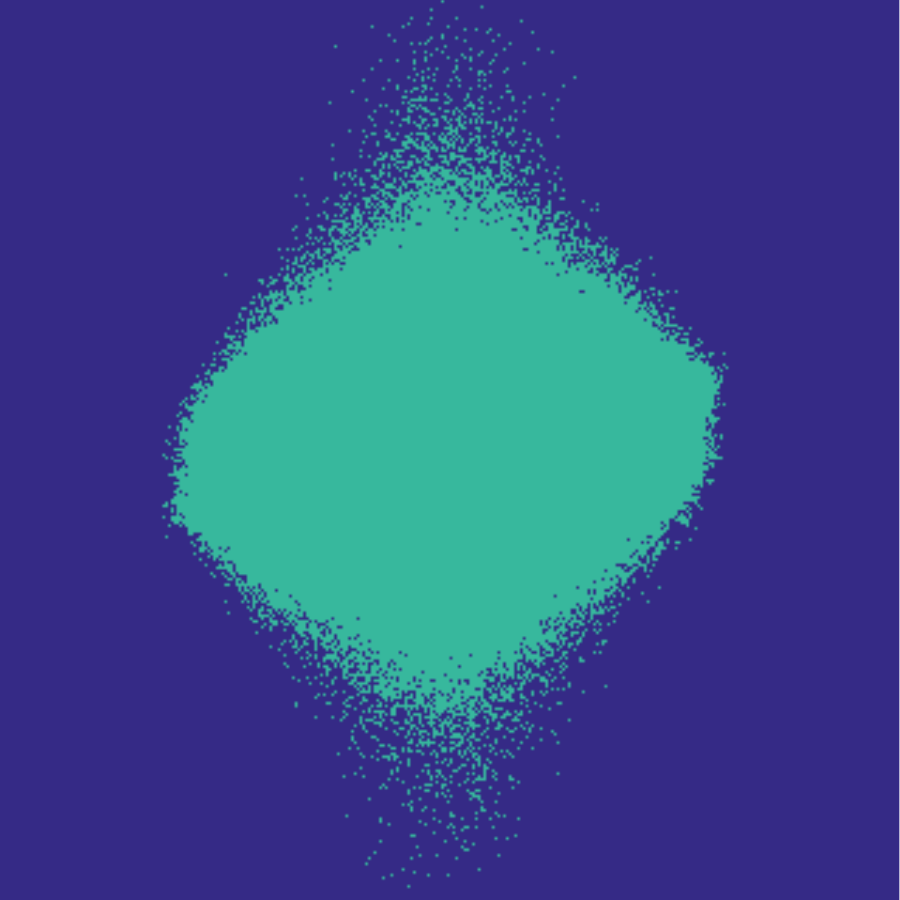} &
\hspace{-4mm}\includegraphics[width=.24\columnwidth]{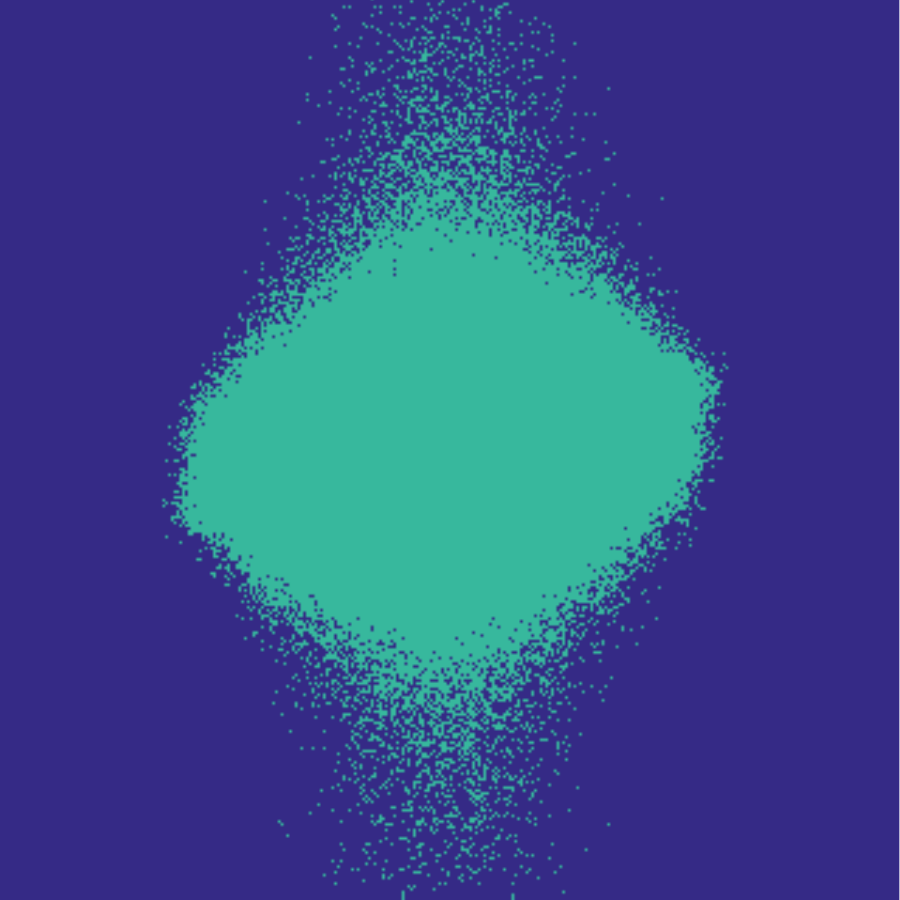} &
\hspace{-4mm}\includegraphics[width=.24\columnwidth]{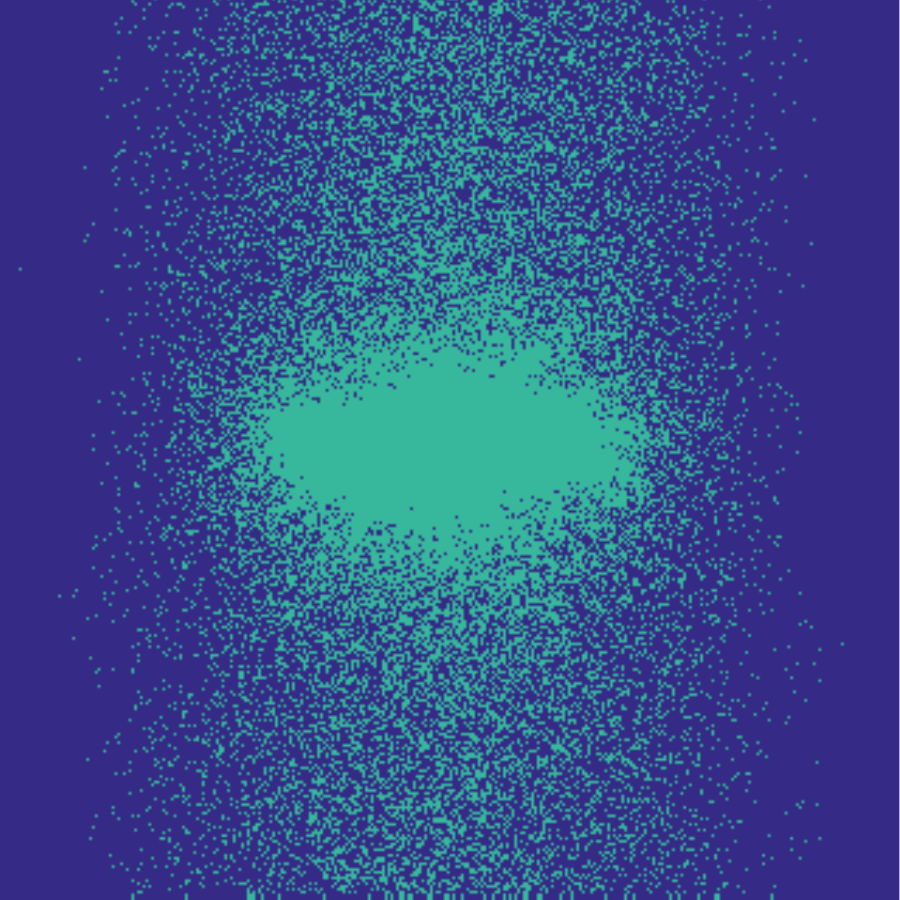} \\[-1mm]
\text{\footnotesize{linear}} &\text{\footnotesize{BP}} & \text{\footnotesize{linear}} & \text{\footnotesize{BP}} & \text{\footnotesize{$f_{\text{avg}}$}} & \text{\footnotesize{$f_{\text{gen}}$}} & \text{\footnotesize{$f_{\text{min}}$}}  \\
\multicolumn{2}{c}{Roman \emph{et al.}}   & \multicolumn{2}{c}{Lustig \emph{et al.}} & & &
\end{tabular}
\caption{The index sets of the tuned random variable sampling schemes \cite{lustig2007sparse,roman2014asymptotic} vs.\ our learned-based approach with various optimization criteria.  The sampling ratios of the three rows are $6.25\%$, $12.5\%$ and $25\%$. }
\label{fig:MRI_maps}
\end{figure*}
%\end{figure}

Figure \ref{fig:MRI_maps} illustrates the best-performing randomized indices vs.\ our learned set of indices in the $k$-space along the $x$ and $y$ directions. When optimized for the linear decoder, the indices of \cite{lustig2007sparse,roman2014asymptotic} concentrate on low frequencies.  While our strategies based on optimizing $\favg$ and $\fgen$ (again using $g(\alpha) = 1 - (1-\alpha)^2$)  also do this to some extent, there is a stark contrast in the shape, since we do not restrict ourselves to patterns exhibiting circular symmetry.  

\begin{figure*}[!b]
\centering
\begin{tabular}{cccc}
\includegraphics[width=.20\textwidth]{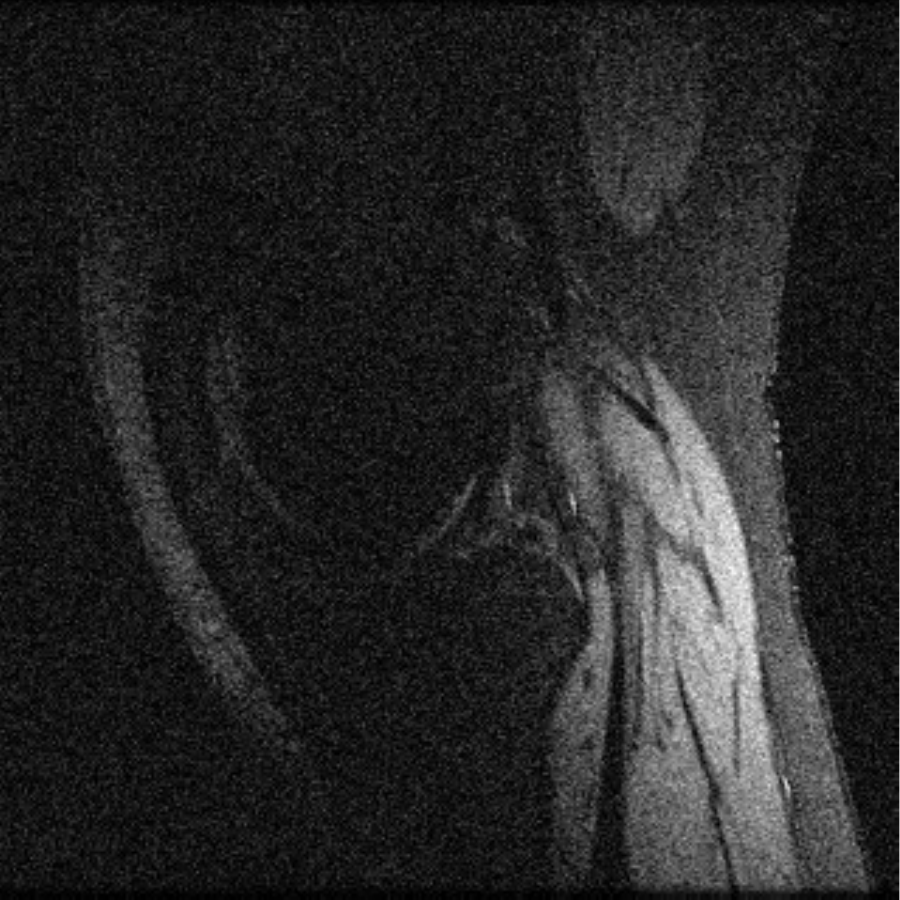} &
\hspace{-4mm}\includegraphics[width=.20\textwidth]{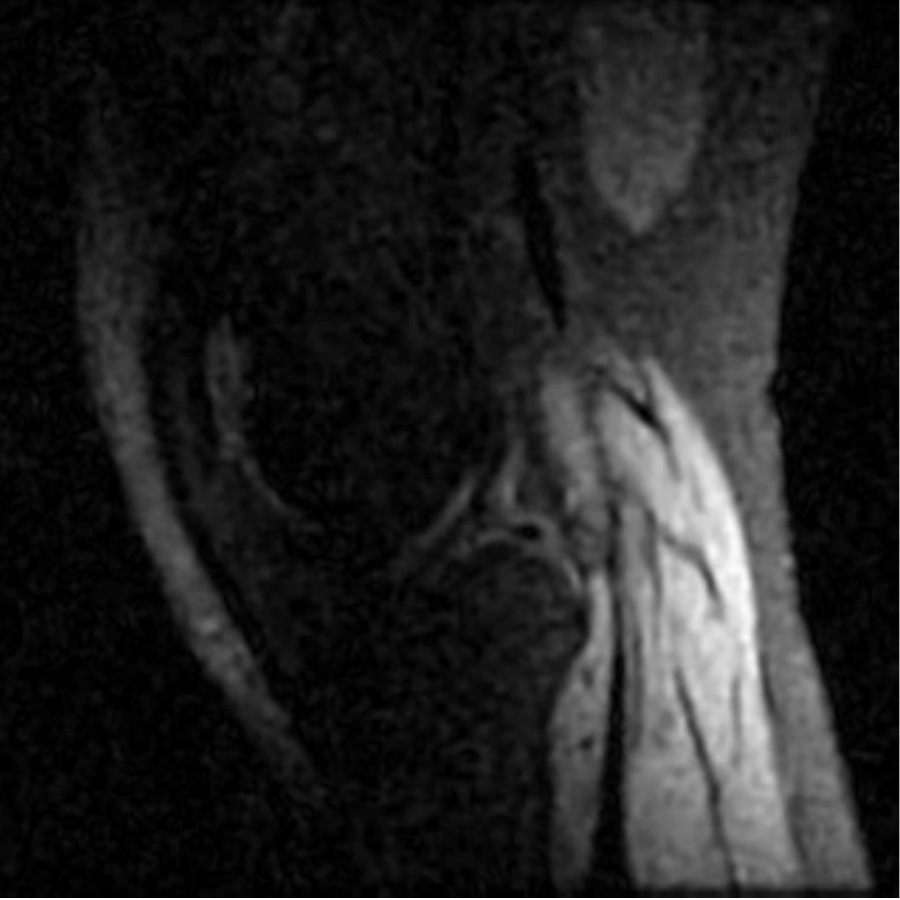} &
\hspace{-4mm}\includegraphics[width=.20\textwidth]{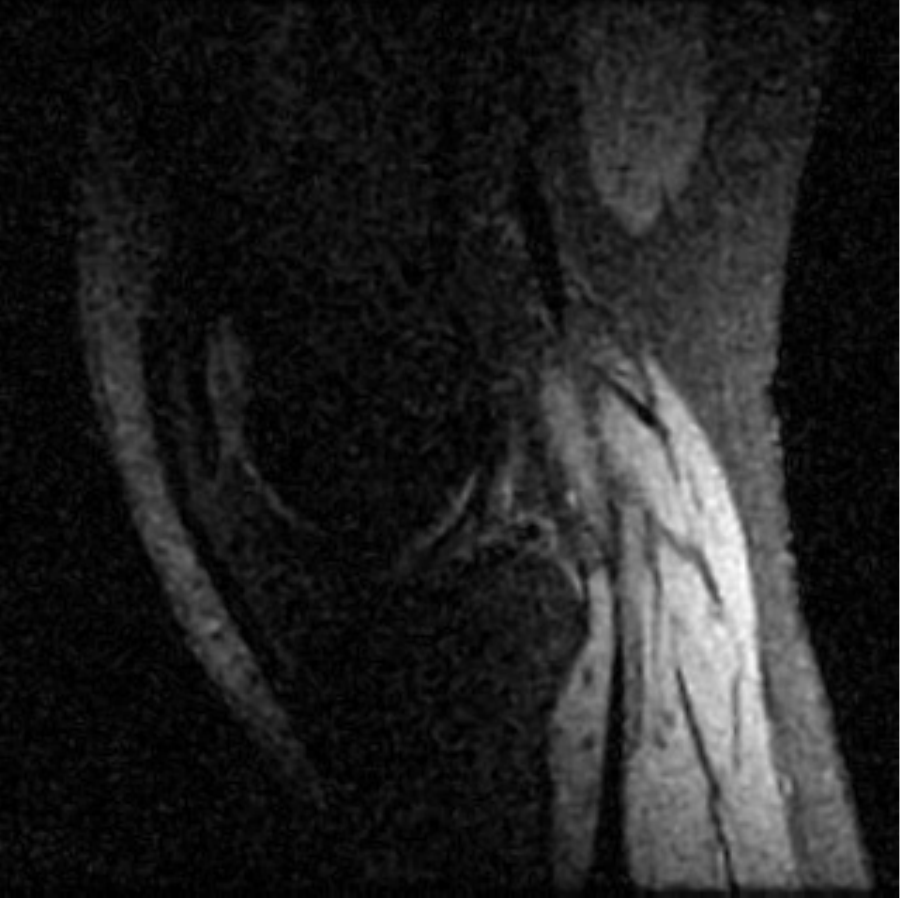} &
\hspace{-4mm}\includegraphics[width=.20\textwidth]{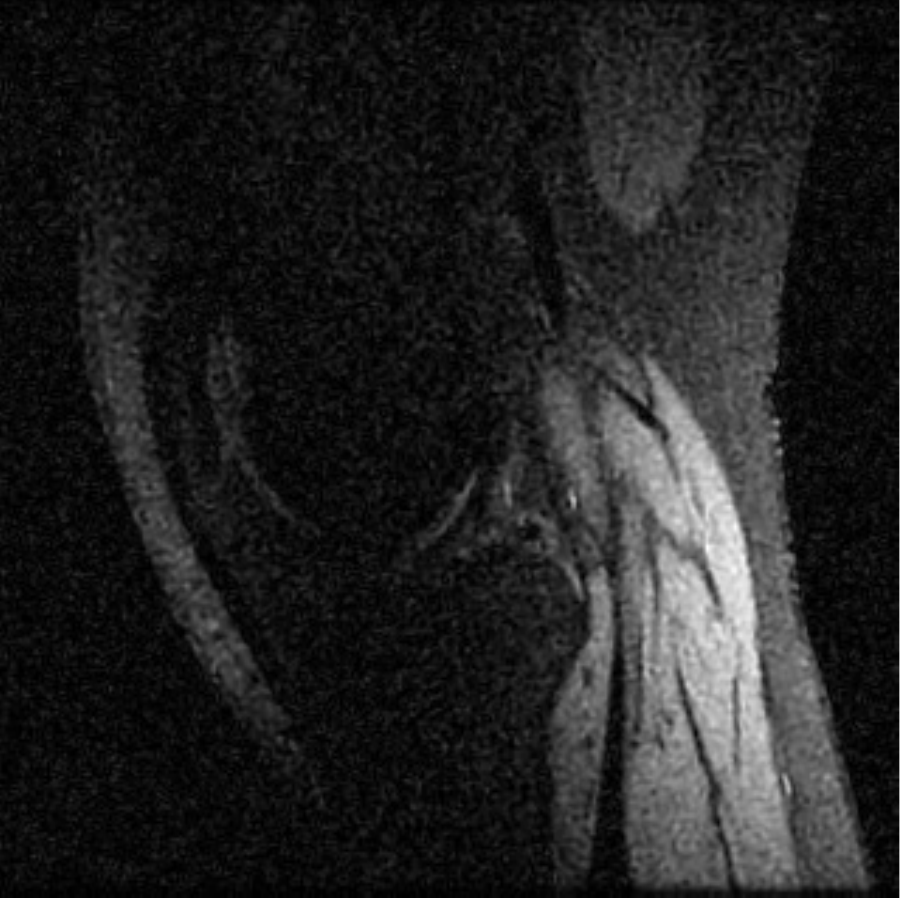} \\ [-1mm]
\text{\footnotesize{Roman et al.} BP} &
\hspace{-4mm}\includegraphics[width=.20\textwidth]{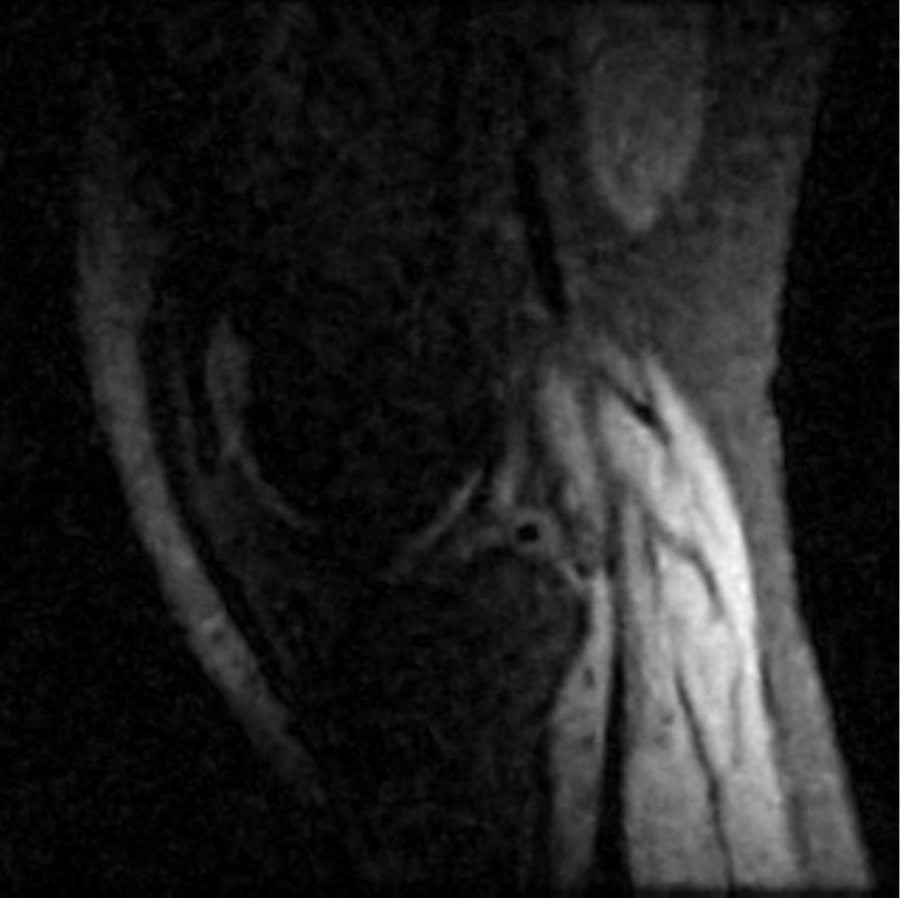} &
\hspace{-4mm}\includegraphics[width=.20\textwidth]{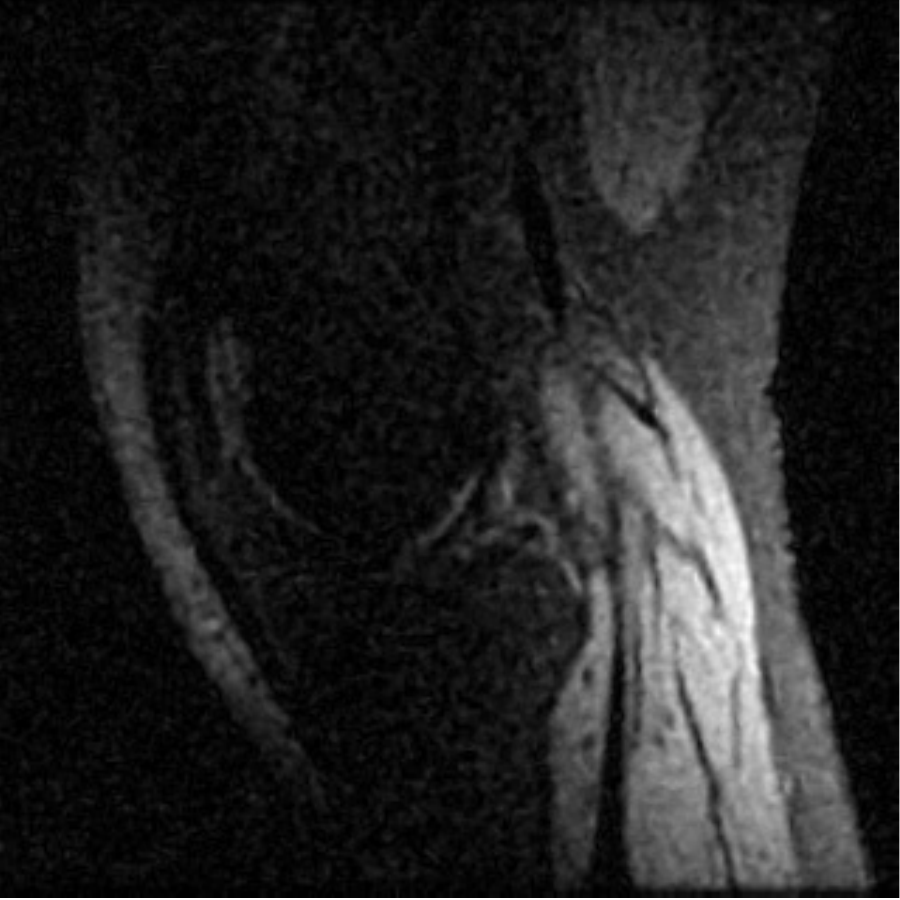} &
\hspace{-4mm}\includegraphics[width=.20\textwidth]{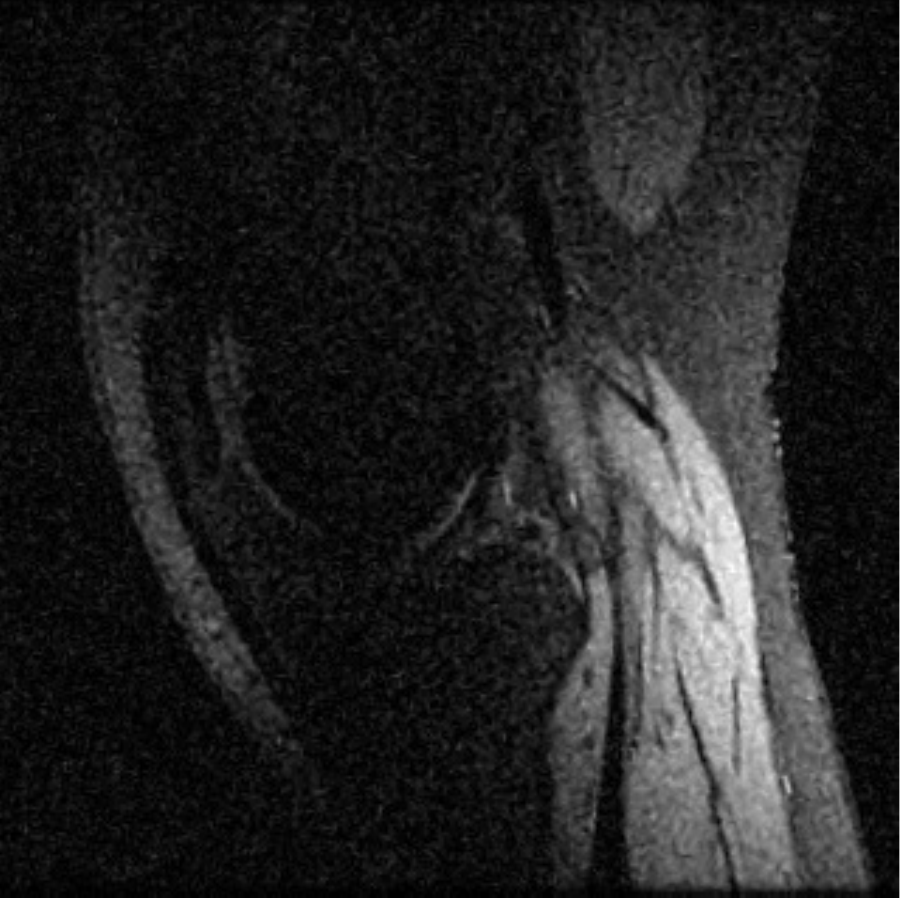} \\ [-1mm]
\text{\footnotesize{Lustig et al.} BP} &
\hspace{-4mm}\includegraphics[width=.20\textwidth]{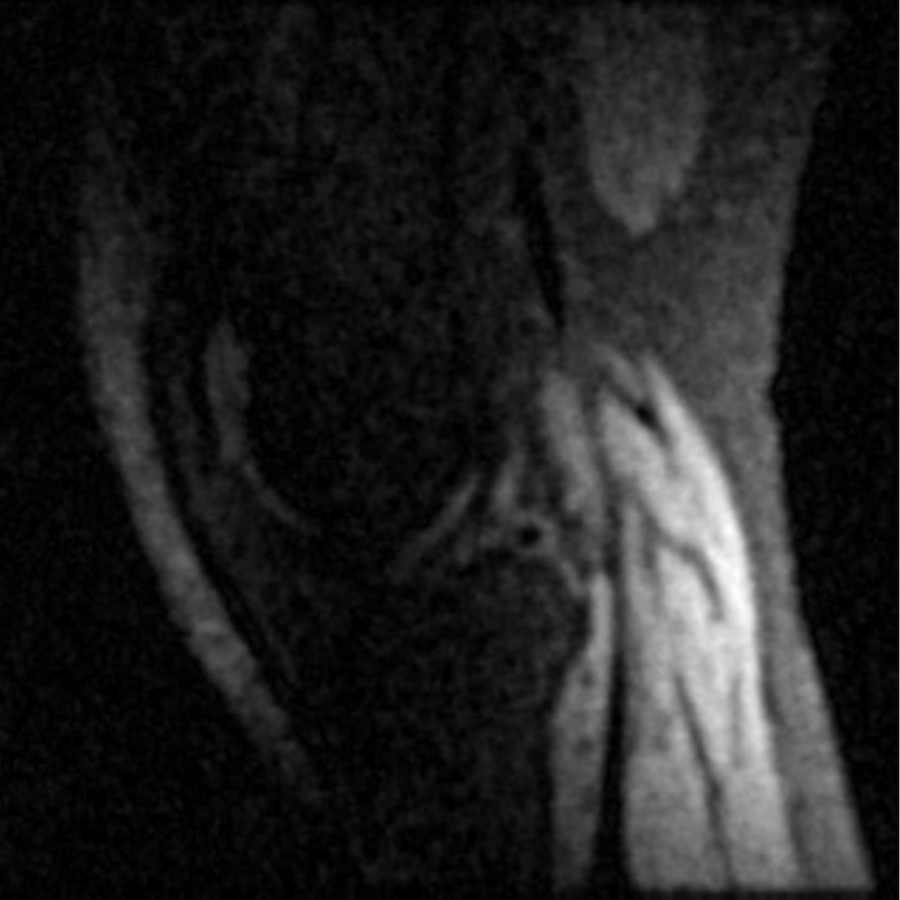} &
\hspace{-4mm}\includegraphics[width=.20\textwidth]{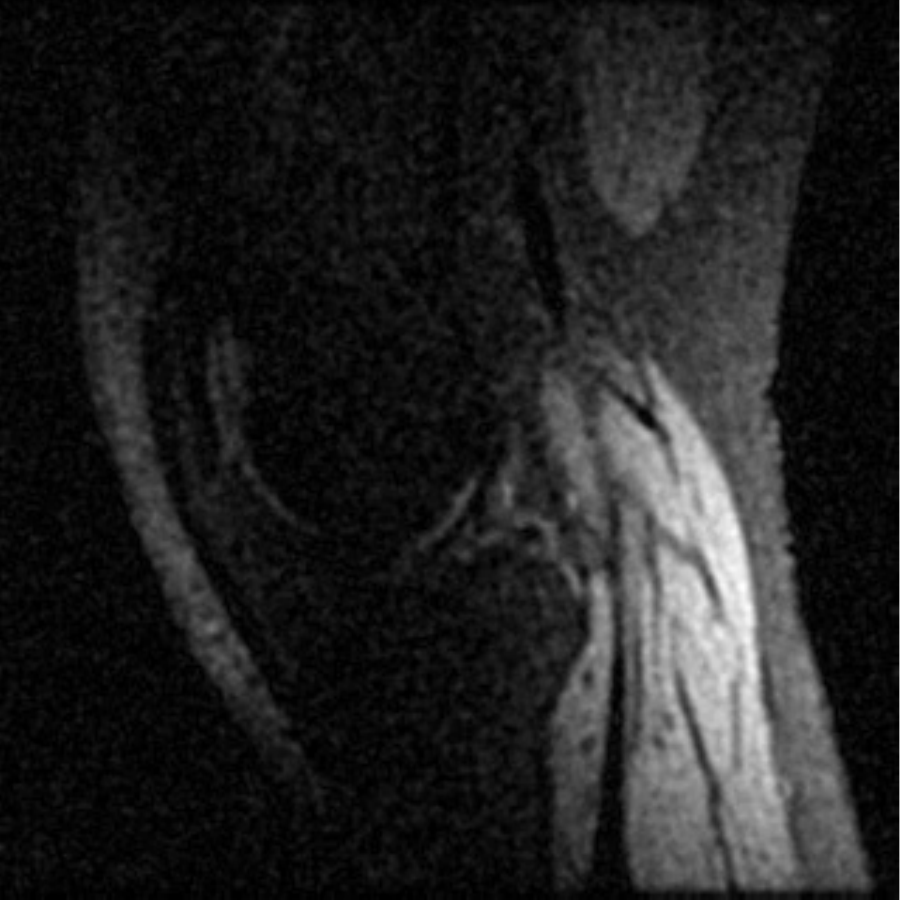} &
\hspace{-4mm}\includegraphics[width=.20\textwidth]{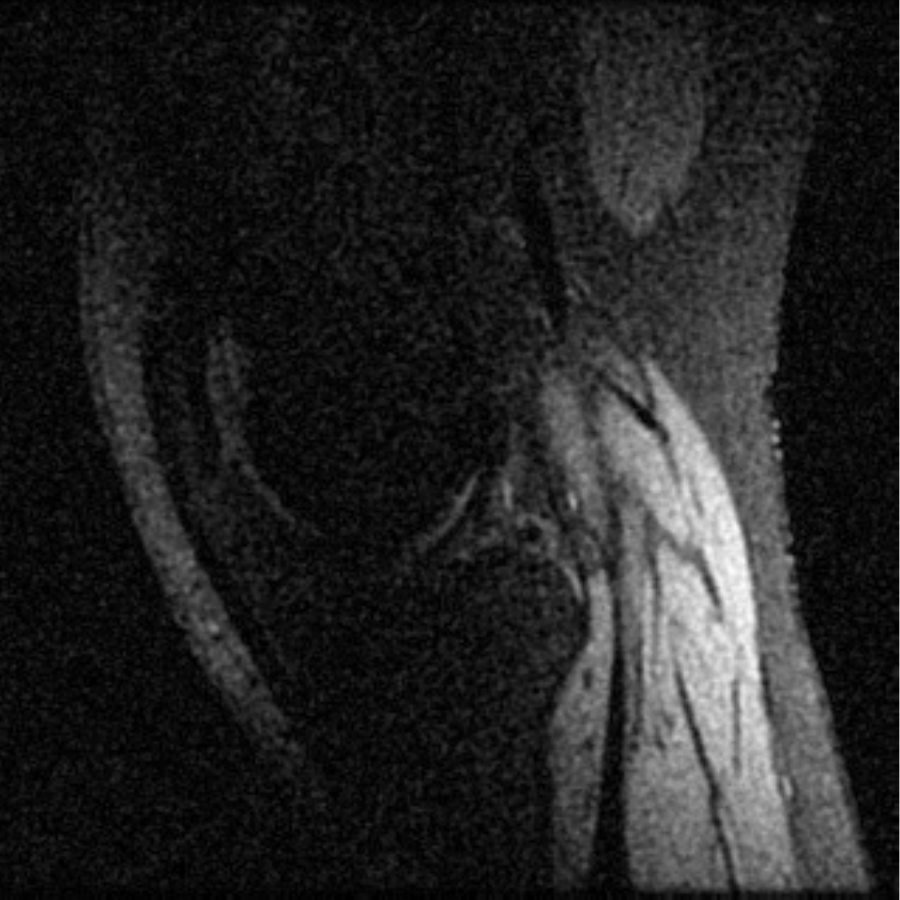} \\ [-1mm]
\text{\footnotesize{$f_{\text{avg}}$ linear}} &
\hspace{-4mm}\includegraphics[width=.20\textwidth]{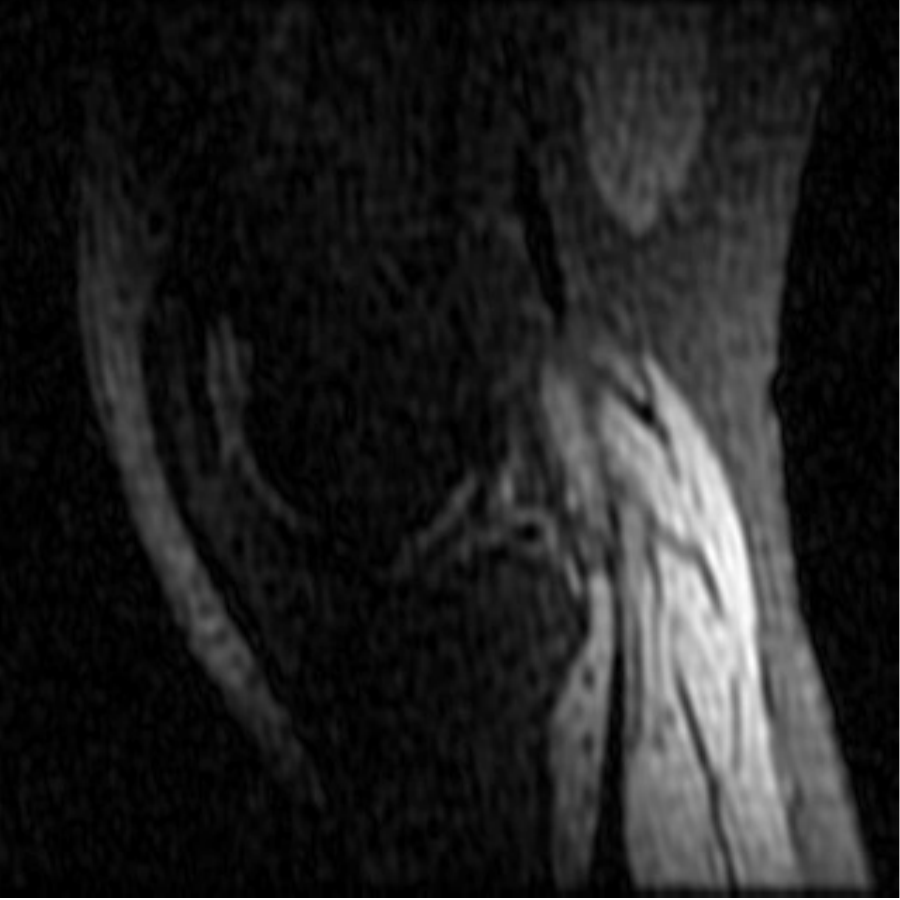} &
\hspace{-4mm}\includegraphics[width=.20\textwidth]{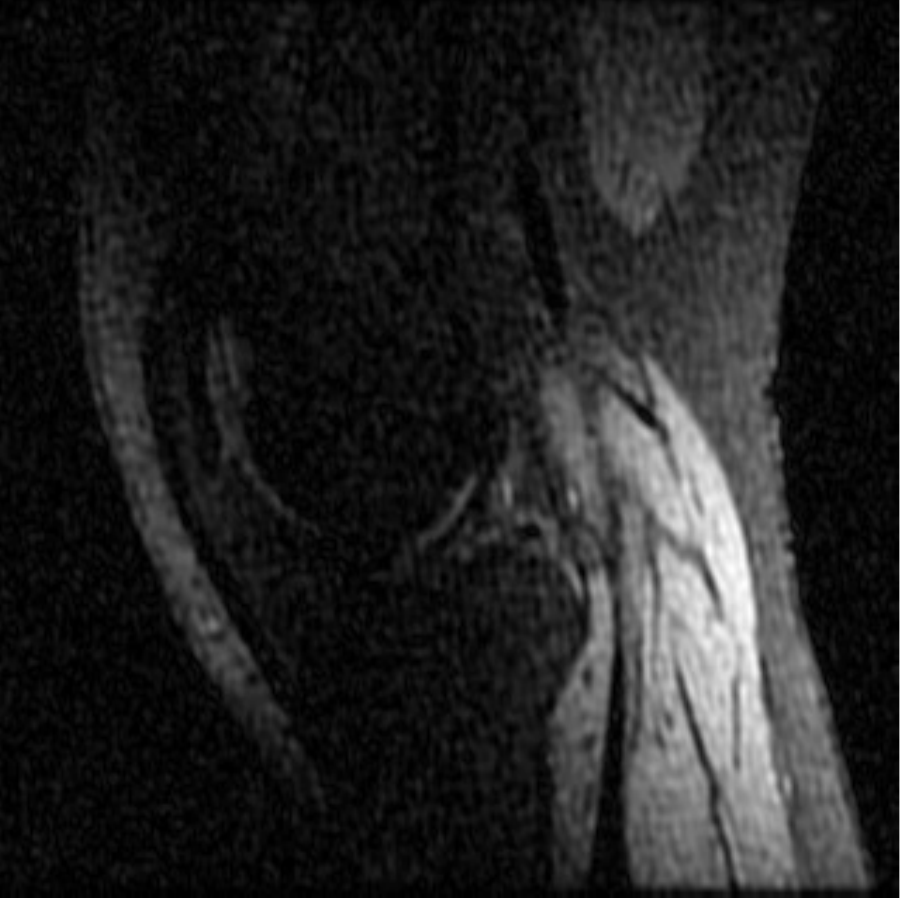} &
\hspace{-4mm}\includegraphics[width=.20\textwidth]{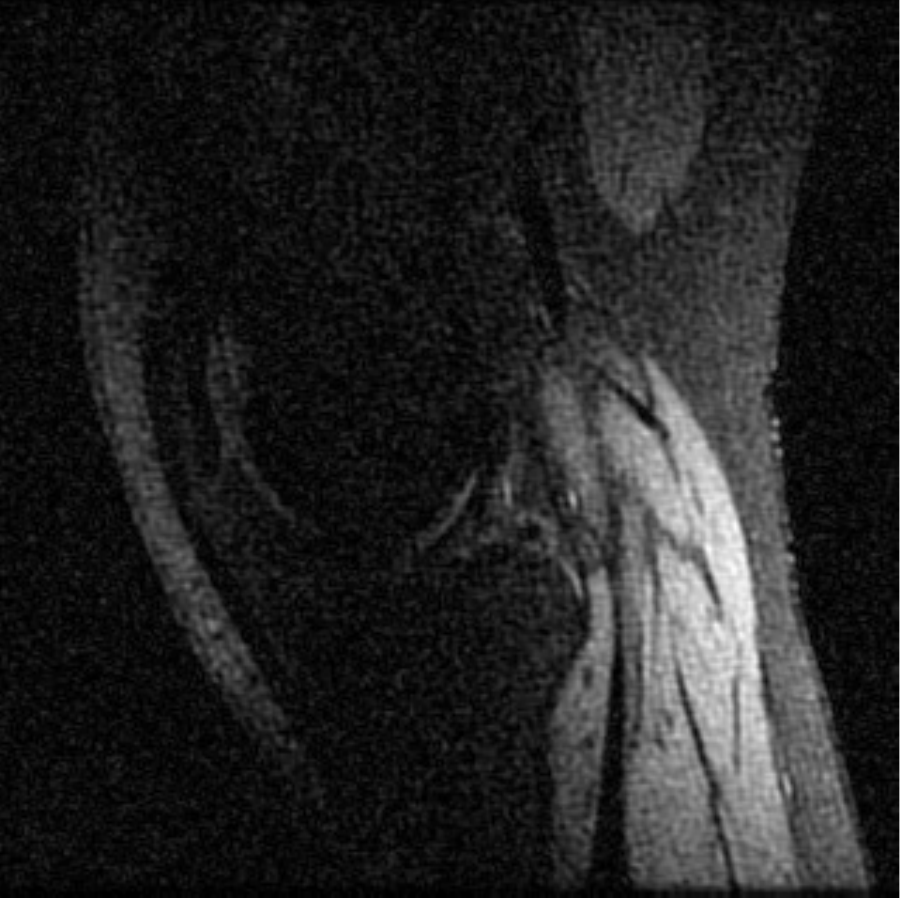} \\ [-1mm]
\text{\footnotesize{$f_{\text{gen}}$ linear}}  &
\hspace{-4mm}\includegraphics[width=.20\textwidth]{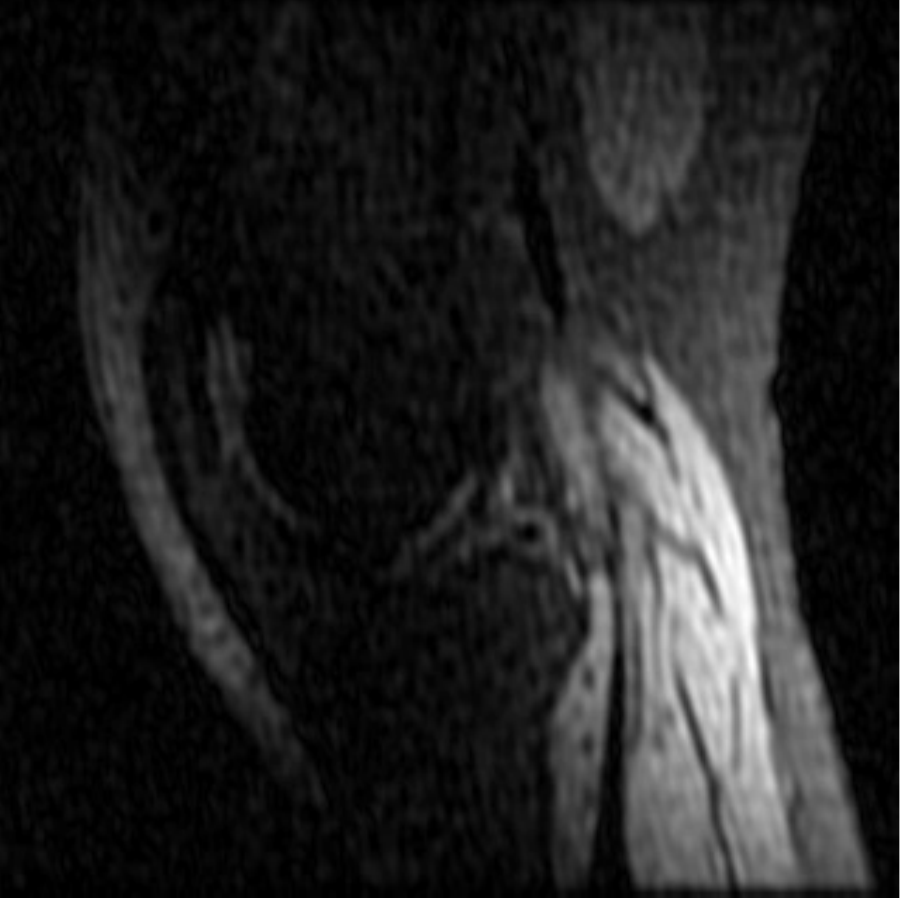} &
\hspace{-4mm}\includegraphics[width=.20\textwidth]{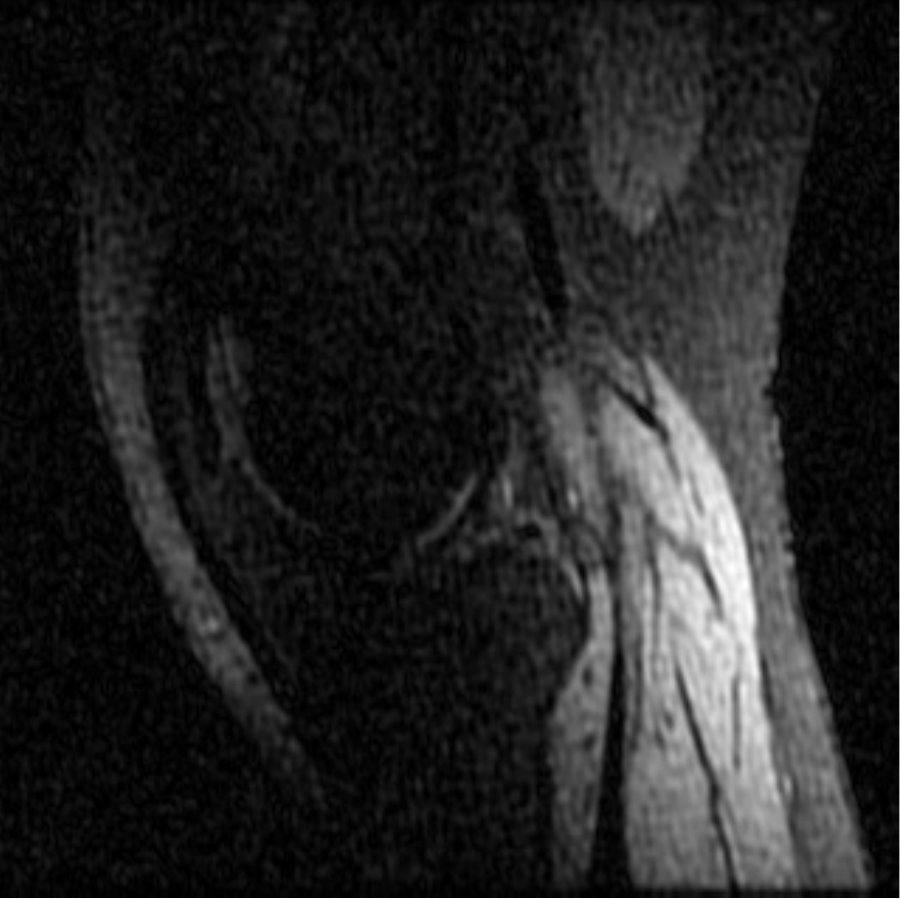} &
\hspace{-4mm}\includegraphics[width=.20\textwidth]{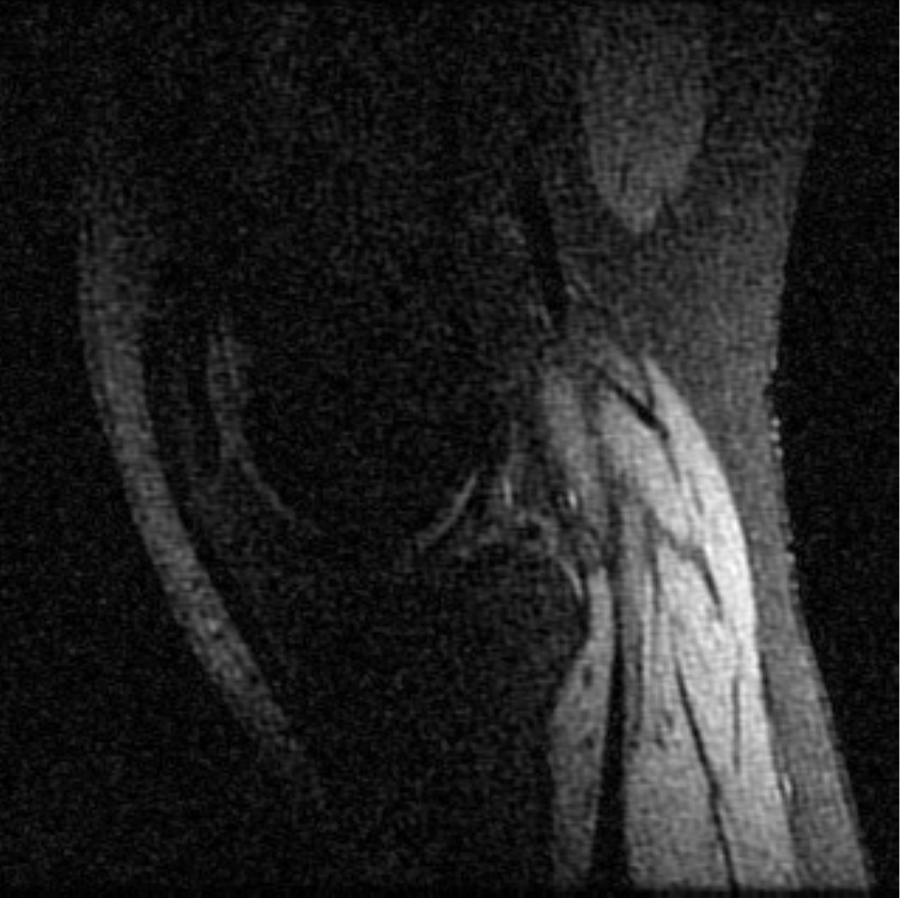} \\ [-1mm]
\text{\footnotesize{$f_{\text{min}}$ linear}} &
\hspace{-4mm}\includegraphics[width=.20\textwidth]{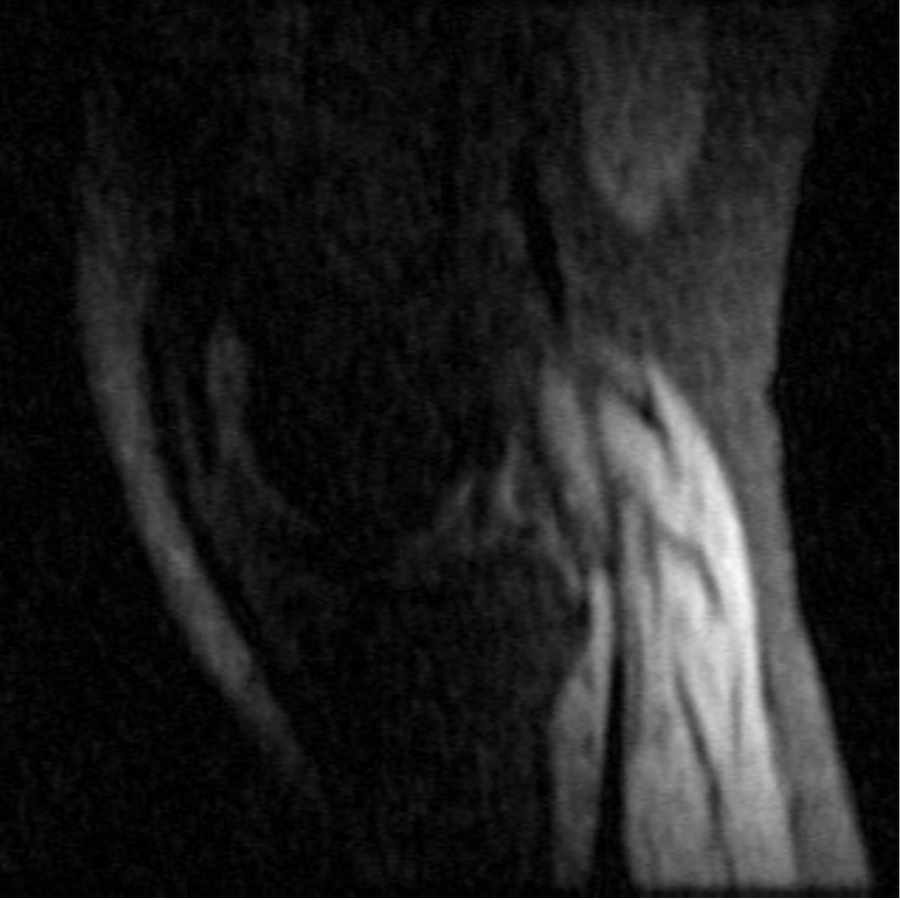} &
\hspace{-4mm}\includegraphics[width=.20\textwidth]{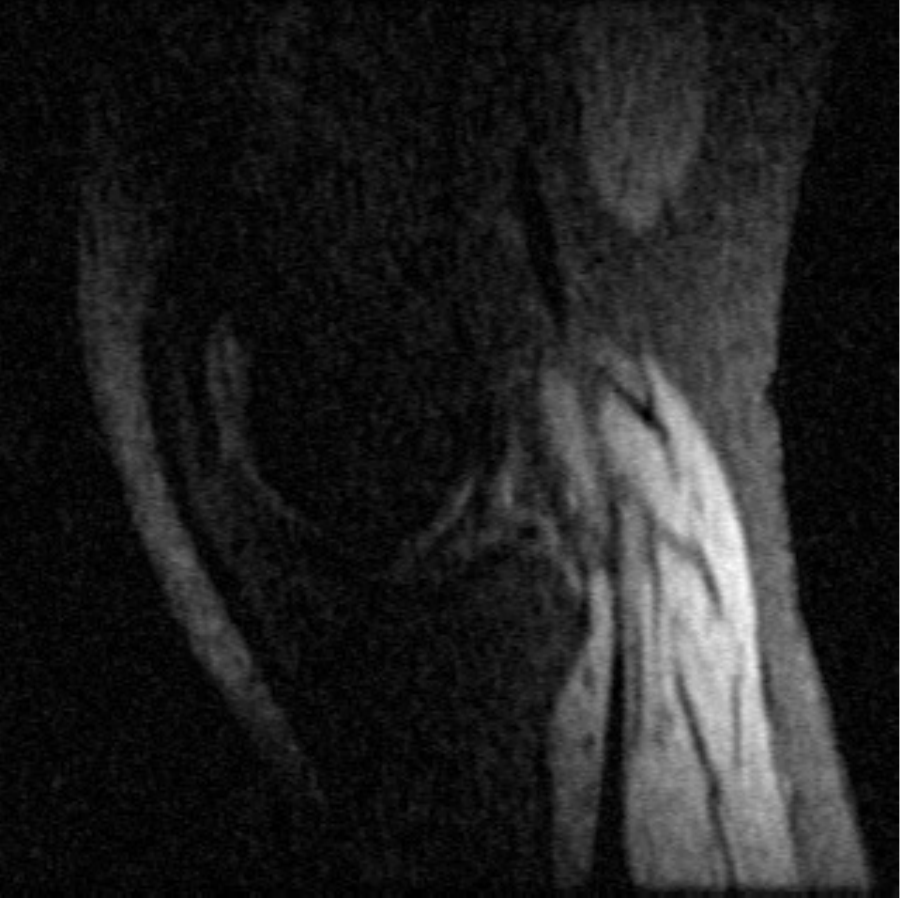} &
\hspace{-4mm}\includegraphics[width=.20\textwidth]{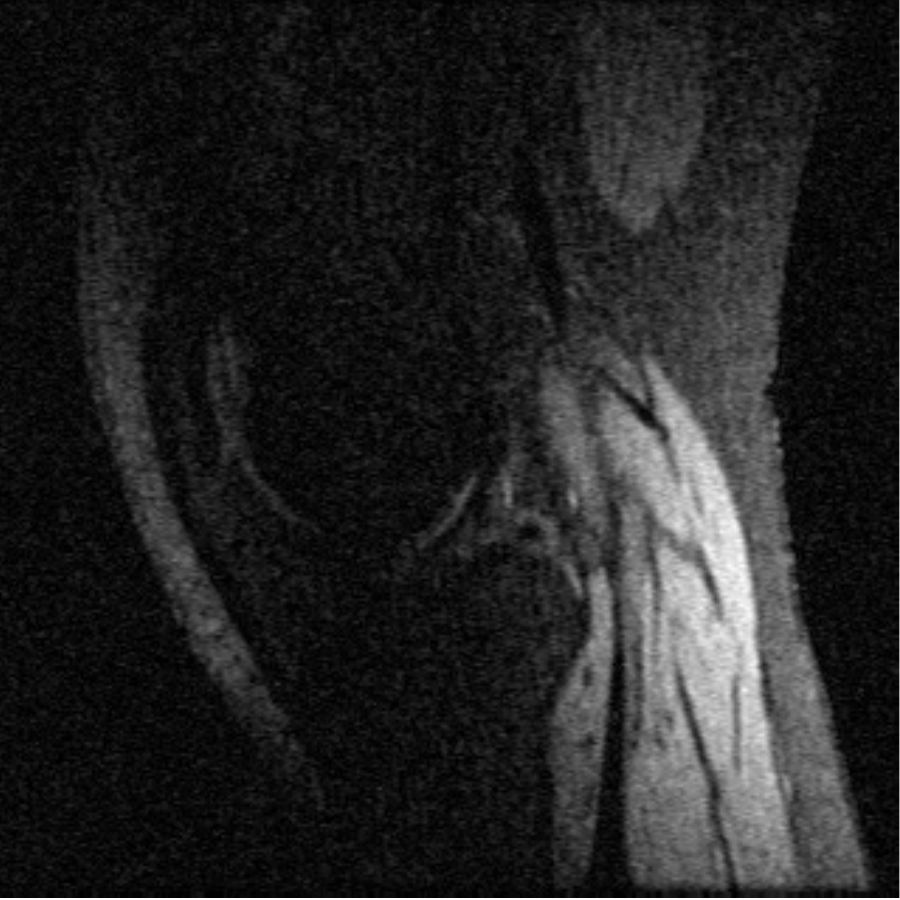} \\ 
& 6.25\% sampling & 12.5\% sampling & 25\% sampling \\ 
\end{tabular}
\caption{MRI  reconstruction example for the test patient 11.  The top-left image corresponds to the original (fully sampled) data, and the remainder of the first row corresponds to using the best $k$ coefficients in an image-adaptive fashion.}
\label{fig:MRI_recon}
\end{figure*}

Table \ref{tab:mri} illustrates the overall test performance of each approach, in addition to the error obtained by the best adaptive (i.e., image dependent) $n$-sample approximations with respect to the $k$-space basis. Based on these numbers, the learning-based approach slightly outperforms the randomized variable-density based approach of Roman \emph{et al.} \cite{roman2014asymptotic}, which in turn slightly outperforms that of Lustig \emph{et al.} \cite{lustig2007sparse}.  The best PSNR in each case is achieved by the indices corresponding to $\favg$; with this choice, even the linear decoder leads to an improvement over \cite{lustig2007sparse} and \cite{roman2014asymptotic} used with BP, while using our indices alongside BP provides a further improvement.  Finally, based on Figure \ref{fig:MRI_recon}, it appears that the improvement of our indices is actually more significant in the relevant parts of the image where the knee is observed, with finer details being seen at $6.25\%$ and $12.5\%$ sampling rates.
%[remove this sentence if it does not finish]The table also shows that the nonlinear $\ell_1$-decoder with the wavelet transform improves the results only marginally over the linear decoder. 

\begin{table}[!h]
\caption{MRI: Average $\ell_2$-errors and average PSNR}\label{tab:mri}
\centering
\begin{tiny}
\begin{tabular}{|c|cc|cc|cc|}	
\hline
\multirow{3}{*}{Indices} 	& \multicolumn{6}{c|}{Sampling rate} 		\\ \cline{2-7}
 					& \multicolumn{2}{c|}{$6.25\%$}	& \multicolumn{2}{c|}{$12.50\%$}	& \multicolumn{2}{c|}{$25\%$} \\ %\cline{2-7}
					& $\ell_2$ & PSNR & $\ell_2$ & PSNR & $\ell_2$ & PSNR \\ \hline
Adaptive				& $0.371$ & $25.299$ dB		& $0.328$ & $26.372$ dB 		& $0.261$ & $28.362$ dB\\ \hline						
$f_\mathrm{avg}$ linear 		& $0.399$ & $24.673$ dB 	& $0.377$ & $25.193$ dB 		& $0.339$ & $26.126$ dB 	\\
$f_\mathrm{gen}$ linear		& $0.404$ & $24.549$ dB 	& $0.386$ & $24.959$ dB 		& $0.352$ & $25.757$ dB \\
$f_\mathrm{min}$ linear	& $0.408$ & $24.462$ dB 	& $0.385$ & $24.969$ dB 		& $0.345$ & $25.950$ dB 	\\ \hline
Lustig et al. linear			& $0.404$ & $24.561$ dB		& $0.378$ & $25.148$ dB 			& $0.340$ & $26.095$ dB \\ \hline
%Knoll et al. linear			& $0.481$ & $21.070$ dB		& $0.447$ & $21.714$ dB 		& $0.399$ & $22.703$ dB \\ \hline
Roman et al. linear			& $0.401$ & $24.625$ dB		& $0.378$ & $25.161$ dB 		& $0.339$ & $26.123$ dB \\ \hline
$f_\mathrm{avg}$ BP	& $0.398$ & $24.699$ dB 	& $0.376$ & $25.210$ dB 		& $0.338$ & $26.145$ dB 	\\
$f_\mathrm{gen}$ BP	& $0.399$ & $24.689$ dB 	& $0.381$ & $25.094$ dB 		& $0.347$ & $25.901$ dB \\
$f_\mathrm{min}$ BP	& $0.402$ & $24.621$ dB 	& $0.380$ & $24.105$ dB 		& $0.342$ & $26.040$ dB 	\\ \hline
Lustig et al. BP			& $0.401$ & $24.640$ dB		& $0.378$ & $25.167$ dB 		& $0.341$ & $26.076$ dB \\ \hline
%Knoll et al. linear			& $0.481$ & $21.070$ dB		& $0.447$ & $21.714$ dB 		& $0.399$ & $22.703$ dB \\ \hline

Roman et al. BP			& $0.405 $ & $24.526$ dB		& $0.381$ & $25.080$ dB 		& $0.340$ & $26.085$ dB \\ \hline

\end{tabular}

\end{tiny}
\end{table}

%Visually, as shown in Figure \ref{fig:MRI_recon}, the details of the reconstructions for the test patient 11 are very similar. 
%It is clear that, without requiring to finely tune density parameters, our learning-based $f_\mathrm{avg}$ method equals the approach of \cite{lustig2007sparse}.
% However, the slight improvements in the numbers are actually accentuated when we look at the details of reconstructions, shown in Figure \ref{fig:MRI_recon}  for the test Patient \#11. We see that the learning-based reconstructions exhibit more details for $6.25\%$ and $12.5\%$ sampling rates.
